\documentclass[preprint]{aastex}
\include{psfig}

\newcommand\eg{{\it e.g.}}
\newcommand\etal{et~al.}
\newcommand\ie{{\it i.e.~}\ }
\newcommand\HI{\ion{H}{1}}

\newcommand\Lya{Ly$\alpha$}
\newcommand\OVIfull{\hbox{O~$\rm VI$}~$\lambda$~1035}
\newcommand\NVfull{\hbox{N~$\rm V$}~$\lambda$~1240}
\newcommand\CIVfull{\hbox{C~$\rm IV$}~$\lambda\lambda$~1549}
\newcommand\HeIIfull{\hbox{He~$\rm II$}~$\lambda$~1640}
\newcommand\CIIIfull{\hbox{C~$\rm III$]}~$\lambda$~1909}
\newcommand\CIIfull{\hbox{C~$\rm II$]}~$\lambda$~2326}
\newcommand\NeIVfull{\hbox{[Ne~$\rm IV$}]~$\lambda$~2424}
\newcommand\OIItfsz{\hbox{O~II}~$\lambda$~2470}
\newcommand\MgIIfull{\hbox{Mg~$\rm II$}~$\lambda$~2800}
\newcommand\OIIItzfs{\hbox{[O~$\rm III$]}~$\lambda$~3047}
\newcommand\OIIItott{\hbox{[O~$\rm III$]}~$\lambda$~3133}
\newcommand\HeIIttzt{\hbox{He~$\rm II$}~$\lambda$~3203}
\newcommand\NeVttfs{\hbox{[Ne~$\rm IV$]}~$\lambda$~3346}
\newcommand\NeVfull{\hbox{[Ne~$\rm IV$]}~$\lambda$~3426}
\newcommand\NeVtfts{\hbox{[Ne~$\rm IV$]}~$\lambda$~3426}
\newcommand\OIIfull{\hbox{[O~$\rm II$]}~$\lambda$~3727}
\newcommand\NeIIIfull{\hbox{[Ne~$\rm III$]}~$\lambda$~3869}
\newcommand\NeIIItesn{\hbox{[Ne~$\rm III$]}~$\lambda$~3869}
\newcommand\NeIIItnss{\hbox{[Ne~$\rm III$]}~$\lambda$~3967}
\newcommand\SIIfull{\hbox{S~$\rm II$}~$\lambda$~4071}
\newcommand\HeIIfses{\hbox{He~$\rm II$}~$\lambda$~4686}
\newcommand\OIIIfnfn{\hbox{[O~$\rm III$]}~$\lambda$~4959}
\newcommand\OIIIfull{\hbox{[O~$\rm III$]}~$\lambda$~5007}
\newcommand\NV{\hbox{N~$\rm V$}}

\newcommand\CIV{\hbox{C~$\rm IV$}}
\newcommand\CIII{\hbox{C~$\rm III]$}}
\newcommand\Ha{H$\alpha$}
\newcommand\Hbeta{H$\beta$}
\newcommand\Hgamma{H$\gamma$}
\newcommand\Hdelta{H$\delta$}
\newcommand\HeII{\hbox{He~$\rm II$}}

\newcommand\OII{[\hbox{O~$\rm II$}]}
\newcommand\OIII{[\hbox{O~$\rm III$}]}
\newcommand\OIV{\hbox{O~$\rm IV$}}
\newcommand\OVI{\hbox{O~$\rm VI$}}
\newcommand\SiIV{\hbox{Si~$\rm IV$}}

\def\spose#1{\hbox to 0pt{#1\hss}}
\def\simlt{\mathrel{\spose{\lower 3pt\hbox{$\mathchar"218$}}
     \raise 2.0pt\hbox{$\mathchar"13C$}}}
\def\simgt{\mathrel{\spose{\lower 3pt\hbox{$\mathchar"218$}}
     \raise 2.0pt\hbox{$\mathchar"13E$}}}
\newcommand\aasup{{A\&AS}}

\shorttitle{Spectroscopy of USS sources}
\shortauthors{De Breuck et al.}

\begin{document}

\title{Spectroscopy of Ultra Steep Spectrum Radio Sources}
\author{Carlos De Breuck\altaffilmark{1,2,3}, Wil van Breugel\altaffilmark{1}, Huub R\"ottgering\altaffilmark{2}, Daniel Stern\altaffilmark{4,5}, George Miley\altaffilmark{2},}
\author{Wim de Vries\altaffilmark{1}, S. A. Stanford\altaffilmark{1,6}, Jaron Kurk\altaffilmark{2} \& Roderik Overzier\altaffilmark{2}}
\email{debreuck@iap.fr, wil@igpp.ucllnl.org, rottgeri@strw.leidenuniv.nl, dan@bigz.berkeley.edu, miley@strw.leidenuniv.nl, wdevries@igpp.ucllnl.org, adam@igpp.ucllnl.org, kurk@strw.leidenuniv.nl, overzier@strw.leidenuniv.nl}

\altaffiltext{1}{Institute of Geophysics and Planetary Physics,
Lawrence Livermore National Laboratory, L$-$413, P.O. Box 808,
Livermore, CA 94550, U.S.A.}
\altaffiltext{2}{Leiden Observatory, P.O. Box 9513, 2300 RA Leiden, The Netherlands}
\altaffiltext{3}{Current address: Institut d'Astrophysique de Paris, 98bis Boulevard Arago, 75014 Paris, France}
\altaffiltext{4}{Astronomy Department, University of California at Berkeley, CA 94720, U.S.A.}
\altaffiltext{5}{Current address: Jet Propulsion Laboratory, California Institute of Technology, Mail Stop 169-327, Pasadena, CA 91109, U.S.A.}
\altaffiltext{6}{Physics Department, University of California, Davis, CA 95616, U.S.A.}

\begin{abstract}
We present optical spectroscopy of 62 objects selected from several samples of ultra steep spectrum (USS) radio sources. 46 of these are from our primary catalog, consisting of 669 sources with radio spectral indices $\alpha < -1.30$ ($S_\nu \propto \nu^\alpha$); this first spectroscopic sub-sample was selected on the basis of their faint optical and near-IR identifications. Most are identified as narrow-lined radio galaxies with redshifts ranging from $z=0.25$ to $z=5.19$. Ten objects are at $z>3$, nearly doubling the number of such sources known to date.
Four of the USS radio sources are identified with quasars, of which at least three have very red spectral energy distributions. The source TN~J0936$-$2242 is identified with an extremely red object (ERO, $R-K>5$); both it and a close companion are at $z=1.479$. The spectrum of the ERO closely resembles that of previously discovered radio galaxies at $z \sim 1.5$. Five sources show continuum emission, but fail to show any clear emission or absorption features, despite integrations of $\sim1$h with the Keck telescope. We suggest that these objects could be (i) radio galaxies with faint emission lines in the ``redshift desert'' at $1.5 \simlt z \simlt 2.3$, (ii) radio galaxies with an obscured AGN, which are dominated by a stellar continuum observed with insufficient S/N, or (iii) pulsars. Three radio sources identified with faint objects in the $K-$band images remain undetected in 50$-$90~min spectroscopic integrations with the Keck telescope, and are possible $z>7$ candidates.
\end{abstract}


\section{Introduction}
The host galaxies of powerful radio sources have been identified with massive galaxies at low and intermediate redshifts \citep[\eg][]{bes98}. The close correlation in the Hubble $K-z$ diagram suggests this is also true out to the highest redshifts \citep[\eg][]{eal97,wvb98}. By studying high redshift radio galaxies (HzRGs), we can thus observe some of the largest mass concentrations on galaxy scales in the early Universe, and put constraints on galaxy formation theories. HzRGs are amongst the most luminous observable objects at very high redshift. 
Although unification theories of radio galaxies and quasars \citep[\eg][]{bar89} imply that we can also use quasars for such studies, radio galaxies offer a much clearer picture of the host galaxy, because of the absence of appreciable contamination from the active galactic nucleus (AGN). In radio galaxies, the cone of bright emission originating from the AGN is believed to be directed away from the observer. In this sideway viewing angle, the AGN emission is presumed to be shielded by optically thick material. This ``natural coronograph'' allows the study of the stellar populations in some HzRGs \citep[\eg][]{spi97,dey97,fos00} and investigations of the morphology of forming massive galaxies unhindered by issues of dynamic range \citep[\eg][]{pen99,pen00}. 

A major advantage of radio galaxies over the large population of Lyman-break galaxies \citep[\eg][]{stei99} is that they are not selected on the basis of their optical properties, but on the basis of radio emission. 
In principle, a radio-selected survey with subsequent near-IR identifications can find galaxies out to $z\simeq7$ (optical spectroscopy) or $z\simeq 16$ (near-IR spectroscopy), un-restricted to specific pass-bands, which limit optical color selection techniques.
Such a continuous redshift range is important to select targets for follow-up observations that depend on atmospheric pass-bands in the near-IR or interference-free frequency ranges in the radio regime. For example, searches for associated \HI\ 21cm line absorption are often hampered by local interference \citep{rot99}, and are only possible in limited redshift ranges.
The absence of any {\it a priori} optical color selection also allows the discovery of objects with anomalous UV/optical spectral energy distributions, and because the radio emission is insensitive to dust absorption, radio-selected samples can provide an unbiased view with respect to the dust properties at high redshift. For example, the importance of dust at high redshift was recently shown by sub-mm observations of a large sample of HzRGs \citep{arc00}, which suggest an increased detection rate at $z>3$, even after $k-$correction effects are taken into account.

Previous searches for HzRGs, using 3-4m class telescopes have been difficult because the galaxies are faint ($R \simgt 24, K \simgt 20$), and fewer than 20 radio galaxies with $z>3$ have been reported in the literature. However, with the availability of modern 8-10m class telescopes, it is now feasible to begin a systematic search for HzRGs at much fainter magnitudes than before.
To start such a program, we constructed a sample of ultra steep spectrum (USS) radio sources \citep[][hereafter paper~I]{deb00a}. Such sources have been used for two decades to find HzRGs. This technique mainly makes use of a ``radio $k-$correction'' effect: at higher redshift, an increasingly steeper part of the generally concave radio spectrum shifts to the fixed observing frequencies of the large radio surveys. 
Previous samples of USS sources have shown the efficiency of this technique in finding $z>2$ radio galaxies: half of the $\alpha < -1.0$ ($S \propto \nu^{\alpha}$) sources are at $z>2$ \citep{cha96,rot97,blu98}. In our USS sample, which is based on several deep, large radio surveys that have recently become available, we select sources with spectral indices $\alpha < -1.30$ in order to enlarge the sample of $z>3$ objects. 

Our sample partially overlaps with the one of \citet{kap00a}, who use a spectral index cutoff of $\alpha < -1.50$, also based on the 365~MHz Texas \citep{dou96} and 1.4~GHz NVSS \citep{con98} radio surveys, to study the distribution of extremely steep spectrum sources. Their sample emphasizes the contribution of pulsars, whose median spectral index is $\sim -1.6$ \citep{lor95}, and could represent a significant fraction of the steepest spectrum radio sources. In our sample, we have excluded the Galactic plane region ($|b|<15\arcdeg$) in order to avoid large amounts of Galactic extinction; this should also reduce the number of pulsars in our sample.

Our observational campaign to find HzRGs from our sample of USS sources consists of three parts: (i) high resolution radio imaging (paper~I\nocite{deb00a}) to select sources with either compact or FR~II \cite{far74} morphologies, and determine radio positions accurate to within $\sim1$\arcsec; (ii) optical and near-IR imaging (De~Breuck \etal, in preparation; hereafter paper~II) to identify the host galaxies of the radio sources; and (iii) optical spectroscopy, as described in this paper. 
The organization of this paper is as follows. In \S2, we present the spectroscopic sample and describe both the observations and data reduction. The results comprise \S3. In \S4, we discuss the redshift distribution of the USS sources, the small population of USS quasars, and speculate on the nature of the sources lacking redshift determination. We summarize our results in \S5. We defer a statistical analysis of the emission line regions in HzRGs to a companion paper \citep{deb00b}. Throughout, we assume $H_0=50$~km~s$^{-1}$Mpc$^{-1}$, $q_0$=0.5, and $\Lambda=0$.

\section{Observations and Data Reduction}
\subsection{Target selection}
We obtained the spectra presented in this paper between 1996 April and 2000 April. At the start of this spectroscopic campaign, our main USS sample was not finalized, as the final versions of the WENSS, Texas, NVSS and FIRST radio source catalogs were not yet published (see paper~I\nocite{deb00a} for a description of these catalogs and our final USS sample, consisting of 669 sources). During this period, we observed objects from two other USS samples, as well as six sources that were excluded from our well-defined ($\alpha < -1.30$) USS sample when the revised flux densities in the radio survey catalogs were published. We refer to these sources using the prefix ``WNR'' and ``TNR'', for sources removed from our WN and TN samples, respectively. We provide the final radio source parameters of these six objects in table \ref{rejwntn}. The entries in this table are similar to those for the sources in paper~I\nocite{deb00a}. In the appendix, we show VLA radio maps of three objects listed in table \ref{rejwntn}. Radio maps from the other three WNR objects can be obtained from the FIRST survey \citep{bec95}.
\begin{table}[t]
\scriptsize
\caption{WNR and TNR radio parameters}
\begin{tabular}{rrrrrrr}
\hline
Name & $S_{low}~^a$ & $S_{1400}$ & $S_{4850}$ & $\alpha_{325}^{1400}$ & $\alpha_{1400}^{4850}$  \\
 & mJy & mJy & mJy &  & \\
\hline
TNR J0516$+$0637 &   269$\pm$  27 &   48.9$\pm$ 1.1 & $  8.60\pm0.26$ & $-1.27\pm$0.08 &  $-1.40\pm0.06$ & \\
WNR J1338$+$3532 &   723$\pm$  29 &  131.0$\pm$ 5.7 &         \nodata & $-0.71\pm$0.18 &         \nodata & \\
WNR J1356$+$3929 &  1440$\pm$  57 &  223.2$\pm$ 9.2 & \nodata         & $-1.28\pm$0.04 &  \nodata        & \\
TNR J1402$-$1510 &  2862$\pm$  53 &  795  $\pm$ 34  & $216.7 \pm 4.8$ & $-0.95\pm$0.04 &  $-1.05\pm0.04$ & \\
WNR J1703$+$3739 &    31$\pm$   6 &    8.3$\pm$ 0.4 & \nodata         & $-0.90\pm$0.11 &  \nodata        & \\
TNR J2254$+$1857 &   175$\pm$  21 &   30.7$\pm$ 1.7 & $  7.2 \pm 0.2$ & $-1.29\pm$0.09 &  $-1.16\pm0.04$ & \\
\hline
\hline
Name & Str$^b$ & LAS & PA & RA & Dec & Pos$^c$ \\
 &  & \arcsec & \arcdeg & J2000 & J2000 & \\
\hline
TNR J0516$+$0637 &   D     &  1.3    &  22     & 05 16 16.22 & $+$06 37 18.0 & V \\
WNR J1338$+$3532 &   D     & 11.6    & 159     & 13 38 15.09 & $+$35 32 03.4 & F \\
WNR J1356$+$3929 &   S     &  1.6    & 164     & 13 56 15.24 & $+$39 29 43.7 & F \\
TNR J1402$-$1510 &   D     & 15.9    &  80     & 14 02 08.66 & $-$15 10 09.5 & V \\
WNR J1703$+$3739 &   T     & 30.1    &   3     & 17 03 32.31 & $+$37 39 00.8 & F \\
TNR J2254$+$1857 &   D     &  2.7    &  36     & 22 54 53.71 & $+$18 57 04.4 & V \\
\hline
\multicolumn{7}{l}{$^a$ $S_{325}$ for WNR sources, $S_{365}$ for TNR sources.} \\
\multicolumn{7}{l}{$^b$ Radio structure code: S=single, D=double, M=multiple morphology.} \\
\multicolumn{7}{l}{$^c$ V=VLA observations, F=FIRST survey.} \\
\end{tabular}
\label{rejwntn}
\end{table}

\begin{table}[t]
\caption{Sources from the \citet{nel95} USS sample}
\footnotesize
\begin{tabular}{rrrrrrrrr}
\hline
Name & RA & Dec & $S_{4860}~^a$ & $\alpha_{365}^{4860}$ & Str$^b$ & LAS & PA & $z$ \\
 & J2000 & J2000 & mJy & & & \arcsec & \arcdeg & \\
\hline
TXS J0137$+$2521 & 01 37 06.81 & $+$25 21 19.2 & 33 & $-1.29$ & D &  6.9 &  82 & 2.896 \\
TXS J1650$+$0955 & 16 50 04.83 & $+$09 55 04.8 & 39 & $-1.25$ & D & 19.4 &  56 & 2.509 \\
TXS J1908$+$7220 & 19 08 23.70 & $+$72 20 11.8 & 42 & $-1.28$ & T & 14.4 &  14 & 3.542 \\
TXS J2036$+$0256 & 20 36 34.78 & $+$02 56 54.8 & 54 & $-1.37$ & D &  3.1 &  34 & 2.129 \\
TXS J2321$+$2237 & 23 21 42.28 & $+$22 37 54.6 & 40 & $-1.28$ & D &  8.1 &  24 & 2.555 \\
TXS J2334$+$1545 & 23 34 58.30 & $+$15 45 51.4 & 25 & $-1.38$ & D &  6.4 & 133 & 2.481 \\
TXS J2355$-$0002 & 23 55 35.57 & $-$00 02 46.6 & 58 & $-1.26$ & T & 33.8 & 127 & 2.590 \\
\hline
\multicolumn{9}{l}{$^a$ Determined from VLA observations, except for TXS J2334$+$1545 and TXS J2351$+$1034} \\
\multicolumn{9}{l}{$\quad$which were obtained from \cite{bec91}.}\\
\multicolumn{9}{l}{$^b$ D=double, T=triple morphology.}\\
\end{tabular} 
\label{kirsten}
\end{table}

We selected seven sources from a sample of $\alpha_{365}^{4850} < -1.25$ sources constructed by \citet{nel95} from the 365~MHz Texas \citep{dou96} and 4.85~GHz Greenbank \citep{bec91} surveys. In table \ref{kirsten}, we reproduce from \citet{nel95} the J2000 coordinates, 4.86~GHz flux densities, spectral indices, angular sizes and position angles of the radio sources we observed spectroscopically. For completeness, we include the previously published Keck spectrum of TXS~J1908$+$7220 \citep{dey99b}
Two sources (USS~J0006$-$0141 and PKS~J1428$-$1502) are USS sources from the sample of \citet{rot94}. We refer to that paper for details on the radio source parameters. In addition, we also present a Keck spectrum of the $z=4.41$ radio galaxy 6C~J0143$+$3253 \citep{raw96}, showing the \CIVfull\ line confirming the redshift.

All remaining sources are part of our main sample of 669 USS sources, described in paper~I\nocite{deb00a}. Here, we report spectroscopic results for 46 objects from this sample. We selected these objects mainly on the basis of observational constraints, preferentially selecting targets with the faintest $K-$band magnitudes, where available. We stress that the spectroscopic sub-sample presented in this paper is not necessarily representative of the entire sample of 669 sources, although we did select objects with a reasonably large spread in radio flux densities.  For completeness, we include the data of three previously published objects from our USS sample, viz. the $z=1.462$ red quasar WN~J0717+4611 \citep{deb98}, the $z=4.11$ radio galaxy TN~J1338$-$1942 \citep{deb99} and the $z=5.19$ radio galaxy TN~J0924$-$2201 \citep{wvb99}. We call the USS sample described in paper~I our 'main' USS sample, while the 16 other sources originate from the commonly referred 'subsidiary' samples.
\begin{table}[t]
\begin{center}
\caption{Spectroscopy Observing Runs}
\small
\begin{tabular}{llrrc}
\hline
Session & UT Date & Telescope & Instrument & photometric? \\
\hline
K1 & 1995 Sep 27    & Keck 10m & LRIS & yes \\
L4 & 1996 Apr 21    & Lick 3m & Kast & yes \\
W1 & 1996 Jul 11--12 & WHT 4.2m & ISIS & yes \\
E1 & 1996 Jul 15--16 & ESO 3.6m & EFOSC1 & yes \\
L6 & 1996 Nov 11    & Lick 3m & Kast & yes \\ 
K2 & 1996 Nov 21    & Keck 10m & LRIS & yes \\
E2 & 1997 Mar 5-6   & ESO 3.6m & EFOSC1 & yes \\
E3 & 1997 Apr 9--11  & ESO 3.6m & EFOSC1 & yes \\
K5 & 1997 Dec 2--3   & Keck 10m & LRIS & no \\
K6 & 1998 Mar 7     & Keck 10m & LRIS & yes \\
L15 & 1998 Oct 18--19 & Lick 3m & Kast & yes \\
K9 & 1998 Dec 19--20 & Keck 10m & LRIS & yes \\
V1 & 1999 Apr 19 & VLT 8.2m & FORS1 & yes \\
K11 & 1999 Jul 12 & Keck 10m & LRIS & yes \\
K14 & 2000 Feb 1  & Keck 10m & LRIS & yes \\
V2 & 2000 Apr 8--9 & VLT 8.2m & FORS1 & no? \\
K15 & 2000 May 5 & Keck 10m & ESI & yes \\
\hline
\end{tabular}
\label{specobs}
\end{center}
\end{table}

\begin{table}[ht]
\vspace{-0.5cm}
\caption{Observational setup}
\small
\begin{tabular}{llcccr}
\hline
Telescope & Instrument & Grism/grating & $\lambda$ coverage  & Slit width & $\lambda$ resolution \\
\hline
Keck     & LRIS   & 300 $\ell$ mm$^{-1}$ & 4100$-$ 8800~\AA & 1\farcs0 & 10~\AA \\
         &        & 150 $\ell$ mm$^{-1}$ & 4100$-$10000~\AA & 1\farcs5 & 15~\AA \\
\\
VLT Antu & FORS1  & 600R                 & 4350$-$ 7250~\AA & 1\farcs3 & 5.5~\AA \\
         &        & 300V                 & 3600$-$ 8950~\AA & 1\farcs0 & 12~\AA \\
         &        & 300I                 & 6350$-$10800~\AA & 1\farcs0 & 11~\AA \\
\\
WHT      & ISIS   & R158B                & 3650$-$ 6000~\AA & 2\farcs5 & 16~\AA \\
         &        & R158R                & 6000$-$ 8950~\AA & 2\farcs5 & 15~\AA \\
\\
ESO 3.6m & EFOSC1 & B300                 & 3700$-$ 7000~\AA & 2\farcs5 & 20~\AA \\
         &        & R300                 & 6000$-$ 9900~\AA & 2\farcs5 & 30~\AA \\
\\
Lick 3m  & Kast   & 452/3306             & 3100$-$ 6150~\AA & 2\farcs0 &  6~\AA \\
         &        & 300/7500             & 5200$-$10600~\AA & 2\farcs0 &  8~\AA \\
\hline
\end{tabular}
\label{specsetup}
\end{table}

\subsection{Observations}
We obtained our spectra at five different observatories. We give an overview of the different observing sessions in table \ref{specobs}, an overview of the instrumental setups used in table \ref{specsetup}, and a journal of the observations in table \ref{specjournal}. We use a consecutive numbering scheme for each observing session, counting also the sessions where we did not obtain any data, or only imaging data (see paper~II). The $R-$ and $K-$band magnitudes listed in table \ref{specobs} are measured in a 4\arcsec\ diameter aperture are reproduced from paper~II, to which we refer for details on the photometry. 

To acquire the faint ($20\simlt R \simlt 25$) identifications of the radio galaxies in the slit of the spectrograph, we followed the same procedure at each observatory: we first centered the slit on a star visible on the Palomar Observatory Sky Survey ($R\lesssim 20$), followed by an offset to the position of the identification of the radio source, which we determined from the optical or $K-$band images presented in paper~II. For some of the faintest objects observed at Keck, we first acquired a fainter offset star that was within the same small field $K-$band image as the identification, before performing a final offset to the object.
At the VLT, we used the FIMS software on the previously obtained $K-$band images to position the identification into the slit.
We list the position angle used for each object in table \ref{specobs}. For objects with resolved radio morphologies, we oriented the slit in the direction of the radio jets. For the other objects, we either aligned the slit through possible companion objects, or used the parallactic angle \citep{fil82}.

Between exposures, we shifted the object by 10\arcsec--20\arcsec\ along the slit to facilitate removal of fringing in the reddest parts of the CCD. During the second exposure, we performed a quick data-reduction of the first exposure, and if this yielded an unambiguous redshift, we shortened the second exposure to save observing time.
We observed two or more standard stars during each night for spectrophotometric calibration, and wavelength calibration lamps either at the beginning and end of the night, or during the night at the position of the object to correct for small gravitational distortion factors on the CCD.
\begin{table}[!ht]
\vspace{-0.6cm}
\caption{Journal of the Spectroscopy Observations}
\tiny
\begin{tabular}{rlcccccrrc}
\hline
Source & $z$ & $m_R$ & $m_K$ & $\alpha_{radio}$ & Sample$^a$ & Session & $t_{exp}$ & Slit PA & Extraction \\
 &  & (mag) & (mag) &  &  &  & (s) &  & width \\
\hline
USS J0006$-$0141 & 1.541  & 23      & \nodata & $-$1.36 & R\"ot  & E1  & 2700   &  46\arcdeg & 6\farcs4 \\
WN  J0040$+$3857 & 2.6055 & $>$24.5 & \nodata & $-$1.36 &    DB  & K9  & 1800   &  42\arcdeg & 2\farcs1 \\
TN  J0121$+$1320 & 3.516  & $>$24   & 19.0    & $-$1.34 &    DB  & K5  & 3080   & 140\arcdeg & 2\farcs3 \\
TXS J0137$+$2521 & 2.8970 & \nodata & \nodata & $-$1.29 &   Nel  & W1  & 3600   &  82\arcdeg & 3\farcs2 \\
6C  J0143$+$3253 & 4.41   & \nodata & \nodata & $-$1.15 &   Raw  & K1  & 2700   & 137\arcdeg & 2\farcs1 \\
TN  J0205$+$2242 & 3.5061 & $>$24   & 19.0    & $-$1.37 &    DB  & K5  & 2400   & 156\arcdeg & 3\farcs8 \\
WN  J0231$+$3600 & 3.079  & 25.0    & \nodata & $-$1.30 &    DB  & K9  & 1800   & 107\arcdeg & 2\farcs1 \\
WN  J0303$+$3733 & 2.506  & 23.2    & \nodata & $-$1.46 &    DB  & K9  & 1200   & 171\arcdeg & 3\farcs4 \\
WN  J0310$+$3644 & \nodata& 23.0    & \nodata & $-$1.70 &    DB  & K9  & 3600   & 249\arcdeg & 1\farcs7 \\
TN  J0402$+$1007 & \nodata& $>$24   & 17.2    & $-$1.38 &    DB  & K9  & 1200   &  67\arcdeg & 1\farcs7 \\
TN  J0452$-$1737 & 2.256  & $>$23   & 18.8    & $-$1.37 &    DB  & K9  & 1800   & 176\arcdeg & 3\farcs8 \\
TNR J0516$+$0637 & 0.3568 & \nodata & \nodata & $-$1.27 & USSrem & E2  & 2700   &  22\arcdeg & 6\farcs1 \\
TN  J0517$-$0641 & \nodata& 24.0    & \nodata & $-$1.46 &    DB  & E2  & 2700   & 132\arcdeg & \nodata \\
WN  J0617$+$5012 & 3.153  & $>$24   & 20.4    & $-$1.37 &    DB  & K9  & 5400   &  10\arcdeg & 8\farcs5 \\
WN  J0717$+$4611 & 1.462  & 21.4    & 16.8    & $-$1.30 &    DB  & K2  & 3600   &  17\arcdeg & 2\farcs7 \\
WN  J0747$+$3654 & 2.992  & $>$23   & 20.5    & $-$1.41 &    DB  & K14 & 5400   & 129\arcdeg & 1\farcs5 \\
WN  J0813$+$4828 & 1.274  & 23.9    & 19.0    & $-$1.42 &    DB  & K9  & 3600   & 132\arcdeg & 4\farcs2 \\
TN  J0910$-$2228 & \nodata& 23.0    & \nodata & $-$1.58 &    DB  & E3  & 1800   & 180\arcdeg & 4\farcs9 \\
TN  J0920$-$0711 & 2.760  & 22.4    & \nodata & $-$1.51 &    DB  & E3  &  900   & 135\arcdeg & 4\farcs9 \\
TN  J0924$-$2201 & 5.19   & $>$24   & 21.7    & $-$1.63 &    DB  & K9  & 10000  &   0\arcdeg & 1\farcs5 \\
TN  J0936$-$2243 & 1.479  & $>$24   & 18.3    & $-$1.45 &    DB  & K9  & 3600   & 168\arcdeg & 1\farcs7 \\
TN  J0941$-$1628 & 1.644  & 23.0    & \nodata & $-$1.33 &    DB  & E2  & 2700   &   3\arcdeg & 4\farcs7 \\
WN  J1015$+$3038 & 0.54:  & 21.4    & 17.8    & $-$1.38 &    DB  & K9  & 1800   & 121\arcdeg & 2\farcs5 \\
TN  J1026$-$2116 & \nodata& $>$24   & 19.8    & $-$1.39 &    DB  & K9  & 3600   &   0\arcdeg & \nodata \\
TN  J1033$-$1339 & 2.425  & $>$24   & 19.2    & $-$1.38 &    DB  & E3  & 2700   & 287\arcdeg & 4\farcs3 \\
TN  J1102$-$1651 & 2.111  & 23.7    & \nodata & $-$1.35 &    DB  & E2  & 2700   &  71\arcdeg & 5\farcs3 \\
TN  J1112$-$2948 & 3.09   & 21.5    & \nodata & $-$1.40 &    DB  & E2  & 2700   & 299\arcdeg & 3\farcs8 \\
WN  J1115$+$5016 & 2.54   & $>$24   & 19.5    & $-$1.32 &    DB  & K14 & 2700   &  30\arcdeg & 0\farcs8 \\
TN  J1123$-$2154 & 4.109  & $>$24.5 & 20.3    & $-$1.55 &    DB  & K14 & 2400   &  79\arcdeg & 1\farcs3 \\
WN  J1123$+$3141 & 3.2174 & 21.3    & 17.8    & $-$1.46 &    DB  & L4  & 1800   &  90\arcdeg & 4\farcs7 \\
TN  J1151$-$3013 & \nodata& $>$24   & 19.8    & $-$1.57 &    DB  & E3  & 2700   & 180\arcdeg & \nodata \\
TN  J1159$-$1629 & \nodata& 23.4    & \nodata & $-$1.35 &    DB  & E3  & 3600   & 215\arcdeg & \nodata \\
WN  J1242$+$3915 & 2.131  & 23.6    & 18,6    & $-$1.33 &    DB  & K14 & 1800   &  51\arcdeg & 1\farcs5 \\
WN  J1314$+$3649 & \nodata& $>$26   & 22.1    & $-$1.41 &    DB  & K11 & 3600   & 126\arcdeg & 2\farcs1 \\
WN  J1333$+$3037 & 1.2127 & 22.5    & \nodata & $-$2.04 &    DB  & K6  &   900  &   0\arcdeg & 2\farcs4 \\
WNR J1338$+$3532 & 2.769  & \nodata & \nodata & $-$0.71 & USSrem & W1  & 3600   & 340\arcdeg & 3\farcs5 \\
TN  J1338$-$1941 & 4.11   & 23.0    & 20.0    & $-$1.31 &    DB  & E3,V1 & 2800 & 210\arcdeg & 4\farcs0 \\
WNR J1356$+$3929 & 0.2525 & \nodata & \nodata & $-$1.28 & USSrem & W1  & 3600   & 156\arcdeg & 2\farcs6 \\
TNR J1402$-$1510 & 0.739  & \nodata & \nodata & $-$0.95 & USSrem & E3  & 1800   & 260\arcdeg & 7\farcs8 \\
PKS J1428$-$1502 & 2.349  & 22      & \nodata & $-$1.18 & R\"ot  & E1  & 2700   & 132\arcdeg & 9\farcs2 \\
WN  J1525$+$3010 & \nodata& 25.3    & 19.5    & $-$1.46 &    DB  & K11 & 5400   &  90\arcdeg & \nodata \\
WN  J1543$+$3512 & 0.7033 & 21.2    & \nodata & $-$1.56 &    DB  & K6  &  600   &   0\arcdeg & 3\farcs3 \\
TXS J1650$+$0955 & 2.510  & 23.5    & \nodata & $-$1.25 &   Nel  & E1  & 2700   &  56\arcdeg & 5\farcs8 \\
WNR J1703$+$3739 & 0.2561 & \nodata & \nodata & $-$0.90 & USSrem & W1  & 1800   & 183\arcdeg & 2\farcs6 \\
MP  J1755$-$6916 & 2.551  & \nodata & 18.3    & $-$1.25 &    DB  & V2  & 1800   &  84\arcdeg & 1\farcs2 \\
MP  J1758$-$6738 & 2.026  & \nodata & 18.3    & $-$1.33 &    DB  & V2  & 1800   & 237\arcdeg & 2\farcs0 \\
WN  J1836$+$5210 & \nodata& \nodata & 17.7    & $-$1.41 &    DB  & K11 & 1200   &   0\arcdeg & 2\farcs1 \\
6C  J1908$+$7220 & 3.536  & 21.2    & 16.5    & $-$1.28 &   Nel  & K1  & 3600   &  90\arcdeg & 2\farcs0 \\
WN  J1911$+$6342 & 3.590  & \nodata & 19.9    & $-$1.42 &    DB  & K11 & 2700   &   0\arcdeg & 2\farcs1 \\
WN  J1917$+$6635 & \nodata& \nodata & 20.0    & $-$1.30 &    DB  & K11 & 3600   &   0\arcdeg & 2\farcs1 \\
MP  J1929$-$3732 & 0.7476 & \nodata & 17.1    & $-$1.29 &    DB  & V2  & 1400   & 352\arcdeg & 3\farcs2 \\
TN  J1954$-$1207 & \nodata& \nodata & 19.9    & $-$1.38 &    DB  & K11 & 3600   &  36\arcdeg & 3\farcs2 \\
MP  J2003$-$8340 & 1.169  & \nodata & 18.4    & $-$1.28 &    DB  & V2  & 1200   &  99\arcdeg & 1\farcs6 \\
TXS J2036$+$0256 & 2.130  & 24      & \nodata & $-$1.37 &   Nel  & E1  & 2700   &  33\arcdeg & 5\farcs2 \\
WN  J2044$+$7044 & \nodata& $>$24   & 19.2    & $-$1.38 &    DB  & K11 & 3000   &  90\arcdeg & \nodata \\
MP  J2045$-$6018 & 1.464  & \nodata & 18.0    & $-$1.30 &    DB  & V2  & 3600   &  37\arcdeg & 4\farcs4 \\
MP  J2048$-$5750 & 1.262  & \nodata & 17.2    & $-$1.22 &    DB  & V2  & 1200   &  32\arcdeg & 2\farcs8 \\
TNR J2254$+$1857 & 2.153  & \nodata & \nodata & $-$1.29 & USSrem & L6  & 3600   &  36\arcdeg & 3\farcs8\\
WN  J2313$+$4053 & 2.99   & \nodata & \nodata & $-$1.50 &    DB  & L15 & 3600   &  90\arcdeg & 6\farcs3 \\
TXS J2321$+$2237 & 2.553  & \nodata & \nodata & $-$1.28 &   Nel  & W1  & 3600   &  27\arcdeg & 6\farcs2 \\
TXS J2334$+$1545 & 2.480  & 23      & \nodata & $-$1.38 &   Nel  & E1  & 2700   & 133\arcdeg & 4\farcs7 \\
TXS J2355$-$0002 & 2.587  & 23      & \nodata & $-$1.26 &   Nel  & E1  & 4500   & 127\arcdeg & 4\farcs8 \\
\hline
\multicolumn{10}{l}{$^a$ DB=Chapter~2; R\"ot=\citet{rot94}; Nel=\citet{nel95}; Raw=\citet{raw96};} \\
\multicolumn{10}{l}{$\quad$USSrem=USS removed sources, this chapter.} \\
\multicolumn{10}{l}{NOTE: Colons denote uncertain redshifts.} \\
\end{tabular}
\vspace{2cm}
\label{specjournal}
\end{table}

\subsubsection{Lick 3m}
We used the Kast double-beam imaging spectrograph \citep{mil94} at the Lick 3m telescope on Mount Hamilton to obtain spectra of objects with $R\lesssim 22$. We used the d55 dichroic to split the light at 5500~\AA. Because the seeing at Mount Hamilton is mostly poor, we consistently used a 2\arcsec\ slit. The main advantage of using Kast for redshift determinations is the large continuous wavelength coverage from $\sim$3150~\AA\ to $\sim$10,000~\AA.

\subsubsection{WHT 4.2m}
During our observing run at the William Herschel 4.2m Telescope at La Palma, we used the ISIS double beam spectrograph \citep{car94} with a dichroic splitting the light at 6100~\AA\ and used a 2\farcs5 wide slit. In the blue arm, we used a recently commissioned LORAL1 CCD, which experienced problems with the flat-field exposures and disk storage, and consequentially lost most of the data at $\lambda < 6000$~\AA. In the red arm, we used a TEK5 CCD and were not subject to these problems. 

\subsubsection{ESO 3.6m}
To extend our spectroscopic sample into the southern hemisphere, we used the ESO faint object spectrograph and camera \citep[EFOSC1;][]{sav97} on the ESO 3.6m telescope at La Silla.  We first observed each object with the B300 grating, covering $\sim$3700~\AA\ to $\sim$7000~\AA\, and re-observed the best high redshift candidates with the R300 grating, which covers $\sim$6000~\AA\ to $\sim$9900~\AA.

\subsubsection{VLT 8.2m}
On two occasions, we used the FORS1 spectrograph \citep{app97} on the ESO VLT Antu telescope. In 1999 April, we obtained a medium deep spectrum covering the \Lya\ line and surrounding wavelength range in TN~J1338$-$1942 \citep{deb99}. In 2000 April, we observed six very southern USS sources as additional targets after the main target of the observing run reached high airmass (the main goal of this run was to detect cluster galaxies around HzRGs, see Kurk \etal\ 2000 \nocite{kur00}).  

\subsubsection{Keck 10m}
To obtain deeper spectra of those objects that failed to yield redshifts on the 3-4m telescopes described above, we first obtained deep K-band images with the near infrared camera \citep[NIRC;][]{mat94} on the Keck~I telescope (see paper~II for details). We then determined accurate offsets from these images for deeper spectroscopy with the Low Resolution Imaging Spectrometer \citep[LRIS;][]{oke95} at the Keck~II telescope on Mauna Kea. During all runs prior to 1998 December, we used the 300 $\ell$ mm$^{-1}$ grating blazed at 5000~\AA\ combined with the 1\arcsec\ wide slit, which results in a resolution of $\sim$ 10~\AA\ (FWHM), with a wavelength coverage from $\sim$4100~\AA\ to $\sim$8800~\AA. Starting from our 1998 December run, we used the 150 $\ell$ mm$^{-1}$ grating blazed at 7500~\AA\ combined with the 1\farcs5 wide slit, which results in a resolution of $\sim$ 15~\AA\ (FWHM) and a wavelength coverage from $\sim$4000~\AA\ to $\sim$10,000~\AA, but with poor S/N at the red end.

\subsection{Data reduction}
We reduced our spectra using the NOAO IRAF package. After bias correction using the overscan region, we flat-fielded the spectra using internal lamps or dome-flats, and removed the cosmic rays using the task `szap', which is based on the cosmic ray cleaning routine `xzap' from the DIMSUM\footnote{DIMSUM is the Deep Mosaicing Software package, developed by P. Eisenhardt, M. Dickinson, A. Stanford, and J. Ward, which is available as a contributing package in IRAF.} package. For objects with only a single exposure, we subtracted the sky emission using a third order polynomial interpolation along the slit at the position of the object. For those objects with more than one exposure, we subtracted one of the bracketing exposures which was shifted by $\simgt$10\arcsec\ along the slit, and subtracted the residual sky contribution which is due to the variation of the sky brightness. After registration using an integer pixel shift determined from other objects along our long slit, we summed both exposures, and extracted a one-dimensional spectrum. In all cases, this $\simgt$10\arcsec\ shift was found to be sufficiently large in order not to affect the flux calibration of the most spatially extended lines (generally \Lya).
Because most of our objects are very faint, we used the aperture that included the maximal amount of line flux for each object instead of a standard aperture (the exact values for each object are listed in table \ref{specjournal}). The use of a wide aperture is justified in HzRG spectra, as the emission line regions are often very extended \citep[\eg][]{rot97}; the main use of our spectra is the redshift determination and relative line flux ratios between the individual emission lines. In most cases, deeper spectra are needed to study the faint continuum flux.
We extracted a calibration lamp spectrum using the same aperture as that for the object, and corrected the derived wavelength solution by applying a small additional offset determined from a comparison with the wavelength of several well-resolved skylines \citep[\eg][]{ost92}. Finally, we flux calibrated the spectra using standard star observations obtained during each observing night. We estimate our flux calibration is accurate to $\sim$20\%, except for the objects which were observed in non-photometric conditions (see table \ref{specjournal}). We include the derived line fluxes of these sources nevertheless for line-ratio studies.

\section{Results}
\subsection{Redshifts and line parameters}
We used the procedures described in \citet{rot97} to determine the central wavelength, total line flux, de-convolved widths, and rest-frame equivalent widths (with their related errors) of all the emission and absorption lines detected in our spectra. For each object, we quote a single redshift, which is an average of all the lines, weighed by the uncertainties due to the line fitting and wavelength calibration \citep[see][]{rot97}. We present the results for the sources in our main and subsidiary USS samples (see \S 2.1) in table \ref{lineparameters}, and show extracted one-dimensional spectra in figure \ref{spectra}.

To determine the redshift and identify the emission and absorption lines, we compared our spectra with composite HzRG spectra \citep{mcc93,ste99}. The spectra at $z\gtrsim 2$ (see WN~J0303+3733, Fig. 1.8 for an example) are generally dominated by extended \Lya\ with a large equivalent width \citep[$W_{\lambda}^{rest} \sim 100$\AA;][]{ste99}. The brightest confirming lines are \CIVfull, \HeIIfull, \CIIIfull, and \CIIfull. In some cases, we also detected \NVfull, \OVIfull\ and the \SiIV/\OIV~$\lambda$~1400 complex. In the lower redshift objects (see MP~J1929$-$3732, Fig. 1.40 for an example), the spectra are generally dominated by \OIIfull, \OIIIfull, or \Ha. Confirming lines in these objects are \NeIVfull, \MgIIfull, \NeVfull, \NeIIIfull, \Hdelta, \Hgamma, and \Hbeta.
In the low redshift objects with sufficiently bright continua (see WN~J1543+3512, Fig. 1.33 for an example), we can also detect the stellar absorption features Ca~II K and H~$\lambda\lambda$3934,3968, G-band~$\lambda$4300 and Mg~I~$b~\lambda\lambda$5174, and continuum break features at 2640~\AA, 2900~\AA\ and 4000~\AA.
\begin{table}[!b]
\caption{Emission and absorption line parameters.}
\label{lineparameters}
\tiny
\begin{tabular}{lcllrrr}
\hline
Source & $z$ & Line & $\lambda_{\rm obs}$ & $10^{-16}\times$ Flux & $\Delta v_{\rm FWHM}$ & $W_{\lambda}^{\rm rest}$ \\
 &  &  & \AA & erg/s/cm$^2$ & km s$^{-1}$ & \AA \\
\hline
USS~0006$-$0141 & 1.541 $\pm$ 0.006 & \CIVfull & 3950 $\pm$ 1 & 5.9 $\pm$ 1.2 & 850 $\pm$ \phn300 & $> 118$ \\
 & & \HeIIfull & 4181 $\pm$ 2 & 3.9 $\pm$ 0.8 & 1050 $\pm$ \phn400 & 35 $\pm$ 11 \\
 & & \CIIIfull & 4862 $\pm$ 1 & 3.4 $\pm$ 0.5 & 900 $\pm$ \phn200 & 23: $\pm$ \phn3 \\
 & & \CIIfull  & 5899 $\pm$ 1 & 1.0 $\pm$ 0.2 & $< 150$ & 6 $\pm$ \phn1\\
 & & \NeIVfull & 6180 $\pm$ 3 & 1.7 $\pm$ 0.3 & 1000 $\pm$ \phn300 & 11 $\pm$ \phn2 \\
\\
WN J0040$+$3857 & 2.6055 $\pm$ 0.0005 & \Lya & 4386 $\pm$ 3 & 4.1 $\pm$ 0.5 & 2600 $\pm$ \phn500 & $>$479 \\
 & &  \CIVfull & 5585 $\pm$ 1 & 1.0: $\pm$ 0.5 &  700: $\pm$  \phn250 & 69: $\pm$ 18 \\
 & & \HeIIfull & 5913 $\pm$ 1 & 0.1 $\pm$ 0.1 & $<$225 & 9 $\pm$  \phn3 \\
 & & \CIIIfull & 6873 $\pm$ 19 & 0.6 $\pm$ 0.1 & 2600 $\pm$ 1700 & 33 $\pm$  \phn9 \\
\\
TN J0121$+$1320 & 3.516 $\pm$ 0.002 & \Lya & 5500 $\pm$ 5 & 178$^{\dag}$ $\pm$ 50 & 1350 $\pm$ \phn400 & 300 $\pm$ 100 \\
 & &  \CIVfull & 6994 $\pm$ 4 & 22.1$^{\dag}$ $\pm$ 2.7 & 1300 $\pm$ \phn500 & 26 $\pm$  \phn5 \\
 & & \HeIIfull & 7411 $\pm$ 6 & 24.9$^{\dag}$ $\pm$ 3.1 & 1600 $\pm$ \phn550 & 40 $\pm$ 10 \\
 & & \CIIIfull & 8617 $\pm$ 8 &  6.2$^{\dag}$ $\pm$ 2.6  & 400 $\pm$ \phn350 & $>$3        \\
\\
TXS J0137+2521 & 2.8970 $\pm$ 0.0005 & \Lya & 4738 $\pm$ 1 & 12.0 $\pm$ 1.2 & 950 $\pm$ \phn150 & $> 326$ \\
\\
6C 0143$+$3253 & 4.413 $\pm$ 0.002 & \Lya & 6582 $\pm$ 1 & 6.38 $\pm$ 0.64 & 1350 $\pm$ \phn100 & 500 $\pm$ 150 \\
 & & \CIVfull & 8399 $\pm$ 2 & 0.15 $\pm$ 0.03 & 300 $\pm$ \phn200 & 9 $\pm$ 3 \\
\\
TN J0205$+$2242 & 3.5061 $\pm$ 0.0004 & \Lya & 5479 $\pm$ 1 & 792$^{\dag}$ $\pm$ 79 & 1300 $\pm$ \phn200 & 650 $\pm$ 100 \\
 & &     \SiIV / \OIV & 6316 $\pm$ 22 & 14.1$^{\dag}$ $\pm$ 2.6 & 2650 $\pm$ 2050 & 14 $\pm$ \phn3 \\
 & &         \CIVfull & 6981 $\pm$  1 & 44.2$^{\dag}$ $\pm$ 4.7 &  850 $\pm$  \phn200 & 40 $\pm$ \phn7 \\
 & &        \HeIIfull & 7392 $\pm$  4 & 30.1$^{\dag}$ $\pm$ 3.7 & 1100 $\pm$  \phn400 & 33 $\pm$ \phn8 \\
 & &        \CIIIfull & 8598 $\pm$ 20 & 19.3$^{\dag}$ $\pm$ 6.2 & $<$1350 & $>$12      \\ 
\\
WN J0231$+$3600 & 3.079 $\pm$ 0.002 & \Lya & 4960 $\pm$ 2 & 1.1 $\pm$ 0.1 & 1950 $\pm$ \phn350 & 100 $\pm$ 25 \\
\\
WN J0303$+$3733 & 2.506 $\pm$ 0.002 & \Lya & 4263 $\pm$ 1 & 10.3 $\pm$ 1.0 & 1950 $\pm$ \phn300 & 550 $\pm$ 250 \\
 & &     \SiIV / \OIV & 4940 $\pm$ 63 &  1.3 $\pm$ 0.3 & \nodata        & \nodata \\
 & &         \CIVfull & 5438 $\pm$  5 &  1.5 $\pm$ 0.2 & 2000 $\pm$ \phn450 & 68 $\pm$ 15 \\
 & &        \HeIIfull & 5749 $\pm$  4 &  0.5 $\pm$ 0.1 & 1100 $\pm$ \phn400 & 21 $\pm$ \phn 5 \\
 & &        \CIIIfull & 6687 $\pm$  3 &  1.4 $\pm$ 0.2 & 1600 $\pm$ \phn400 & 65 $\pm$ 13 \\
 & &         \CIIfull & 8140 $\pm$ 10 &  0.4 $\pm$ 0.1 &  800 $\pm$ \phn600 & \nodata \\
\\
TN J0452$-$1737 & 2.256 $\pm$ 0.003 & \HeIIfull & 5348 $\pm$ 2 & 0.3 $\pm$ 0.1 & 1000$\pm$ \phn400 & 6 $\pm$ \phn1 \\
 & &  \CIIIfull & 6212: $\pm$ 1 & 0.1: $\pm$ 0.1 & $<$250 & 3 $\pm$ \phn2 \\
 & &   \CIIfull &  7587 $\pm$ 3 & 0.4 $\pm$ 0.1 & 550 $\pm$ \phn250 & 9 $\pm$ \phn2 \\
\\
TNR J0516+0637 & 0.3568 $\pm$ 0.0002 & \OIIfull & 5056 $\pm$ 1 & 2.1 $\pm$ 0.3 & 500 $\pm$ \phn150 & 24 $\pm$ \phn3 \\
 & & G-band & 5819 $\pm$ 16 & $-$2.7 $\pm$ 0.6 & 3000 $\pm$ 1700 & \nodata \\
\\
WN J0617$+$5012 & 3.153 $\pm$ 0.002 & \Lya & 5050 $\pm$ 2 & 0.8 $\pm$ 0.2 & 1600 $\pm$ \phn350 & $>$400 \\
\\
WN J0717$+$4611 & 1.462 $\pm$ 0.001 & \CIVfull & 3816 $\pm$ 2 & 30.2 $\pm$ 4.0 & 1300 $\pm$ \phn400 & 28 $\pm$ 12 \\
 & & \CIIIfull & 4700 $\pm$ 2 & 0.9 $\pm$ 0.1 & 1350 $\pm$ \phn300 & 6 $\pm$ \phn1 \\
 & & \CIIfull  & 5726 $\pm$ 2 & 0.4 $\pm$ 0.1 & 1300 $\pm$ \phn400 & 3.2 $\pm$ 0.4 \\
 & & \NeIVfull & 5968 $\pm$ 2 & 0.6 $\pm$ 0.1 & 1150 $\pm$ \phn300 & 2.8 $\pm$ 0.4 \\
 & & \MgIIfull & 6900 $\pm$ 2 & 6.6 $\pm$ 0.6 & 11000 $\pm$ \phn500 & 13 $\pm$ \phn1 \\
 & & \NeVttfs  & 8242 $\pm$ 3 & 0.5 $\pm$ 0.3 & 500 $\pm$ \phn350 & 1.2 $\pm$ 0.4 \\
 & & \NeVtfts  & 8437 $\pm$ 2 & 0.7 $\pm$ 0.3 & 600 $\pm$ \phn200 & 1.2 $\pm$ 0.4 \\
\\
WN J0747$+$3654 & 2.992 $\pm$ 0.004 & \Lya & 4863 $\pm$ 3 & 0.78 $\pm$ 0.09 & 2050 $\pm$ \phn450 & 400 $\pm$ 150 \\
 & &\SiIV/\OIV & 5579 $\pm$ 19 & 0.13 $\pm$ 0.04 & 2350 $\pm$1250 &  14 $\pm$  \phn4 \\
 & &  \CIVfull & 6178 $\pm$  7 & 0.17 $\pm$ 0.03 & 1350 $\pm$ \phn600 &  31 $\pm$ 11 \\
 & & \HeIIfull & 6541 $\pm$  4 & 0.22 $\pm$ 0.04 & 1050 $\pm$ \phn600 &  38 $\pm$ 12 \\
 & & \CIIIfull & 7625 $\pm$ 33 & 0.11 $\pm$ 0.05 & $<$2150        & $>$48        \\
 & &  \CIIfull & 9286 $\pm$  6 & 0.20 $\pm$ 0.04 &  850 $\pm$ \phn500 & $>$26        \\
\end{tabular}
\end{table}
\begin{table}
\tiny
\begin{tabular}{lcllrrr}
\hline
Source & $z$ & Line & $\lambda_{\rm obs}$ & $10^{-16}\times$ Flux & $\Delta v_{\rm FWHM}$ & $W_{\lambda}^{\rm rest}$ \\
 &  &  & \AA & erg/s/cm$^2$ & km s$^{-1}$ & \AA \\
\hline
WN J0813$+$4828 & 1.274 $\pm$ 0.001 & \CIIIfull & 4330 $\pm$ 13 & 0.2 $\pm$ 0.1 & 1500 $\pm$ 1300 & 1291 $\pm$ \phn8 \\
 & &  \CIIfull & 5300 $\pm$  9 & 0.4 $\pm$ 0.1 & 1950 $\pm$ 1150 & 28 $\pm$  \phn7 \\
 & & \NeIVfull & 5510 $\pm$  8 & 0.2 $\pm$ 0.1 & $<$1100 &  9 $\pm$ \phn4 \\
 & & \MgIIfull & 6376 $\pm$ 20 & 0.1 $\pm$ 0.1 & $<$1250 & $>$5 \\
 & &  \NeVtfts & 7795 $\pm$ 18 & 0.4 $\pm$ 0.2 & $<$1950 & 18 $\pm$ \phn8 \\
 & &  \OIIfull & 8476 $\pm$  2 & 2.2 $\pm$ 0.3 &  \phn800 $\pm$  \phn250 & 74 $\pm$ 15 \\
 & &\NeIIIfull & 8798: $\pm$ 59 & 1.3: $\pm$ 0.4 & 3350: $\pm$ 2800 & 32: $\pm$ 10 \\
\\
TN J0920$-$0711 & 2.760 $\pm$ 0.003 & \Lya & 4573 $\pm$ 1 & 44.0 $\pm$ 4.4 & 2050 $\pm$ \phn150 & 350 $\pm$ 60 \\
 & & \SiIV / \OIV & 5281 $\pm$ 8 & 0.6 $\pm$ 0.3 & 1100 $\pm$ \phn600 & 5 $\pm$ \phn2 \\
 & &     \CIVfull & 5831 $\pm$ 1 & 3.6 $\pm$ 0.5 & 1100 $\pm$ \phn200 & 37 $\pm$ \phn7 \\
 & &    \HeIIfull & 6155 $\pm$ 1 & 3.0 $\pm$ 0.4 & 950 $\pm$ \phn200 & 29 $\pm$ \phn5 \\
\\
TN J0924$-$2201 & 5.195 $\pm$ 0.007 & \Lya & 7531 $\pm$ 9 & 0.35 $\pm$ 0.06 & 1500 $\pm$ \phn200 & $>$180 \\
\\
TN J0941$-$1628 & 1.644 $\pm$ 0.001 & \CIVfull & 4095 $\pm$ 3 & 3.2 $\pm$ 0.5 & 2100 $\pm$ \phn600 & $> 140$ \\
 & &  \HeIIfull & 4331 $\pm$ 3 & 0.9 $\pm$ 0.2 & 1250 $\pm$ \phn450 & 29 $\pm$ \phn9 \\
 & &  \CIIIfull & 5051 $\pm$ 2 & 2.0 $\pm$ 0.3 & 1900 $\pm$ \phn400 & 72 $\pm$ 15 \\
 & &   \CIIfull & 6149 $\pm$ 2 & 1.9 $\pm$ 0.2 & 1500 $\pm$ \phn250 & 59 $\pm$ \phn9 \\
\\
TN J1033$-$1339 & 2.425 $\pm$ 0.002 & \Lya & 4164 $\pm$ 1 & 9.8 $\pm$ 1.0 & 2000 $\pm$ \phn200 & $> 885$ \\
 & & \SiIV / \OIV & 4806: $\pm$ 17 & 4.1: $\pm$ 0.5 & 62000 $\pm$ 2300 & \nodata \\ 
 & &     \CIVfull & 5308 $\pm$ 2 & 2.3 $\pm$ 0.3 & 1500 $\pm$ \phn300 & 140 $\pm$ 50 \\
 & &    \HeIIfull & 5623 $\pm$ 3 & 0.8 $\pm$ 0.1 & 950 $\pm$ \phn200 & $> 128$ \\
 & &    \CIIIfull & 6527 $\pm$ 4 & 0.7 $\pm$ 0.1 & 900 $\pm$ \phn250 & 56 $\pm$ 27 \\
\\
TN J1102$-$1651 & 2.111 $\pm$ 0.003 & \Lya & 3781 $\pm$ 1 & 2.7: $\pm$ 0.5 & 800 $\pm$ \phn250 & $> 129:$ \\
 & &  \CIVfull & 4824 $\pm$ 1 & 1.0 $\pm$ 0.2 & 700 $\pm$ \phn250 & 35 $\pm$ 11 \\
 & & \HeIIfull & 5090 $\pm$ 3 & 1.3 $\pm$ 0.2 & 1350 $\pm$ \phn400 & 57 $\pm$ 20 \\
 & & \CIIIfull & 5927 $\pm$ 7 & 1.1 $\pm$ 0.2 & 1750 $\pm$ \phn550 & 53 $\pm$ 20 \\
\\
TN J1112$-$2948 & 3.09 $\pm$ 0.01 & \Lya & 4976 $\pm$ 1 & 2.9 $\pm$ 0.3 & 1300 $\pm$ \phn150 & 93 $\pm$ 21 \\
 & &  \CIVfull & 6307: $\pm$ 1 & 0.7: $\pm$ 0.1 & 500: $\pm$ \phn150 & 42: $\pm$ 17 \\
 & & \HeIIfull & 6717 $\pm$ 1 & 1.2 $\pm$ 0.1 & 800 $\pm$ \phn150 & 38 $\pm$ \phn8 \\
\\
WN J1115$+$5016 & 2.54 $\pm$ 0.01 & \Lya & 4323 $\pm$ 2 & 2.03 $\pm$ 0.21 & 2550 $\pm$ \phn400 & 350 $\pm$ 150 \\
 & &  \CIVfull & 5413 $\pm$  8 &-0.18 $\pm$ 0.03 & 1950 $\pm$ \phn850 &-4 $\pm$ \phn2 \\
 & & \HeIIfull & 5801 $\pm$ 12 & 0.03 $\pm$ 0.01 & 1150 $\pm$ \phn900 & 3 $\pm$ \phn2 \\
 & & \CIIIfull & 6749 $\pm$  5 & 0.15 $\pm$ 0.02 & 2450 $\pm$ \phn500 &26 $\pm$ \phn3 \\
 & &  \CIIfull & 8223 $\pm$ 46 & 0.06 $\pm$ 0.03 & $<$2350        & $>$22     \\
\\
TN J1123$-$2154 & 4.109 $\pm$ 0.004 & \Lya & 6213 $\pm$ 4 & 0.18 $\pm$ 0.05 & 550 $\pm$ \phn350 & $>$47 \\
\\
WN J1123$+$3141 & 3.2174 $\pm$ 0.0004 & \OVIfull & 4370 $\pm$  6 & 4.7 $\pm$ 0.6 & 2550 $\pm$ 1050 & 22 $\pm$ \phn4 \\
 & &      \Lya & 5128 $\pm$ 1 & 6.2 $\pm$ 0.6 & 400 $\pm$ \phn150 & 25 $\pm$ \phn3 \\
 & &  \CIVfull & 6532 $\pm$ 2 & 3.4 $\pm$ 0.4 & 900 $\pm$ \phn250 & 14 $\pm$ \phn2 \\
 & & \HeIIfull & 6917 $\pm$ 2 & 2.4 $\pm$ 0.3 & 700 $\pm$ \phn250 & 10 $\pm$ \phn1 \\
\\
WN J1242$+$3915 & 2.131 $\pm$ 0.001 & \CIVfull & 4873 $\pm$ 38 & 0.09 $\pm$ 0.05 & \nodata & \nodata \\
 & & \HeIIfull & 5135 $\pm$ 2 & 0.36 $\pm$ 0.07 &  500 $\pm$ \phn400 & 32 $\pm$ 11 \\
 & & \CIIIfull & 5975 $\pm$ 7 & 0.76 $\pm$ 0.11 & 2300 $\pm$ \phn800 & 85 $\pm$ 26 \\
 & &  \CIIfull & 7286 $\pm$29 & 0.25 $\pm$ 0.07 & 2500 $\pm$1300 & 29 $\pm$ 10 \\
\\
WN J1333$+$3037 & 1.2127 $\pm$ 0.0006 & \CIIIfull & 4219 $\pm$ 4 & 0.3 $\pm$ 0.1 & 850 $\pm$ \phn500 & 20 $\pm$ \phn6 \\
 & &   \CIIfull & 5149 $\pm$ 6 & 0.4 $\pm$ 0.1 & 1600 $\pm$ \phn750 & 33 $\pm$  \phn7 \\
 & &  \MgIIfull & 6190 $\pm$ 7 & 0.9 $\pm$ 0.1 & 2400 $\pm$ \phn800 & 52 $\pm$  \phn8 \\
 & &   \OIIfull & 8247 $\pm$ 1 & 3.5 $\pm$ 0.4 &  850 $\pm$ \phn150 &180 $\pm$ 40 \\
 & & \NeIIIfull & 8553 $\pm$ 5 & 1.1 $\pm$ 0.2 & 1000 $\pm$ \phn550 & 44 $\pm$ 10 \\
\\
WNR J1338+3532 & 2.769 $\pm$ 0.003 & \Lya & 4586 $\pm$ 1 & 17.8 $\pm$ 1.8 & 1800 $\pm$ \phn150 & $> 693$ \\
 & &  \CIVfull & 5832 $\pm$ 1 & 1.3 $\pm$ 0.2 & \phn400 $\pm$ 150 & $> 2$ \\
 & & \HeIIfull & 6189 $\pm$ 1 & 3.0 $\pm$ 0.3 & \phn950 $\pm$ 200 & $> 19$ \\
 & & \CIIIfull & 7196 $\pm$ 1 & 2.2 $\pm$ 0.3 & \phn750 $\pm$ 200 & $>  16$ \\
\\
TN J1338$-$1941 & 4.11 $\pm$ 0.01 & \Lya & 6211 $\pm$ 1 & 13.63 $\pm$ 1.37 & 1000 $\pm$ \phn100& 200 $\pm$ 60 \\
 & &  \CIVfull & 7899 $\pm$ 3 & 1.0 $\pm$ 0.1 &  1450 $\pm$ \phn350 & $> 237$ \\
 & & \HeIIfull & 8382 $\pm$ 35 & 0.2 $\pm$ 0.1 & 1500 $\pm$ \phn900 & $> 37$ \\
\\
WNR J1356+3929 & 0.2525 $\pm$ 0.0002 & \OIIIfull & 6272 $\pm$ 1 & 1.3 $\pm$ 0.2 & 400 $\pm$ \phn100 & 8 $\pm$ \phn1 \\
 & & \Ha & 8218 $\pm$ 1 & 4.5 $\pm$ 0.5 & 500 $\pm$ \phn100 & 44 $\pm$ \phn6 \\
\\
TNR J1402-1510 &  0.739 $\pm$ 0.001 & \MgIIfull & 4829 $\pm$ 40 & 6.88 $\pm$ 1.23 & 7650 $\pm$ 3750 & 80 $\pm$ 17 \\
 & &   \NeVfull & 5953 $\pm$ 1 & 3.2 $\pm$ 0.4 & 1550 $\pm$ \phn200 & 26 $\pm$ \phn2 \\
 & &   \OIIfull & 6475 $\pm$ 1 & 35.1 $\pm$ 3.5 & 1950 $\pm$ \phn100 & 320 $\pm$ 35 \\
 & & \NeIIIfull & 6732 $\pm$ 1 & 3.8 $\pm$ 0.4 & $< 180$ & 32 $\pm$ \phn3 \\
 & &    \Hdelta & 7131 $\pm$ 3 & 2.2 $\pm$ 0.4 & 950 $\pm$ \phn350 & 14 $\pm$ \phn2 \\
 & &    \Hgamma & 7557 $\pm$ 12 & 4.1 $\pm$ 0.9 & 1900 $\pm$ \phn800 & 30 $\pm$ \phn6 \\
 & &    \Hbeta  & 8448 $\pm$ 2 & 7.1 $\pm$ 1.1 & 750 $\pm$ \phn200 & 31 $\pm$ \phn5 \\
 & & \OIIIfnfn  & 8626 $\pm$ 1 & 10.8 $\pm$ 1.3 & 400 $\pm$ \phn100 & 47 $\pm$ \phn6 \\
 & & \OIIIfull  & 8705 $\pm$ 1 & 56.8 $\pm$ 5.8 & 1000 $\pm$ \phn100& 400 $\pm$ 60 \\
\end{tabular}
\end{table}
\begin{table}
\tiny
\begin{tabular}{lcllrrr}
\hline
Source & $z$ & Line & $\lambda_{\rm obs}$ & $10^{-16}\times$ Flux & $\Delta v_{\rm FWHM}$ & $W_{\lambda}^{\rm rest}$ \\
 &  &  & \AA & erg/s/cm$^2$ & km s$^{-1}$ & \AA \\
\hline
PKS J1428-1502 & 2.349 $\pm$ 0.004 & \Lya & 4076 $\pm$ 1 & 20.7 $\pm$ 2.2 & 2650 $\pm$ \phn350 & $> 785$ \\
 & &  \CIVfull & 5189 $\pm$ 3 & 2.3 $\pm$ 0.4 & 1300 $\pm$ \phn450 & 37 $\pm$ 11 \\
 & & \HeIIfull & 5495 $\pm$ 4 & 2.3 $\pm$ 0.4 & 1550 $\pm$ \phn550 & 35 $\pm$ 10 \\
 & & \CIIIfull & 6378 $\pm$ 3 & 1.0 $\pm$ 0.3 & 750 $\pm$ \phn450 & 18 $\pm$ \phn6 \\
\\
WN J1543$+$3512 & 0.7033 $\pm$ 0.0006 & \OII & 6348 $\pm$ 2 & 0.8 $\pm$ 0.1 & 550 $\pm$ \phn300 & 22 $\pm$ \phn4 \\
 & & CaII K & 6706 $\pm$ 2 & $-$0.4 $\pm$ 0.1 & \nodata & $-$11 $\pm$ \phn4 \\
 & & Mg b & 8788 $\pm$ 10 & $-$4.3 $\pm$ 2.0 & \nodata & $-$61 $\pm$ 30 \\
\\
TXS J1650+0955 & 2.510 $\pm$ 0.002 & \Lya & 4267 $\pm$ 1 & 20.9 $\pm$ 2.1 & 1850 $\pm$ \phn250 & \nodata \\
 & &  \CIVfull & 5441 $\pm$ 1 & 3.2 $\pm$ 0.4 & 1200 $\pm$ \phn200 & 36 $\pm$ \phn5 \\
 & & \HeIIfull & 5760 $\pm$ 2 & 2.7 $\pm$ 0.3 & 1700 $\pm$ \phn400 & 26 $\pm$ \phn3 \\
 & & \CIIIfull & 6692 $\pm$ 6 & 1.2 $\pm$ 0.3 & 1350 $\pm$ \phn600 & 20 $\pm$ \phn5 \\
\\
WNR J1703+3739 & 0.2561 $\pm$ 0.0003 & \OIIIfull & 6291 $\pm$ 1 & 1.1 $\pm$ 0.2 & 500 $\pm$ \phn150 & 7 $\pm$ \phn1 \\
 & & \Ha & 8242 $\pm$ 1 & 2.4 $\pm$ 0.3 & 400 $\pm$ \phn100& 25 $\pm$ \phn4 \\
\\
MP J1755$-$6916 & 2.551 $\pm$ 0.004 & \Lya & 4317 $\pm$ 1 & 9.0 $\pm$ 0.9 & 1700 $\pm$ \phn250 & 370 $\pm$ 100 \\
 & &          \NVfull & 4404 $\pm$12 & 0.05 $\pm$ 0.03 & \nodata & \nodata \\
 & & \SiIV / \OIV     & 4989 $\pm$ 2 & 0.21 $\pm$ 0.04 & $<$400 & 10 $\pm$ \phn3 \\
 & &         \CIVfull & 5505 $\pm$ 3 & 0.61 $\pm$ 0.08 & 1450 $\pm$ \phn400 & 37 $\pm$ \phn8 \\
 & &        \HeIIfull & 5820 $\pm$ 6 & 0.39 $\pm$ 0.06 & 1500 $\pm$ \phn450 & 22 $\pm$ \phn5 \\
 & &        \CIIIfull & 6773 $\pm$ 4 & 0.57 $\pm$ 0.08 & 1100 $\pm$ \phn500 & 29 $\pm$ \phn7 \\
\\
MP J1758$-$6738 & 2.026 $\pm$ 0.002 & \CIVfull & 4688 $\pm$ 1 & 1.0 $\pm$ 0.1 & 1750 $\pm$ \phn300 & 39 $\pm$ \phn4 \\
 & &        \HeIIfull & 4969 $\pm$ 2 & 0.60 $\pm$ 0.06 & 1800 $\pm$ \phn300 & 24 $\pm$ \phn3 \\
 & &        \CIIIfull & 5772 $\pm$ 2 & 0.42 $\pm$ 0.05 & 1350 $\pm$ \phn300 & 23 $\pm$ \phn3 \\
\\
6C J1908+7220 & 3.5356$\pm$ 0.0003& \Lya & 5516 $\pm$0.3&  9.0  $\pm$ 0.9  & 1200 $\pm$ \phn 90 &  87 $\pm$ 9  \\
 & & \HeIIfull & 7437.9$\pm$0.4 & 1.3  $\pm$ 0.1  & 1350 $\pm$ \phn 70 &  9 $\pm$ 1  \\
\\
WN J1911$+$6342 & 3.590 $\pm$ 0.002 & \Lya & 5582: $\pm$ 1 & 1.4: $\pm$ 0.1 & 1350: $\pm$ \phn250 & 220: $\pm$ 40 \\ 
 & &         \CIVfull & 7109 $\pm$ 3 & 0.08 $\pm$ 0.01 &  950 $\pm$ \phn350 & 17 $\pm$  \phn3 \\
 & &        \HeIIfull & 7517 $\pm$ 4 & 0.08 $\pm$ 0.01 & 1000 $\pm$ \phn400 & 40 $\pm$ 12 \\
\\
MP J1929$-$3732 & 0.7476 $\pm$ 0.0006 & \CIIfull & 4062 $\pm$ 3 & 2.05 $\pm$ 0.29 & 1550 $\pm$ \phn600 & 46 $\pm$ \phn9 \\
 & &        \NeIVfull & 4234 $\pm$ 1 &   1.75 $\pm$ 0.23 &  700 $\pm$ \phn300 &  42 $\pm$  \phn8 \\
 & &         \OIItfsz & 4320 $\pm$ 1 &   0.8  $\pm$ 0.2  & 1300 $\pm$ \phn300 &  80 $\pm$ 20 \\
 & &        \MgIIfull & 4892 $\pm$ 1 &   3.12 $\pm$ 0.33 & 1550 $\pm$ \phn300 &  61 $\pm$  \phn7 \\
 & &        \OIIItzfs & 4955 $\pm$ 1 &   0.43 $\pm$ 0.20 & $<750$             & \nodata      \\
 & &        \OIIItott & 5475 $\pm$ 2 &   0.93 $\pm$ 0.12 & 1200 $\pm$ \phn300 &  16 $\pm$  \phn2 \\
 & &        \HeIIttzt & 5600 $\pm$ 1 &   0.79 $\pm$ 0.30 & $<750$             & \nodata      \\
 & &         \NeVttfs & 5848 $\pm$ 1 &   2.07 $\pm$ 0.40 & 1150 $\pm$ \phn200 &  62 $\pm$  \phn9 \\
 & &         \NeVtfts & 5988 $\pm$ 1 &   4.97 $\pm$ 0.50 & 1100 $\pm$ \phn200 &  63 $\pm$  \phn6 \\
 & &         \OIIfull & 6511 $\pm$ 1 &  15.97 $\pm$ 1.60 & 1050 $\pm$ \phn150 & 200 $\pm$ 20 \\
 & &       \NeIIItesn & 6761 $\pm$ 1 &   9.06 $\pm$ 0.91 & 1100 $\pm$ \phn150 &  90 $\pm$  \phn9 \\
 & &       \NeIIItnss & 6934 $\pm$ 1 &   4.00 $\pm$ 1.20 & 1400 $\pm$ \phn300 & 116 $\pm$ 18 \\
 & &         \SIIfull & 7113 $\pm$ 2 &   2.45 $\pm$ 1.10 & 1650 $\pm$ \phn400 &  52 $\pm$ 15 \\
 & &      \Hdelta     & 7167 $\pm$ 3 &   1.48 $\pm$ 0.90 & 1300 $\pm$ \phn350 &  34 $\pm$ 20 \\
 & &      \Hgamma     & 7582 $\pm$ 1 &   2.50 $\pm$ 0.29 &  800 $\pm$ \phn200 &  33 $\pm$  \phn4 \\
 & &        \HeIIfses & 8188 $\pm$ 6 &   2.24 $\pm$ 0.33 & 1150 $\pm$ \phn500 &  29 $\pm$  \phn4 \\
 & &       \Hbeta     & 8494 $\pm$ 1 &   7.44 $\pm$ 1.75 & 1300 $\pm$ \phn450 & 178 $\pm$ 25 \\
 & &        \OIIIfnfn & 8667 $\pm$ 1 &  37.41 $\pm$ 3.75 &  850 $\pm$ \phn150 & 273 $\pm$ 31 \\
 & &        \OIIIfull & 8750 $\pm$ 1 & 112.88 $\pm$11.29 & 1100 $\pm$ \phn150 & 850 $\pm$100 \\
\\
MP J2003$-$8340 & 1.169 $\pm$ 0.001 & \OIIfull & 8085 $\pm$ 2 & 3.28 $\pm$ 0.36 & 850 $\pm$ \phn200 & 140 $\pm$ 50 \\
\\
TXS J2036+0256 & 2.130 $\pm$ 0.001 & \Lya & 3805 $\pm$ 1 & 6.8: $\pm$ 1.0 & 700 $\pm$ \phn250 & \nodata \\
 & &  \CIVfull & 4859 $\pm$ 12 & 0.6 $\pm$ 0.3 & 1550 $\pm$ 1100 & $> 12$ \\
 & & \HeIIfull & 5131 $\pm$ 5 & 0.7 $\pm$ 0.2 & 1100 $\pm$ \phn800 & 23 $\pm$ 10 \\
 & & \CIIIfull & 5965 $\pm$ 5 & 1.2 $\pm$ 0.3 & 1200 $\pm$ \phn650 & 29 $\pm$ 10 \\
\\
MP J2045$-$6018 & 1.464 $\pm$ 0.001 & \CIVfull & 3817 $\pm$ 1 & 5.59 $\pm$ 0.57 & 650 $\pm$ \phn300 & 35 $\pm$ \phn4 \\
 & &        \HeIIfull & 4042 $\pm$ 1 & 4.72 $\pm$ 0.48 & $<$250             & 33 $\pm$ \phn4 \\
 & &        \CIIIfull & 4701 $\pm$ 1 & 2.68 $\pm$ 0.27 &  850 $\pm$ \phn250 & 19 $\pm$ \phn2 \\
 & &         \CIIfull & 5734 $\pm$ 1 & 0.48 $\pm$ 0.30 & $<$200             & \nodata    \\
 & &        \NeIVfull & 5972 $\pm$ 1 & 1.72 $\pm$ 0.17 &  750 $\pm$ \phn200 & 13 $\pm$ \phn1 \\
 & &         \OIItfsz & 6091 $\pm$ 1 & 0.19 $\pm$ 0.15 &  400 $\pm$ \phn150 &  4 $\pm$ \phn1 \\
 & &        \MgIIfull & 6901: $\pm$ 3 & 0.92: $\pm$ 0.11 & 1350: $\pm$ \phn450 &  7: $\pm$ \phn1 \\
 & &        \OIIItott & 7721 $\pm$ 4 & 0.91 $\pm$ 0.13 &  900 $\pm$ \phn350 &  9 $\pm$ \phn1 \\
 & &         \NeVttfs & 8246 $\pm$ 1 & 1.32 $\pm$ 0.25 &  650 $\pm$ \phn150 & 38 $\pm$ \phn9 \\
 & &         \NeVfull & 8444 $\pm$ 1 & 2.90 $\pm$ 0.30 &  450 $\pm$ \phn150 & 27 $\pm$ \phn3 \\
\\
MP J2048$-$5750 & 1.262 $\pm$ 0.002 & \OIIItott & 7082 $\pm$ 17 & 2.61 $\pm$ 0.33 & 3150 $\pm$ 1450 & 14 $\pm$ \phn2 \\
 & &         \NeVfull & 7748 $\pm$ 4 & 3.39 $\pm$ 0.38 & 1400 $\pm$ \phn350 & 16 $\pm$  \phn2 \\
 & &         \OIIfull & 8439 $\pm$ 1 &46.50 $\pm$ 4.66 & 1500 $\pm$ \phn150 &200 $\pm$ 20 \\
 & &       \NeIIIfull & 8750 $\pm$ 2 &11.73 $\pm$ 1.20 & 1150 $\pm$ \phn200 & 37 $\pm$  \phn4 \\
\\
TNR J2254+1857 & 2.153 $\pm$ 0.002 & \Lya & 3833 $\pm$ 2 & 33.0 $\pm$ 3.8 & 2050 $\pm$ \phn400 & $> 86$ \\
 & & \HeIIfull & 5178 $\pm$ 21 & 3.2 $\pm$ 1.5 & 1550 $\pm$ 1250 & $>25$ \\
\end{tabular}
\end{table}
\begin{table}[t]
\tiny
\begin{tabular}{lcllrrr}
\hline
Source & $z$ & Line & $\lambda_{\rm obs}$ & $10^{-16}\times$ Flux & $\Delta v_{\rm FWHM}$ & $W_{\lambda}^{\rm rest}$ \\
 &  &  & \AA & erg/s/cm$^2$ & km s$^{-1}$ & \AA \\
\hline
WN J2313$+$4053 & 2.99 $\pm$ 0.01 & \OVIfull & 4005 $\pm$ 3 & $-$26 $\pm$ 4 & 3300 $\pm$ \phn500 & $-$33 $\pm$ 10 \\ 
 & &         \Lya     & 4846 $\pm$  5 & 330 $\pm$ 50 & 4100 $\pm$ 1000 & 350 $\pm$ 50 \\
 & & \SiIV / \OIV     & 5570 $\pm$ 19 &  17 $\pm$  3 & 4150 $\pm$ 1450 &   4 $\pm$  \phn1 \\
 & &         \CIVfull & 6178 $\pm$  3 & 164 $\pm$ 17 & 5100 $\pm$  \phn300 &  31 $\pm$  \phn3 \\
 & &        \CIIIfull & 7614 $\pm$  8 &  66 $\pm$  7 & 4850 $\pm$  \phn650 &  14 $\pm$  \phn1 \\
\\
TXS J2321+2237 & 2.553 $\pm$ 0.001 & \Lya & 4320 $\pm$ 1 & 8.1 $\pm$ 0.8 & 1050 $\pm$ \phn200 & 44 $\pm$ \phn7 \\
\\
TXS 2334+1545 & 2.480 $\pm$ 0.003 & \Lya & 4231 $\pm$ 3 & 3.1 $\pm$ 0.5 & 1900 $\pm$ \phn450 & $> 65$ \\
 & &  \CIVfull & 5402 $\pm$ 36 & 0.1 $\pm$ 0.1 & $< 550$ & $> 1$ \\
 & & \CIIIfull & 6644 $\pm$ 15 & 0.8 $\pm$ 0.3 & 1600 $\pm$ 1200 & 12 $\pm$ \phn5 \\
\\
TXS J2355-0002 & 2.587 $\pm$ 0.003 & \Lya & 4366 $\pm$ 1 & 1.2 $\pm$ 0.2 & 900 $\pm$ \phn300 & 27 $\pm$ \phn8 \\
 & & \NVfull  & 4455 $\pm$ 4 & 0.8 $\pm$ 0.3 & 1200 $\pm$ \phn600 & 22 $\pm$ \phn8 \\
 & & \CIVfull & 5551 $\pm$ 2 & 0.3: $\pm$ 0.1 & 450 $\pm$ \phn250 & 6: $\pm$ \phn2 \\
\\
\hline
\multicolumn{7}{l}{$^{\dag}$ Estimated from non-photometric conditions, only to be used to determine line ratios.} \\
\multicolumn{7}{l}{NOTE: Measurements affected by sky-lines are indicated with a colon. See \S 4.3.2 for more information.} \\
\end{tabular}
\end{table}

The redshifts of five objects (TXS~J0137+2521, WN~J0231+3600, TN~J0924$-$2201, TN~J1123$-$2154, and MP~J2003$-$8340) are based on only a single emission line identification (\Lya\ or \OIIfull). We believe these identifications are correct (with a possible exception of TN~J1123$-$2154 due to the faintness of the line), because the alternative explications would have predicted other emission lines in the observed region of the spectrum, which were not detected \citep[see][for a discussion]{wvb99}. Moreover, the derived redshifts are consistent with the predictions from the Hubble $K-z$ diagram \citep{wvb98}. A final argument is the large equivalent width of the line identified with \Lya, although this argument alone is insufficient to exclude a line as \OIIfull\ \citep{ste00}.

The above line identifications yielded redshifts for 50 of the 62 objects that we observed spectroscopically. All 16 objects from the subsidiary USS samples yielded redshifts, while this was possible for 74\% of the 46 objects selected from our main USS sample. This difference is most like likely due to the lower average redshift and higher radio flux densities in the subsidiary samples, as compared to our main sample. In \S 3.4, we discuss the sources which defied redshift determination. Of the objects with redshifts, 94\% are radio galaxies, while 6\% are quasars (see WN~J2313+4053 ,Fig. 1.46 for an example).

\subsection{Notes on individual sources}
We now discuss the redshift determination of the individual sources. We list all sources from our main and subsidiary samples in order of right ascension. We show the extracted one-dimensional spectra in figure \ref{spectra}.

{\it USS J0006$-$0141 (Fig. 1.1), at $z=1.541$,} exhibits a classic HzRG spectrum. The shape and flux level of the continuum is uncertain due to problems with internal reflections.

{\it WN~J0040$+$3857 (Fig. 1.2), at $z=2.606$,} exhibits a classic HzRG spectrum, with the \Lya\ emission slightly weaker than normal. The \CIVfull\ line is severely affected by the strong [OI]~$\lambda$~5577~\AA\ skyline. 

{\it TN~J0121$+$1320 (Fig. 1.3), at $z=3.516$,} displays a clear associated \HI\ absorption system in the \Lya\ line. The observations were made in non-photometric conditions with heavy cirrus.

{\it TXS J0137$+$2521 (Fig. 1.4), at $z=2.897$,} displays a single emission line, and no continuum emission. We identify this line as \Lya\ on the basis of the absence of confirming lines if the line were \OIIfull.

{\it 6C  J0143$+$3253 (Fig. 1.5), at $z=4.413$,} was the highest redshift radio galaxy at the time of discovery \citep{raw96}. Our Keck spectrum shows a weak \CIVfull\ line, confirming the redshift reported by \citet{raw96}. The \Lya\ line shows strong blueward absorption, like the $z=4.11$ galaxy TN~J1338$-$1942 \citep{deb99}. This sources was not included in our main sample because the spectral index $\alpha_{325}^{1400}=-1.15$ does not qualify our extreme $\alpha<-1.30$ criterion.

{\it TN~J0205$+$2242 (Fig. 1.6), at $z=3.506$,} has a very luminous and extended (8\arcsec) \Lya\ halo. The velocity structure of the \Lya\ line repeats at lower S/N in \CIV. The observations were made in non-photometric conditions with heavy cirrus.

{\it WN~J0231$+$3600 (Fig. 1.7), at $z=3.079$,} has a redshift based on only one emission line, which we identify with \Lya. It is extremely unlikely that the identification of this line is \OIIfull, because the corresponding \OIIIfull\ is not detected, while this line is 2.4 times brighter than \OIIfull\ in the composite HzRG spectrum of \citet{mcc93}. The presence of a small continuum break is consistent with the identification of the line as \Lya.

{\it WN~J0303$+$3733 (Fig. 1.8), at $z=2.504$,} exhibits a classic HzRG spectrum with strong \Lya\ emission.

{\it WN~J0310$+$3644 (Fig. \ref{nozfig})} shows continuum emission, but no emission or absorption lines. The high resolution radio map of Kaplan (2000b)\nocite{kap00b} shows a complex morphology reminiscent of a core-jet or head-tail FR-I type radio source.

{\it TN~J0402$+$1007 (no spectrum shown)} is an ERO with $R-K>6.8$ and $K=17.2$ in a 4\arcsec\ aperture. The optical spectrum shows a marginal detection of a broad emission line around 7074~\AA\ in one of the two exposures, but this needs re-confirmation. If this source would fall on the Hubble $K-z$ relation, we could identify this possible line with \MgIIfull\ at $z=1.52$. A near-IR $H-$band spectrum would be able to detect the \Ha\ line in this case.

{\it TN~J0452$-$1737 (Fig. 1.9), at $z=2.256$,} has a low-ionization spectrum with \CIIfull\ stronger than the confirming \HeIIfull\ and \CIIIfull\ lines. This object shows the difficulty to identify the redshift of objects in the ``redshift desert'' (see \S 4.2.5).

{\it TNR J0516$+$0637 (Fig. 1.10), at $z=0.357$,} is dominated by an old stellar population. The redshift is confirmed by the presence of the 4000~\AA\ break, and the \OIIfull\ emission line.

{\it TN~J0517$-$0641 (no spectrum shown)} remains undetected in our ESO 3.6m spectrum. 

{\it WN~J0617$+$5012 (Fig. 1.11), at $z=3.153$,} has the faintest observed K-band flux in our sample ($K=21.9 \pm 0.4$ in a 2\arcsec\ aperture). Figure \ref{glue0617} shows the region of the spectrum containing the only emission line we detect, which we interpret as \Lya\ at $z=3.15$, based on its large spatial extent and equivalent width, and on the absence of confirming lines. The line consists of a brighter ``core'' which coincides with the region between the radio lobes, and two fainter ``wings'', extending 2\arcsec\ (top) and 4\arcsec\ (bottom) beyond the radio lobes. Also note the asymmetry in the radio lobes with respect to the K-band identification. This asymmetry is also seen in TN~J1338$-$1942 \citep{deb99}, and might reflect density variations in the ambient medium of radio galaxies at the highest redshifts.

\vspace{-3cm}
\centerline{\psfig{file=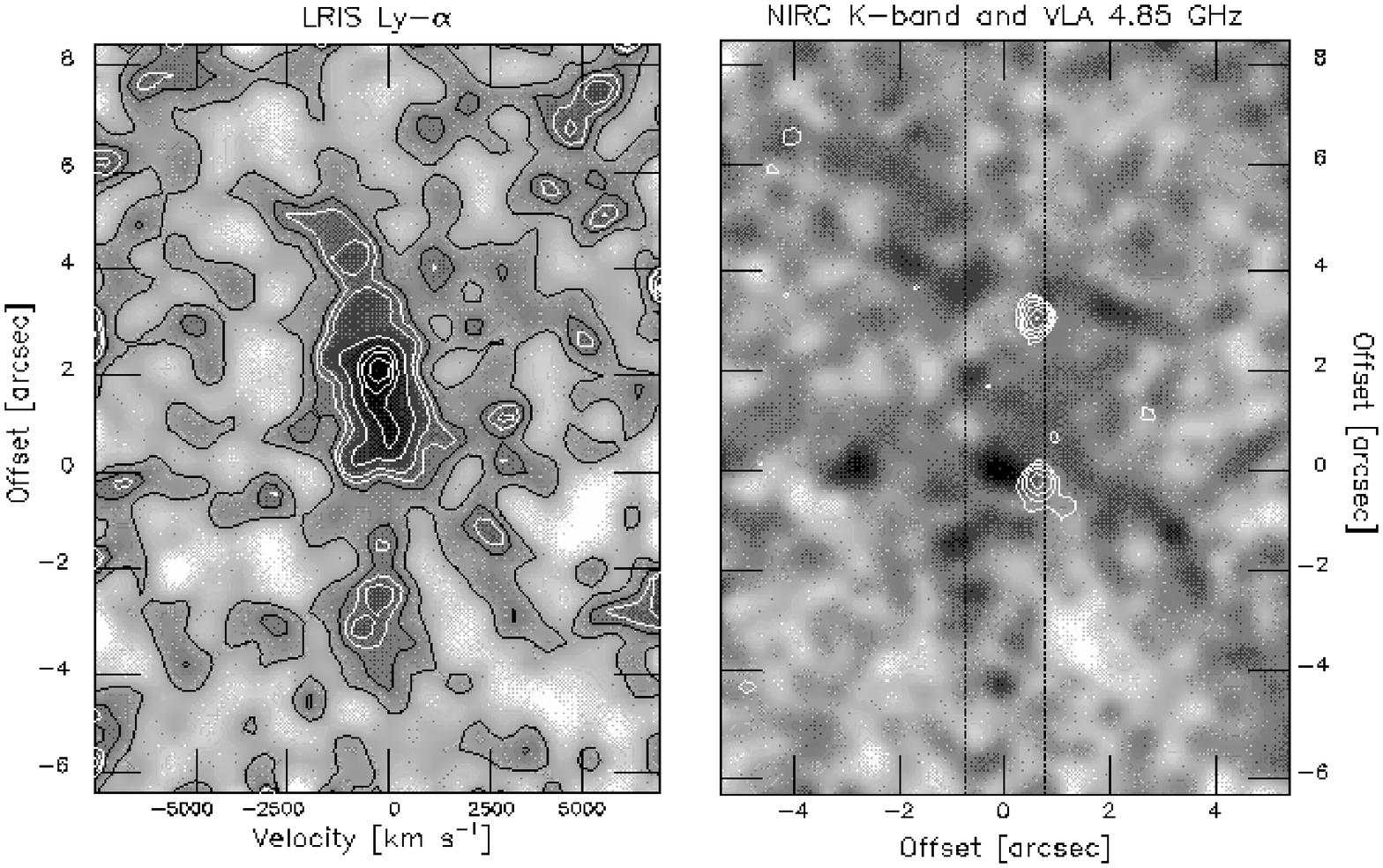,width=13cm}}
\addtocounter{figure}{1}
\figcaption[glue0617]{\small {\it Left:} two dimensional spectrum of the \Lya\ line at $z=3.153$ of WN~J0617$+$5012 obtained with LRIS at Keck II ($t_{int}=90$ min), showing the ``core'' of the line and two faint ``wings'' extending 4\arcsec\ to the bottom at the velocity of the core and 2\arcsec\ to the top with blueshifts of up to 2000 km s$^{-1}$. {\it Right:} K-band image obtained with NIRC at Keck I ($t_{int}=64$ min) with VLA 4.85~GHz radio contours overlaid (subject to an astrometric uncertainty of $\simlt$1\arcsec). The position of the spectroscopic slit in the left panel is indicated with dotted lines. Both panels are at the same scale and orientation (position angle 10\arcdeg\ East of North). Note that the \Lya\ ``core'' falls in between the 3\farcs4 double radio source, but the wings extend further out than the radio source. \label{glue0617}} 

{\it WN~J0717$+$4611 (Fig. 1.12), at $z=1.462$,} is a red quasar. Keck spectro-polarimetry has given strong support for dust scattering as the origin of the redness. The object is described in detail by \citet{deb98}.

{\it WN~J0747$+$5611 (Fig. 1.13), at $z=2.992$,} has a weak emission-line spectrum. The blue wing of the \Lya\ line is strongly attenuated by associated \HI\ absorption.

{\it WN~J0813$+$4828 (Fig. 1.14), at $z=1.274$,} exhibits the classical emission lines from a lower redshift radio galaxy. The \NeIIIfull\ emission falls in a region that is severely affected by fringing of the CCD.

{\it TN~J0910$-$2228 (Fig. \ref{esonozfig})} shows continuum emission but no emission or absorption lines in our ESO 3.6m spectrum. A slightly deeper spectrum with a larger spectral coverage will probably be sufficient to determine the redshift.

\centerline{\psfig{file=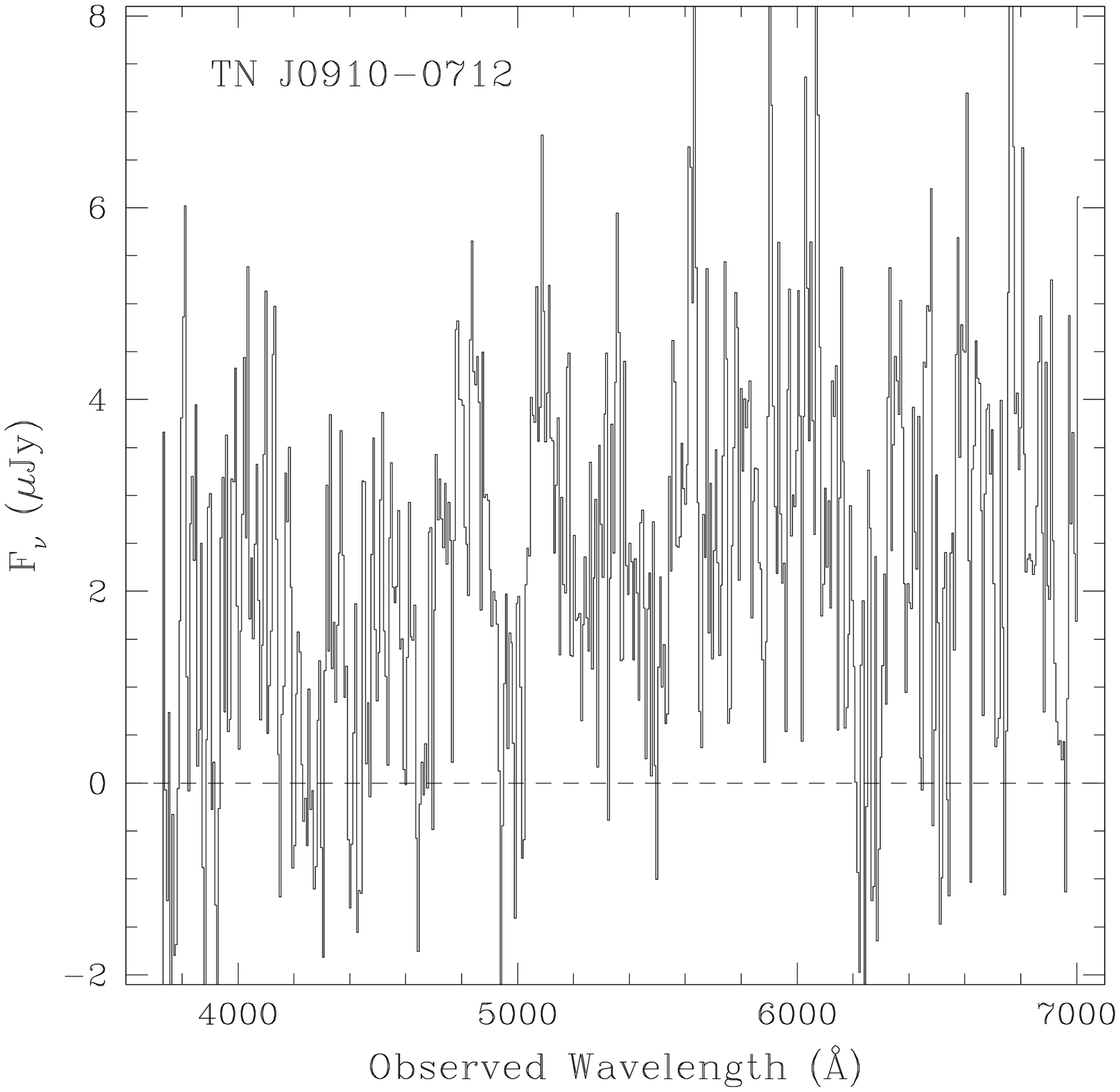,width=8cm}}
\figcaption[esonozfig]{\small Spectrum of TN~J0910$-$0712 obtained with EFOSC1 on the ESO~3.6m telescope showing continuum emission, but no identifiable features. \label{esonozfig}} 

{\it TN~J0920$-$0712 (Fig. 1.15), at $z=2.760$,} has strong absorption at the blue side of the \Lya\ line. Such absorption by associated \HI\ is often seen in HzRGs \citep[\eg][]{deb00b}, and other high-redshift galaxies \citep[\eg][]{dey98}.

{\it TN~J0924$-$2201 (Fig. 1.16), at $z=5.19$,} is the highest redshift source from our sample. It is described in detail by \citet{wvb99}.

{\it TN~J0936$-$2243 (Fig. 1.17 and Fig. \ref{tn0936spec}), at $z=1.479$,} is an object consisting of two components, one of them showing an old stellar population, and the other showing a brighter emission line spectrum. We discuss this object in greater detail in \S4.3.

{\it TN~J0941$-$1628 (Fig. 1.18), at $z=1.644$,} displays a low ionization spectrum, which might well have an important contribution from shock ionization \citep[see][]{deb00b}. Although the object is in the ``redshift desert'', the emission lines are bright enough to be easily detected.

{\it WN~J1015$+$3038 (Fig. 1.19), at $z=0.54$,} is an ERO with $R-K=5.2$ in a 4\arcsec\ diameter aperture. The source has faint optical continuum emission for an object at this redshift, but the $K=17.8$ is consistent with $z \sim 0.5$ in the $K-z$ diagram. The spectrum shown in figure \ref{spectra} is smoothed by a factor of three to increase S/N. The redshift is based on the 4000~\AA\ break and on tentative detections of an unresolved Ca~II~K+H doublet and Mg~I~$b$.

{\it TN~J1026$-$2116 (no spectrum shown)} remains undetected in our 1h Keck spectrum, with the possible exception of the marginal detection of an emission line at 6646~\AA\ that requires confirmation.

{\it TN~J1033$-$1339 (Fig. 1.20), at $z=2.427$,} exhibits a classic HzRG spectrum. The shape and flux level of the continuum is uncertain due to problems with internal reflections.

{\it TN~J1102$-$1651 (Fig. 1.21), at $z=2.111$,} exhibits a classic HzRG spectrum with \Lya\ at the very edge of the spectrum, where the flux calibration is more uncertain.

{\it TN~J1112$-$2948 (Fig. 1.22), at $z=3.09$,} exhibits a classic HzRG spectrum. The \CIVfull\ emission is severely affected by the strong [OI]~$\lambda$ 6300~\AA\ skyline 

{\it WN~J1115$+$5016 (Fig. 1.23), at $z=2.54$,} displays a \CIVfull\ broad absorption line system at $z=2.494 \pm 0.005$ that completely attenuates the \CIVfull\ emission. This is only the second radio galaxy showing a BAL system, after TXS~J1908$+$7220 \citep[][Fig. 1.38]{dey99b}. The region of the spectrum where we predict the \NeIVfull\ line is affected by a cosmic ray hit in one of our exposures, so we do not attempt to fit the parameters of this line.

{\it TN~J1123$-$2154 (Fig. 1.24), at $z=4.109$,} has no continuum detected, but only one faint, slightly extended emission line, which is repeated in two individual exposures. We provisionally identify this line as \Lya\ on the basis of the absence of confirming \OIIIfull\ and \Ha\ lines if the identification had been \OIIfull\ or \OIIIfull. The $z=4.109$ would be consistent with the faint $K=20.4$ identification. The spectrum shown in figure \ref{spectra} is smoothed by a factor of 3 to increase the signal to noise ratio.

{\it WN~J1123$+$3141 (Fig. 1.25), at $z=3.217$,} shows clear signs of interaction in the $K-$band image presented in paper~II. The noisy area in the spectrum around 5500~\AA\ is due to the dichroic.

{\it TN~J1151$-$3013 (no spectrum shown)} remains undetected in our ESO 3.6m spectrum.

{\it TN~J1159$-$1629 (no spectrum shown)} remains undetected in our ESO 3.6m spectrum.

{\it WN~J1242$+$3915 (Fig. 1.26), at $z=2.131$,} is an example source in the `redshift desert', where the redshift must be based on weaker emission lines.

{\it WN~J1314$+$3649 (Fig. \ref{nozfig})} displays a faint continuum in our 1h Keck spectrum, but no emission or absorption lines.

{\it WN~J1333$+$3037 (Fig. 1.27), at $z=1.213$,} has a hint of broad \MgIIfull\ emission, suggestive of a scattered quasar component \citep[\eg][]{tra98}.

{\it WNR J1338$+$3532 (Fig. 1.28), at $z=2.769$,} exhibits a classic HzRG spectrum, with indications for associated \HI\ absorption near the peak of the \Lya\ line. The apparent discontinuity at $\sim$6000~\AA\ is due to the dichroic.

{\it TN~J1338$-$1942 (Fig. 1.29), at $z=4.11$,} was at the time of discovery the highest redshift radio galaxy in the southern hemisphere. It has one of the most luminous \Lya\ halos observed. More details can be found in \citet{deb99}. The \Lya\ flux listed in table \ref{lineparameters} is from VLT observations, while the \CIVfull\ and \HeIIfull\ measurements are from the discovery ESO 3.6m spectrum, which covers a larger wavelength range.

{\it WNR J1356$+$3929 (Fig. 1.30), at $z=0.2525$,} shows a bright continuum which is at some places severely affected by fringing in our single exposure.

{\it TNR J1402$-$1510 (Fig. 1.31), at $z=0.739$,} shows a rich emission line spectrum and a 4000~\AA\ continuum break.

{\it PKS J1428$-$1502 (Fig. 1.32), at $z=2.349$,} exhibits a classic HzRG spectrum. The continuum break across \Lya\ is uncertain because it is at the edge of the CCD flux calibration.

{\it WN~J1525$+$3010 (no spectrum shown)} remains undetected in our deep Keck spectrum. 

{\it WN~J1543$+$3512 (Fig. 1.33), at $z=0.703$,} is dominated by an old stellar population, with \OIIfull\ observed weakly in emission. The spectrum in appendix A is smoothed by a factor 3 to increase the S/N.

{\it TXS J1650$+$0955 (Fig. 1.34), at $z=2.510$,} exhibits a classic HzRG spectrum. The shape and flux level of the continuum is uncertain due to problems with internal reflections.

{\it WNR J1703$+$3739 (Fig. 1.35), at $z=0.256$,} shows a bright continuum which is at some places severely affected by fringing in our single exposure.

{\it MP J1755$-$6916 (Fig. 1.36), at 2.551,} exhibits a classic HzRG spectrum. The \Lya\ emission is spatially extended over $\simgt 7\arcsec$, and has a complex morphology.
 
{\it MP J1758$-$6738 (Fig. 1.37), at 2.026,} is an example source in the `redshift desert', where the redshift must be based on weaker emission lines.

{\it WN~J1836$+$5210 (Fig. \ref{nozfig})} displays a relatively bright continuum with several changes in the slope, but no emission or absorption lines.

{\it TXS J1908$+$7220 (Fig. 1.38), at $z=3.536$,} we reproduce the deep Keck spectrum by \citet{dey99b} which shows the first broad absorption line system detected in a radio galaxy. This source is also known as 6C~1908+72.

{\it WN~J1911$+$6342 (Fig. 1.39), at $z=3.590$,} falls at a redshift where \Lya\ coincides with the strong [OI]~$\lambda$ 5577~\AA\ skyline. However, the high redshift is confirmed by two other emission lines and a continuum break across \Lya.

{\it WN~J1917$+$6635 (Fig. \ref{nozfig})} displays a faint continuum in our 1h Keck spectrum, but no emission or absorption lines.

{\it MP J1929$-$3732 (Fig. 1.40), at 0.7476,} has a very rich emission line spectrum. At least 19 emission lines can be identified in our spectrum.

{\it TN~J1954$-$1207 (Fig. \ref{nozfig})} displays a faint continuum in our 1h Keck spectrum, but no emission or absorption lines.

{\it MP J2003$-$8340 (Fig. 1.41), at 1.169,} is the southernmost source from our spectroscopic sample. We detect only a single bright emission line and continuum emission on both sides, which excludes \Lya\ as a possible identification. We can also exclude \OIIIfull\ and \Ha\ on the basis of the absence of confirming lines within our observed wavelength range.

{\it TXS J2036$+$0256 (Fig. 1.42), at $z=2.130$,} exhibits a classic HzRG spectrum. The shape and flux level of the continuum is uncertain due to problems with internal reflections.
 
{\it WN~J2044$+$7044 (no spectrum shown)} remains undetected in our Keck spectrum. This source has a higher galactic extinction than most others: $A(R)=1.26$, as determined from the extinction maps of \citet{schl98}, and using the extinction curve of \citet{car89}.

{\it MP J2045$-$6018 (Fig. 1.43), at 1.464,} is an extremely luminous radio source. With $P_{408}=2 \times 10^{36}$~erg~s$^{-1}$~Hz$^{-1}$, it surpasses even the most luminous 3C sources, but not a number of radio galaxies at very high redshift, such as 8C~1435$+$635 at $z=4.25$ \citep{lac94}. The parameters for the \MgIIfull\ emission are uncertain because this line coincides with atmospheric B-band. The ionization spectrum is typical of a large, powerful radio source, showing a clear dominance of photo-ionization \citep[\eg][]{bes00,deb00b}. A detailed analysis of this source is deferred to a future publication.

{\it MP J2048$-$5750 (Fig. 1.44), at 1.262,} has very broad emission lines, around 1500~km~s$^{-1}$ deconvolved FWHM for \OIIfull. Such line widths would be slightly higher than all other powerful radio galaxies around this redshift \citep[\eg][]{bau00}. However, this object does not appear to be a quasar, because we do not detect broad hydrogen lines (\Hdelta, \Hgamma, \Hbeta), and the $K-$band identification is slightly resolved (see paper~II).

{\it TNR J2254$+$1857 (Fig. 1.45), at $z=2.153$,} shows a bright \Lya\ line with associated \HI\ absorption.
 
{\it WN~J2313$+$4053 (Fig. 1.46), at $z=2.99$,} is the optically brightest object we observed spectroscopically. It is a quasar with a continuum slope of $\alpha =-2.2$ ($F_{\nu} \propto \nu^{\alpha}$). The broad absorption near 4000~\AA\ is consistent with a detached \OVIfull\ broad absorption line (BAL) system with a velocity width of $3300 \pm 500$~km~s$^{-1}$ and a $z$ offset by 36,000~km~s$^{-1}$. No absorption system is seen in any other lines. A deeper spectrum would be needed to confirm the \OVIfull\ BAL. An object that shows an \OVIfull\ BAL but not a \CIVfull\ one is quite unique.
Normally BALQSOs are identified by their \CIVfull\ BAL \citep{wey91}. The significance of this discovery is that it may be the tip of the iceberg for an under-detected population of BALQSOs, ones that show BALs only from very high ionized species (\OVI\ and above).
The radio spectrum of this source is also exceptional, and shows a strong convex curvature from $\alpha_{151}^{325}=-2.28$ to $\alpha_{325}^{1400}=-1.50$.

{\it TXS J2321$+$2237 (Fig. 1.47), at $z=2.553$,} shows the \Lya\ and \SiIV/\OIV\ complex; the \CIV, \HeII\ and \CIII\ lines are too weak to detect in the blue arm, which experienced technical problems (see \S 2.2.2).

{\it TXS J2334$+$1545 (Fig. 1.48), at $z=2.480$,} has a relatively bright continuum and weak emission lines.

{\it TXS J2353$-$0002 (Fig. 1.49), at $z=2.587$,} shows \NVfull\ emission that is almost as bright as \Lya. Such high \NV/\Lya\ ratios have been detected before in HzRGs \citep[\eg]{oji94,dey95}, and have been explained by the strong attenuation of \Lya\ by dust. The \CIVfull\ line is close to the strong [OI]~$\lambda$ 5577~\AA\ skyline, and the flux might well be affected. The \HeIIfull\ line is not detected. A deeper and higher resolution spectrum will be needed to confirm these high \NV/\CIV\ and \NV/\HeII\ ratios, which are more likely indicative of a high nitrogen abundance \citep{vil99,ver99} than of attenuation of \Lya\ by large amounts of dust. 

\subsection{Sources without redshifts}
For 35\% of the spectroscopically observed sources from our main USS sample, we could not determine the redshift, because either no distinctive emission or absorption features were detected, or no emission was detected in the spectrum. In table \ref{noz}, we list all 12 such sources with the reason for the failure of the redshift determination. We only include sources in this table that have been observed sufficiently long and under good conditions (\ie not during twilight, heavy obscuration or bad seeing). We find that half of the spectra do not have any emission detected, but this fraction drops to a third when we consider only the deep Keck spectra. In Figure \ref{nozfig}, we show the spectra where we did detect continuum emission, but no identifiable emission or absorption features.

\begin{table}
\caption{Attempted spectroscopic observations not yielding a redshift}
\tiny
\begin{tabular}{rrrrrrlll}
\hline
Source & $S_{1400}$ & $\alpha$ & LAS & $m_R~^a$ & $m_K~^a$ & run & $t_{exp}$ & comments \\
 & mJy &  & \arcsec & mag & mag &  & s & \\
\hline
WN J0310$+$3644 &  24 & $-$1.70 &  2.0 &    23.0$^b$ & \nodata & K9  & 3600 & flat continuum, no lines \\
TN J0402$+$1007 &  75 & $-$1.38 &  0.8 & $>$24   & 17.2    & K9  & 1200 & undetected, possible line at 7074~\AA \\
TN J0517$-$0641 &  40 & $-$1.46 &  1.0 &    24.0 & \nodata & E2  & 2700 & undetected \\
TN J0910$-$2228 &  55 & $-$1.58 &  2.0 &    23.0 & \nodata & E3  & 1800 & continuum, no lines \\
TN J1026$-$2116 &  62 & $-$1.39 &  1.0 & $>$24   & 19.8    & K9  & 3600 & undetected, possible line at 6646~\AA \\
TN J1151$-$3013 &  53 & $-$1.57 &  1.0 & $>$24   & 19.8    & E3  & 2700 & undetected \\
TN J1159$-$1629 &  53 & $-$1.35 &  1.4 &    23.4 & \nodata & E3  & 3600 & undetected \\
WN J1314$+$3649 &  36 & $-$1.41 &  1.3 & $>$26   & 22.1    & K11 & 3600 & flat continuum, no lines \\
WN J1525$+$3010 &  17 & $-$1.46 &  1.2 &    25.3 & 19.5    & K11 & 5400 & undetected \\
WN J1836$+$5210 &  24 & $-$1.41 &  1.4 & \nodata & 17.7    & K11 & 1200 & wavy continuum, no lines \\
WN J1917$+$6635 &  11 & $-$1.30 &  1.9 & \nodata & 20.0    & K11 & 3600 & flat continuum, no lines \\
TN J1954$-$1207 & 103 & $-$1.38 & $<$7 & \nodata & 19.9    & K11 & 3600 & flat continuum, no lines \\
WN J2044$+$7044 &  42 & $-$1.38 &  1.3 & $>$24   & 19.2    & K11 & 3000 & undetected \\
\hline
\multicolumn{9}{l}{$^a$ 4\arcsec\ diameter aperture, see Chapter~3 for details.} \\
\multicolumn{9}{l}{$^b$ $I-$band magnitude.} \\
\end{tabular}
\label{noz}
\end{table}

\centerline{\psfig{file=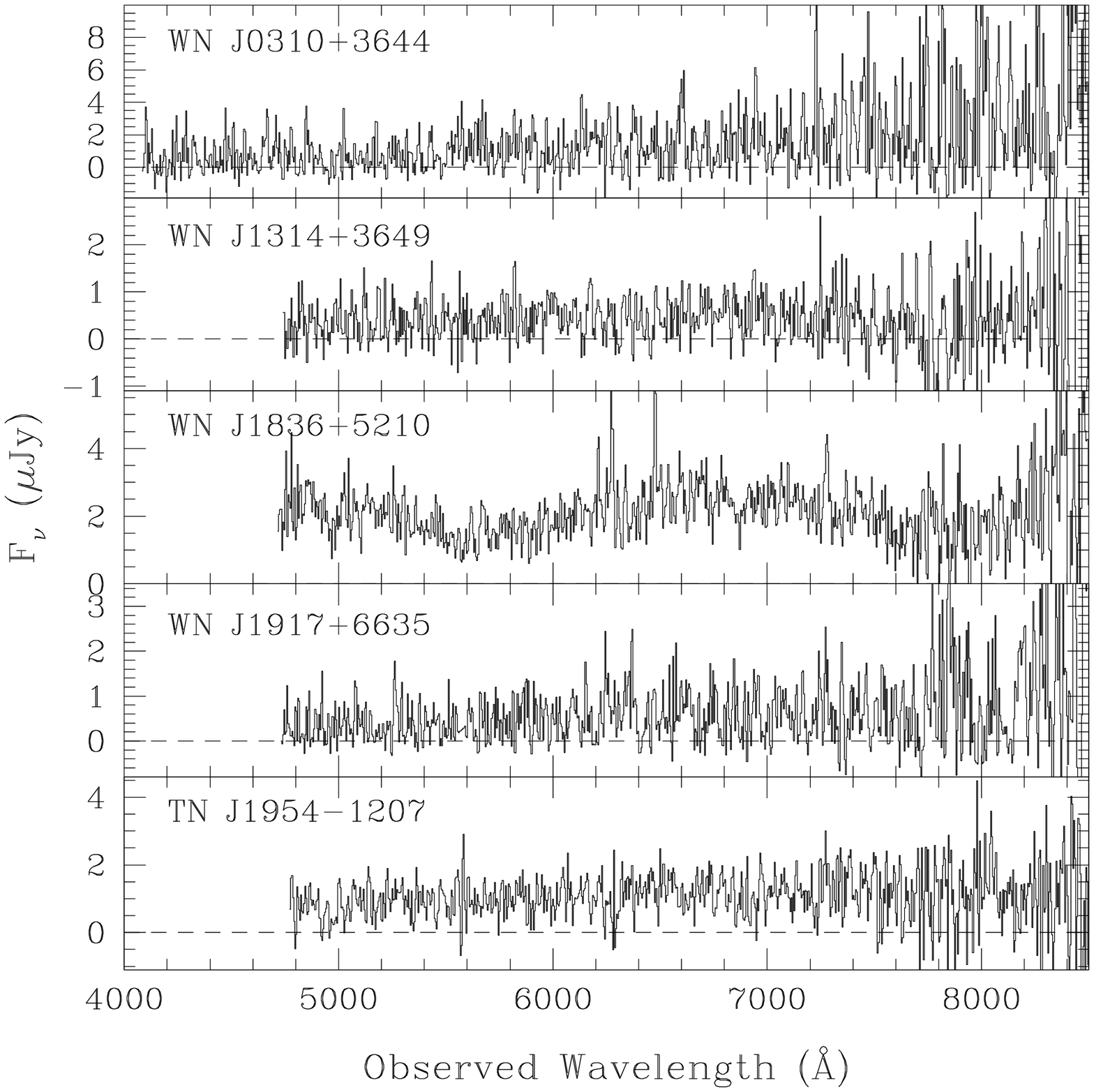,width=19cm}}
\figcaption[nozfig]{\small Spectra obtained with LRIS at Keck showing continuum emission, but no identifiable features. Although the spectral coverage extends out to $\sim 1 \mu$m, we show only the part out to 8500~\AA\ because the S/N at longer wavelengths is too low to provide useful information. \label{nozfig}}

The most obvious reason for the failure to detect any optical emission could be an incorrect offset to the object. However, we are confident this is not the case for the sources observed with the Keck telescope because (i) we used the same procedures and quality of images to determine the offsets for both the detected and undetected sources, and (ii) other objects detected along our 2\farcm6 long slit provided us with accurate spectro-astrometry with which we could verify the proper slit placement, based on the optical and/or near-IR images of the field.

We also exclude the possibility that the identification we determined is not the host galaxy, but an optical counterpart of an individual component of the radio source because: 
(i) our USS sample was constructed to exclude objects $>$1\arcmin, and follow-up high resolution radio images did not show other components in the field; 
(ii) the position differences between the different frequency and resolution radio observations agree within $\sim$1\arcsec\, indicating we are not missing any radio source component with a significantly different radio spectrum; 
(iii) the identifications of the sources in table \ref{noz} are within 1\arcsec\ from the predicted radio positions, and do not show a different near-IR morphology or $R-K$ color than the confirmed HzRG sources;
(iv) the near-IR identifications of 85 sources from our main USS sample are virtually complete, with no large position differences between the radio and near-IR positions.

We discuss the possible nature of these sources lacking redshifts in \S4.2.

\centerline{\psfig{file=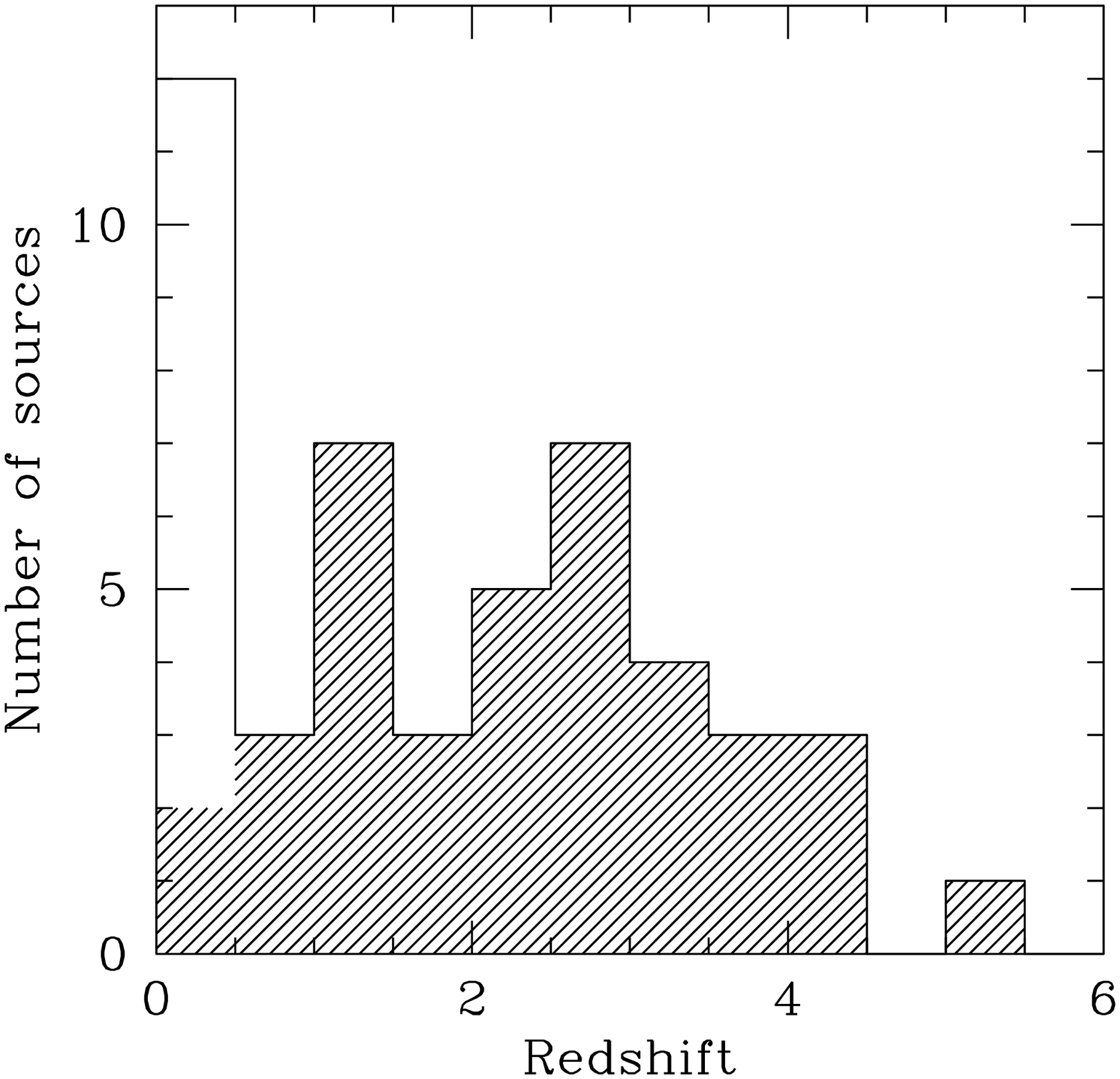,width=8cm}}
\figcaption[zhis]{\small Redshift distribution of the WN/TN sources described in this paper. The unshaded histogram includes the low-redshift cluster identifications. Note the dip at $1.5<z<2.5$ coinciding with the 'redshift desert' (see text). \label{zhis}} 

\centerline{\psfig{file=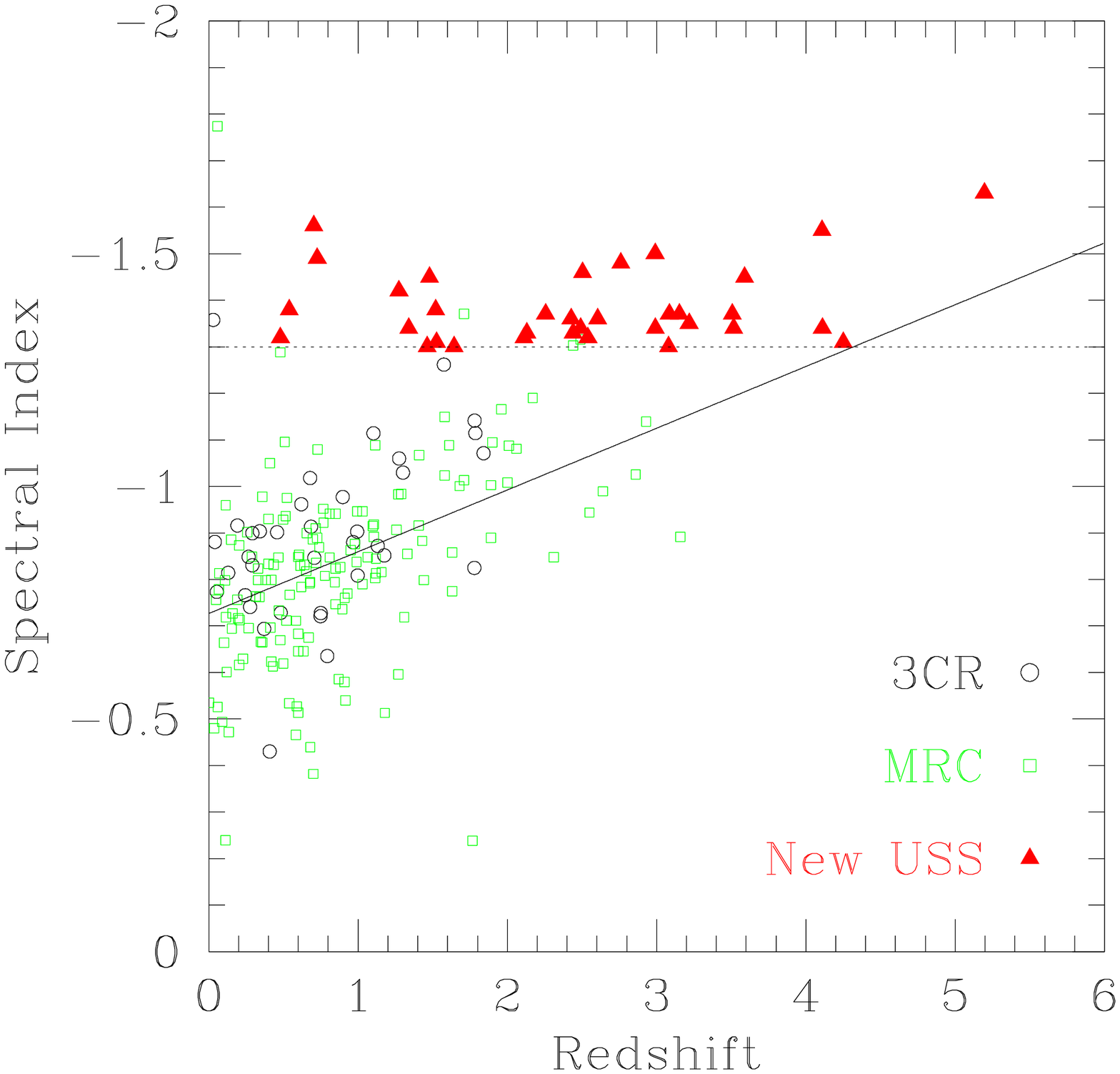,width=9cm}}
\figcaption[zalpha]{\small Spectral index from the 325~MHz WENSS or 365~MHz Texas and 1.4~GHz NVSS survey plotted against redshift for 36 3CR \citep{spi85}, 152 MRC \citep{mcc96}, and 35 WN/TN sources. The solid line shows a linear fit to the 3CR and MRC data while the dashed horizontal line indicates the spectral index cutoff we used in our WN/TN USS samples. \label{zalpha}} 

\section{Discussion}
\subsection{Redshift distribution of USS sources}
Figure \ref{zhis} shows the redshift distribution of all sources with known redshifts in our USS sample (we exclude the 12 sources observed from the other samples in this analysis). We find that the redshift distribution is roughly flat from $z=0.5$ to $z=4$, with a mean of $\bar{z}=2.5$.
Two features deserve special attention: (i) the peak in the lowest redshift bin, and (ii) the slight drop at $1.5<z<2.5$ (see also Fig. \ref{zalpha}). 
The peak at low redshift is due to cluster sources found from the literature. We refer to our discussion in paper~I\nocite{deb00a}, where we argued that a USS sample is an efficient way to select cluster galaxies.
The lack of sources at $1.5<z<2.5$ is probably due to the ``redshift desert''. In this redshift range, \OII\ has shifted outside of the observable spectroscopic window, while \Lya\ has not yet entered, and the redshift has to be based on weaker emission lines. We return to this issue in \S4.2.5.

To examine the relation between our {\it observed} spectral index and redshift in a consistent way, we determined the spectral indices for those spectroscopically identified radio galaxies from the complete 3CR \citep{spi85} and MRC \citep{mcc96} samples. We used spectral indices determined from the same radio surveys as in our USS sample (see paper~I\nocite{deb00a} for a discussion on our sample selection and spectral index completeness issues).
In practice, we determined the $\alpha_{325}^{1400}$ for the 3CR sources, using fluxes from the 325~MHz WENSS \citep{ren97} and 1.4~GHz NVSS \citep{con98}, and the $\alpha_{355}^{1400}$ from the 365~MHz Texas survey \citep{dou96} and NVSS, using a 60\arcsec\ search radius to correlate the survey catalogs. We only used the single component ('S'-type) sources in WENSS to avoid  problems due to different resolutions. From the Texas survey, we only used those sources that were well modeled and not subject to lobe-shift problems \citep['+++' flag, see ][]{dou96}. The complicated beam of the Texas survey leads to problems in the determination of the fluxes for the 3CR survey (which contains many more large sources than the MRC), so we opted not to include Texas-NVSS spectral indices from the 3CR sources in our analysis.

This procedure yields spectral indices for 36 3CR and 152 MRC sources (Fig.~\ref{zalpha}).  Inspection of figure~\ref{zalpha} suggests a correlation between redshift and spectral index which we verify statistically: the Spearman rank correlation coefficient is $r_S=-0.44$, implying a $>$99.99\% likelihood that redshift and spectral index are correlated for this sample.

To obtain an approximate estimate of the number of sources we excluded by the application of our $\alpha < -1.30$ criterion, we fitted a linear function to the $z-\alpha$ relationship (Fig. \ref{zalpha}). The fit has an offset of $\alpha(z=0) = -0.73$, and a slope of $-$0.13. We stress that this fit is not well constrained, especially at $z>2$, where few radio galaxies from flux limited samples are known. A linear fit is probably inadequate at high redshift, and the relation should probably curve up at high redshift because of effects such as (i) additional steepening of the spectra due to inverse Compton losses against the cosmological background, and (ii) a relation between the radio spectral curvature and radio power \citep[\eg][]{blu99,car99}; at the highest redshifts, the most powerful radio sources are more dominant in flux limited surveys, which would lead to a steeper rest-frame spectral index, in addition to the steepening due to $k-$correction.
The combined result would be that our crude linear fit predicts too flat spectral indices at the highest redshifts, which would suggest that by $z \sim 4$, our sample includes roughly half of the sources that would have been in a sample without a spectral index cutoff. This number should be considered {\it very} tentative; a sample of high redshift sources with a much shallower spectral index cutoff would be needed to provide a more reliable measure. 

After the preferential selection of sources with relatively faint $K-$band magnitudes, we find roughly equal numbers of objects from $z=0$ to $z=5$. This is in contrast with pure flux limited surveys, where the fraction of sources above a certain radio luminosity at the highest redshifts decreases when the flux limit is lowered. This decrease is seen, even when there is no redshift cutoff in the co-moving space density \citep{blu98,jar99}. In practice, this means that pure flux limited samples will find much less galaxies at $z \simgt 3$ than at $z \simlt 3$. Because our $K-$band filtered USS sample does find many $z \simlt 3$ sources, it is clearly more efficient in finding the highest redshift objects. 
However, we note that our efficiency may depend to a large amount on the selection of sources with faint $K-$band magnitudes. A $K-$band imaging campaign not only increases the chances of finding very high redshift radio galaxies, but also increases the efficiency of the spectroscopic observations by providing accurate offset positions, that can be used to point the spectroscopic slit at the optimal position. To obtain such images in the optical would require much longer observing times, than the typical 16$-$32~min integrations we used (see paper~II), and the seeing will likely be poorer.

Counting the 19 previously determined redshifts from the literature (see paper~I\nocite{deb00a} and table \ref{quasars}), we now have spectroscopic information for 65 (=10\%) sources in our USS sample. The diffuse radio morphologies for 26 additional sources indicate low redshifts, these 91 sources are still insufficient to draw any firm conclusions on the co-moving space density of radio sources. For example, a sample that extends out to flatter spectral indices ($\alpha > -1.3$) would be needed to better estimate the fraction of sources excluded by our spectral index selection as a function of redshift.

\subsection{Undetermined redshifts}
We now consider the sources of which we could not determine the redshifts either because we detected no spectral features or because no optical emission was detected.
For three of the non-detections, we obtained the spectra with the ESO~3.6m telescope. These sources might well be detectable with 8-10m class telescopes. For example, a 45min integration with the ESO~3.6m telescope did not detect any emission in TN~J0924$-$2201 and TN~J0936$-$2243, while the deeper Keck spectra were sufficient to allow the redshift determination. The only non-Keck spectrum that had a continuum detection, but no redshift (TN~J0910$-$2228, Fig. \ref{esonozfig}) had a spectral coverage from 4000~\AA\ to 7000~\AA. Such a limited wavelength coverage could easily miss some of the brighter emission lines, especially when \Lya\ nor \OII\ are in this wavelength window (from $z=0.9$ to $z=2.3$). In the following, we therefore consider only our deep Keck spectra with a continuous spectral coverage from 4000~\AA\ or 4775~\AA\ to 1~$\mu$m.

We note that sources with only continuum emission have been detected with 4m class telescopes in previous samples of radio sources (\eg\ the 7C and 8C sample of \citet{lac99}, and the USS sample constructed from the WENSS survey by \cite{ren98}). These authors attributed the failure to determine the redshift to a lack of sensitivity prohibiting the detection of emission lines in the $1.2 < z < 1.8$ 'redshift desert' or absorption lines in a stellar continuum dominated spectrum. It is remarkable that the latter class contains mainly compact USS sources, much like the objects in table \ref{noz}, but they are several magnitudes brighter in the optical \citep[$R \sim 22.5$;][]{ren98}.

In the following, we discuss various possibilities for the identifications for the sources in figure \ref{nozfig}.

\subsubsection{Stars}
The fraction of mJy-level radio sources that are identified with the coronal magnetic activity of stars brighter than $V=11$ is $<$0.1\%, while this fraction decreases linearly with increasing $V-$ magnitude \citep{hel99}. Even if radio stars would have the same spectral index distribution as extra-galactic radio sources, we would expect from figure~1 of \citet{hel99} to find $<$1 star in our USS sample, while we have already detected one bright ($V=8.76$) star in our sample (paper~I\nocite{deb00a}). In fact, there are indications that radio (proto-)stars have generally small positive spectral indices $0 < \alpha < 2$ (\citep{luc00}), in which case, we would expect to find even less stars in a USS sample. We conclude that it is unlikely that many more identifications in our sample are stars.
We also exclude the possibility that the spectra are chance alignments of foreground stars, because the radial profiles from the optical or near-IR images in paper~I\nocite{deb00a}I are resolved, and our spectra in figure \ref{nozfig} do not show characteristic stellar absorption lines. 

\subsubsection{BL Lacs or un-beamed quasars}
One of the main characteristics of BL Lac objects is the low equivalent width of their emission lines \citep[$W_{\lambda}^{\rm rest}<$5~\AA; ][]{sto91}. This would be an obvious interpretation for most of our featureless spectra. However, the radio to optical flux ratio is several orders of magnitude higher than seen in BL Lac objects \citep{lau99}, and we therefore reject this interpretation. 

\citet{fan99} reported the discovery of SDSSp~J53259.96-003944.1, an object at $z=4.62$ which shows all the characteristics of a quasar (unresolved optical morphology and $M_B<-23$), except for the broad emission lines, which are missing. The absence of radio and X-ray emission, optical polarization and variability are unlike BL Lac objects, and lead to the suggestion that this object could represent a rare class of quasars without broad emission line regions.
If such an object were observed at $z \lesssim 2.3$, we would not detect the \Lya\ break, and such an object would appear as a featureless continuum. However, the objects with continuum detected in Table \ref{noz} are clearly resolved in the $K-$band images and are significantly fainter than SDSSp~J53259.96-003944.1. They do not qualify as quasars on the basis of absolute magnitude (for $z \simlt2.6$) or compactness. However, if SDSSp~J53259.96-003944.1 does indeed represent a rare type of AGN without broad emission line regions, very little is known about the nature of these objects, and on their radio properties. Our extreme radio spectral index selection might then preferentially select the radio loud counterparts of such objects. Nevertheless, the very steep radio to optical flux ratio and absence of an unresolved component in the near-IR images remain difficult to reconcile with a direct view of the central AGN continuum emission.

\subsubsection{Stellar continuum dominated galaxies with obscured AGN}
Due to the extreme faintness of the identifications, the S/N of our spectra could be insufficient to identify stellar absorption features or continuum breaks at 2640~\AA, 2900~\AA\ or 4000~\AA. In low-redshift objects ($z\lesssim 1$), we would expect to detect these breaks from an old stellar population within our $\gtrsim$4000~\AA\ wavelength coverage, such as in WN~1015+3038 or WN~J1543+3512 (see appendix A). 
Even in the two spectra with slightly better S/N spectra (WN~J1836+5210 and TN~J1954$-$1207) do not see evidence for breaks or absorption features, suggesting that either the continuum is completely non-stellar, or the stellar population is significantly diluted by an underlying non-stellar component. In WN~J0310+3644, WN~J1314+3649 and WN~J1917+6635, even deeper spectra would be needed to determine the possible presence of an old stellar population in the continuum emission. A strong indication for the presence of an old stellar population could come from a large $R-K$ color, but we only have one object with continuum emission that has both $R-$ and $K-$band photometry. However, the distribution of $R-K$ colors of 81 sources from our USS sample shows that at least one third, and possibly more than half of the sources are EROs with $R-K>5$. \citet{dad00} argued that EROs are mainly composed of $z\simgt 1$ ellipticals. Our sample indeed contains at least one such object, TN~J0936$-$2243 (see \S 4.3), and it is well possible that some of the objects in table \ref{noz} also belong to this class of objects.

Alternatively, these objects could be dominated by a relatively young stellar population which is observed after the starburst. Such objects would not show any strong stellar emission lines, and would not yet display strong continuum breaks. Their redshift determination would then depend on weak interstellar absorption lines. The absence of the Lyman break would limit their redshift to $z<2.9$ or $z<2.3$ (WN~J0310$+$3644). However, the generally large $R-K$ colors in our sample are not consistent with the expected blue continuum from the young stars.

\citet{mus00} found that 75\% of the hard X-ray background is composed of AGN, of which half have faint or even optically undetectable counterparts presumed due to an adverse orientation of the molecular torus that surrounds the active nucleus in the context of the unified schemes of AGN. Because our radio survey is also insensitive to dust obscuration, it is therefore conceivable that our sample contains a number of such obscured AGN. Sensitive X-ray imaging of these sources with Chandra or XMM-Newton could reveal these AGN.

Such an obscuration would also be consistent with the second class of ERO identifications, which consists of strongly dust-reddened, star-forming AGN or starbursts \citep[\eg][]{dey99a,sma99}. Large dust masses have been detected in HzRGs \citep[\eg][]{arc00}. If the narrow line region were located behind a thick dust layer, this could obscure much of the emission lines. The objects in table \ref{noz} all have compact radio structures, and are probably still confined within the host galaxy, where the dust would be located. As a result, the narrow emission line regions could still be shielded from our view by a dust layer on the outside of the host galaxy. 
In larger radio sources, extending beyond the bounds of the host galaxy the radio jets could have cleared the dust away, allowing the detection of narrow emission line regions ionized by a central photo-ionizing source, or by shock ionization.

The presence of a large amount of dust should be revealed by bright sub-mm emission, which could be detected by instruments like SCUBA on the JCMT. If these objects are indeed obscured AGN, like the objects that compose the hard X-ray background, they should be detectable in sensitive X-ray surveys with Chandra or XMM-Newton. \citet{bar00} have made deep optical, near-IR, submillimeter, and 20cm radio observations of a deep Chandra field. Their high quality LRIS spectroscopy of 20 sources also failed to provide the redshift of six out of 19 X-ray selected objects. Two of these sources lacking redshift determination have possible submm detections. It might well be that some our sources are very similar objects, but with $\sim$1000 times brighter radio flux densities.

\subsubsection{Intrinsically weak AGN in the redshift desert}
If the object is at a redshift where only the faintest emission lines fall into the observable window, the S/N ratio might still be insufficient, despite our long integrations.
The most probable redshift in this case would lie in the so-called ``redshift desert'' at $1.5 \simlt z \simlt 2.3$, where the brightest restframe optical emission lines (\OII, \OIII, \Ha) have shifted out of the observed range, but the break across \Lya\ is still too blue to be observed. For most sources, except WN~J1314+3649, such redshifts are low, but still consistent within the scatter of the Hubble $K-z$ diagram \citep[\eg][]{wvb98}.

The faintest detectable emission lines in integrations of $\sim$1~hour with LRIS on the Keck telescope have integrated flux densities of $\sim 1 \times 10^{-17}$~erg~cm$^{-2}$~s$^{-1}$ \citep[see also ][]{ste99}. At redshifts within the redshift desert, this corresponds with line luminosities in the range $\sim 1-4 \times 10^{41}$~erg~s$^{-1}$, and radio powers in the range $\log$($P_{325}$/erg s$^{-1}$~Hz$^{-1}$) = 34 - 35. Such values do fall on the extrapolations of the correlations between line luminosities and radio powers for all UV lines, except \Lya\ \citep{deb00b}.
 
Further evidence that we are missing a significant number of sources in this redshift range comes from the redshift distribution of our spectroscopic sub-sample (Fig. \ref{zhis}). Although based on very incomplete data which is subject to complicated non-optical selection effects (\eg\ selecting preferably objects with fainter $K-$band magnitudes), figure \ref{zhis} suggests an under-representation of $\sim 6$ sources with redshifts in the redshift desert. We therefore consider it likely that about half of the sources lacking redshifts are intrinsically weaker AGN with redshifts $1.5 \simlt z \simlt 2.3$. This redshift range falls near the epoch where the co-moving space density of steep spectrum radio sources peaks or flattens off to a constant high level \citep{jar99}, so we expect to find a substantial number of radio galaxies in complete samples of radio sources. Our findings predict it will prove very difficult to obtain complete spectroscopic redshift information for samples of flux densities $\lesssim 100$~mJy. Larger aperture optical telescopes would be needed to obtain redshifts in this difficult redshift range.

\subsubsection{Pulsars}
The radio spectral indices of pulsars show a broad distribution with the mean at $\sim -1.6$ \citep{lor95}, suggesting a fraction of our USS sources could be pulsars. The detection of 3 previously known pulsars in our USS sample and the over-density of $\alpha < -1.6$ sources near the Galactic plane \citep{deb00a} are strong evidence for the presence of a significant number of pulsars in our USS sample. To avoid excessive Galactic extinction, we have excluded the Galactic plane ($|b|<15$\arcdeg) from our sample, which should reduce the number of Galactic pulsars in our USS sample, but it is possible that it still contains a number of pulsars, especially for objects having the steepest spectral indices.

Optical spectroscopy of pulsars has proven difficult, even with large aperture telescopes. We have found optical spectra of three pulsars in the literature: the Crab pulsar \citep{nas96}, PSR~B0540$-$69 \citep{hil97} and Geminga \citep{mar98}. All three spectra show flat continua with possible broad absorption and emission (in the case of PSR~B0540$-$69) features. These spectra are qualitatively similar to most of the spectra in Figure \ref{noz}, retaining these objects as pulsar candidates. If a significant number of these indeed turn out to be pulsars, a sample of extremely steep spectrum sources \citep[\eg][]{kap00a} would be an efficient pulsar search method, especially when concentrated on the Galactic plane. Several objects from our sample have already been observed with radio pulsar search programmes, but without success to date (M.\ Kouwenhoven, private communication; D.\ Backer, private communication).

\subsubsection{Extremely high redshift objects}
An intriguing possible explanation for the objects we did not detect even after integration times of 1 hour or longer with the Keck telescope is that they are at such high redshifts that even the \Lya\ line has redshifted out of the optical window ($z \gtrsim 7$). The $R$=25.3 detection in WN~J1525$+$3010 excludes this object as a potential extremely high redshift galaxy, but the non-detections in $R$ of TN~J1026$-$2116 and WN J2044$+$7044 are consistent with this interpretation, although our optical imaging is too shallow, and cannot be considered as strong evidence that the $R-$band samples a rest-frame wavelength range below the Lyman break. We are planning near-IR spectroscopy to search for possible emission lines in these objects.

\centerline{\psfig{file=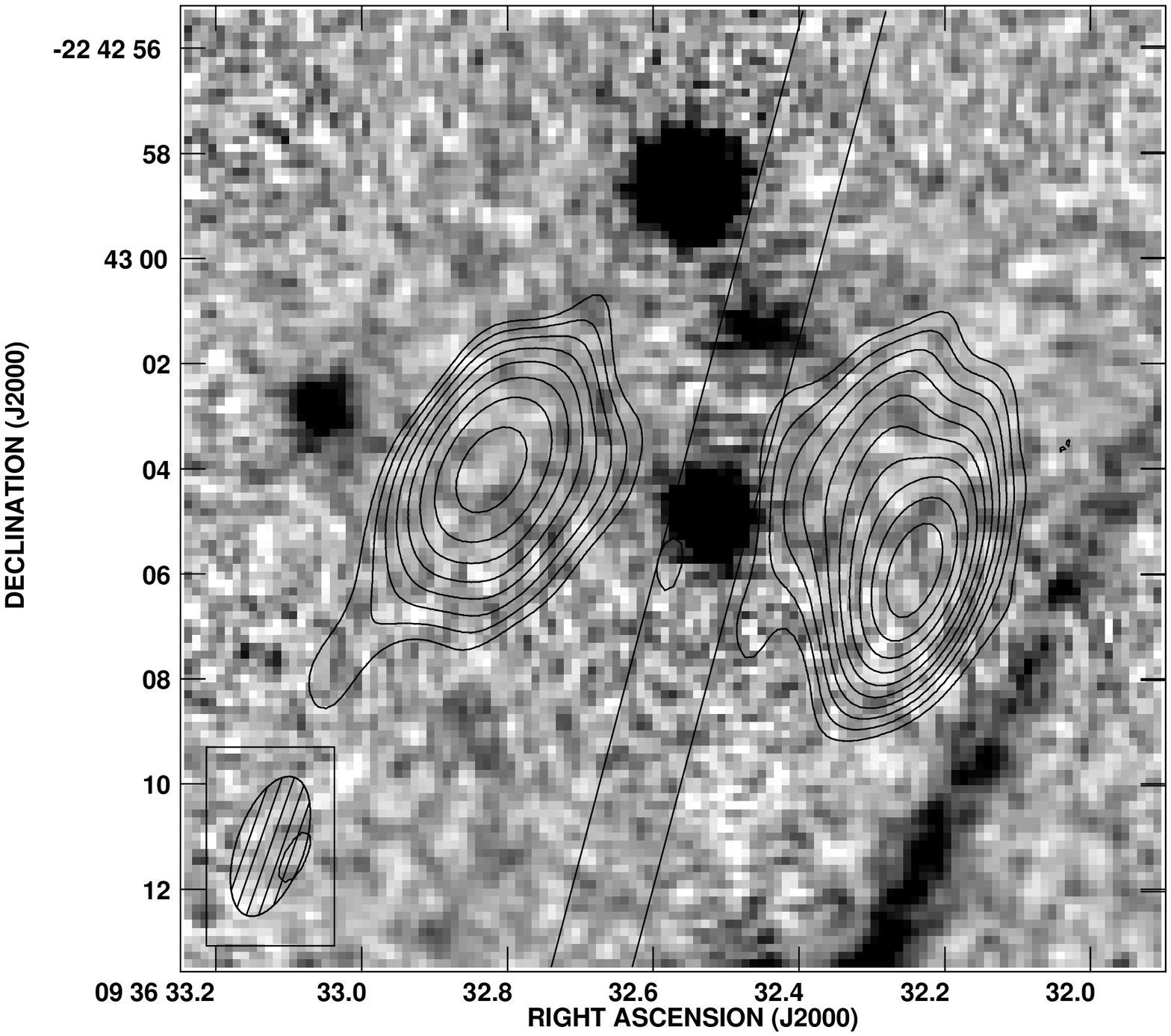,width=9.4cm,angle=-90}}
\figcaption[tn0936kc]{\small {\it Greyscales:} NIRC $K-$band image; {\it contours:} VLA 4.85~GHz radio image of TN~J0936$-$2243. Contour levels are 0.24, 0.34, 0.48, 0.68, 0.96, 1.4, 1.9, 2.7, 3.8 and 5.4 mJy~beam$^{-1}$. The VLA beam is indicated as a hatched ellipse in the lower left corner. The two parallel lines indicate the position of the 1\farcs5 wide long slit used to obtain the spectra shown in figure \ref{tn0936spec}. Note the two objects in the slit, viz. the bright identification at the midpoint of the radio lobes, and the diffuse object due north near the extension of the western radio lobe. The astrometric uncertainty in the registration of the radio and near-IR images is $<$0\farcs4. The dark streak in the bottom right corner is a bleed-out trail of a nearby bright star. \label{tn0936kc}} 

\subsection{TN~J0936$-$2243: an old, red galaxy}
One of the most peculiar sources we observed spectroscopically, is TN~J0936$-$2243. We show an overlay of our VLA radio map (see paper~I\nocite{deb00a}) on our NIRC K-band image (see paper~II) in figure \ref{tn0936kc}. The astrometric solution was based on the USNO-A2.0 (Hipparcos-based) catalog. The error in the relative astrometry is dominated by the uncertainty between the optical and radio reference frames, which is $\sim$0\farcs4 \citep[90\% confidence limit; ][]{deu99}. 
The near-IR identification is a marginally resolved $K=18.3 \pm 0.3$ object at the geometric midpoint of the radio lobes, but the North-Eastern extension of the western radio lobe suggests that the extended $K=19.4 \pm 0.3$ object 3\farcs5 North of the midpoint between the radio lobes might well be related to this object. Higher resolution and deeper radio data will be needed to clarify the radio-optical morphology, and if the companion object should be considered part of the main galaxy or not. In the $R-$band image obtained with the ESO~3.6m telescope (see paper~II), we marginally detect the northern object, but do not detect the southern object down to $R=24$, classifying this object as an ERO with $R-K > 5$. We obtained an LRIS spectrum with the slit passing through the two near-IR objects, as indicated in figure \ref{tn0936kc}.

\centerline{\psfig{file=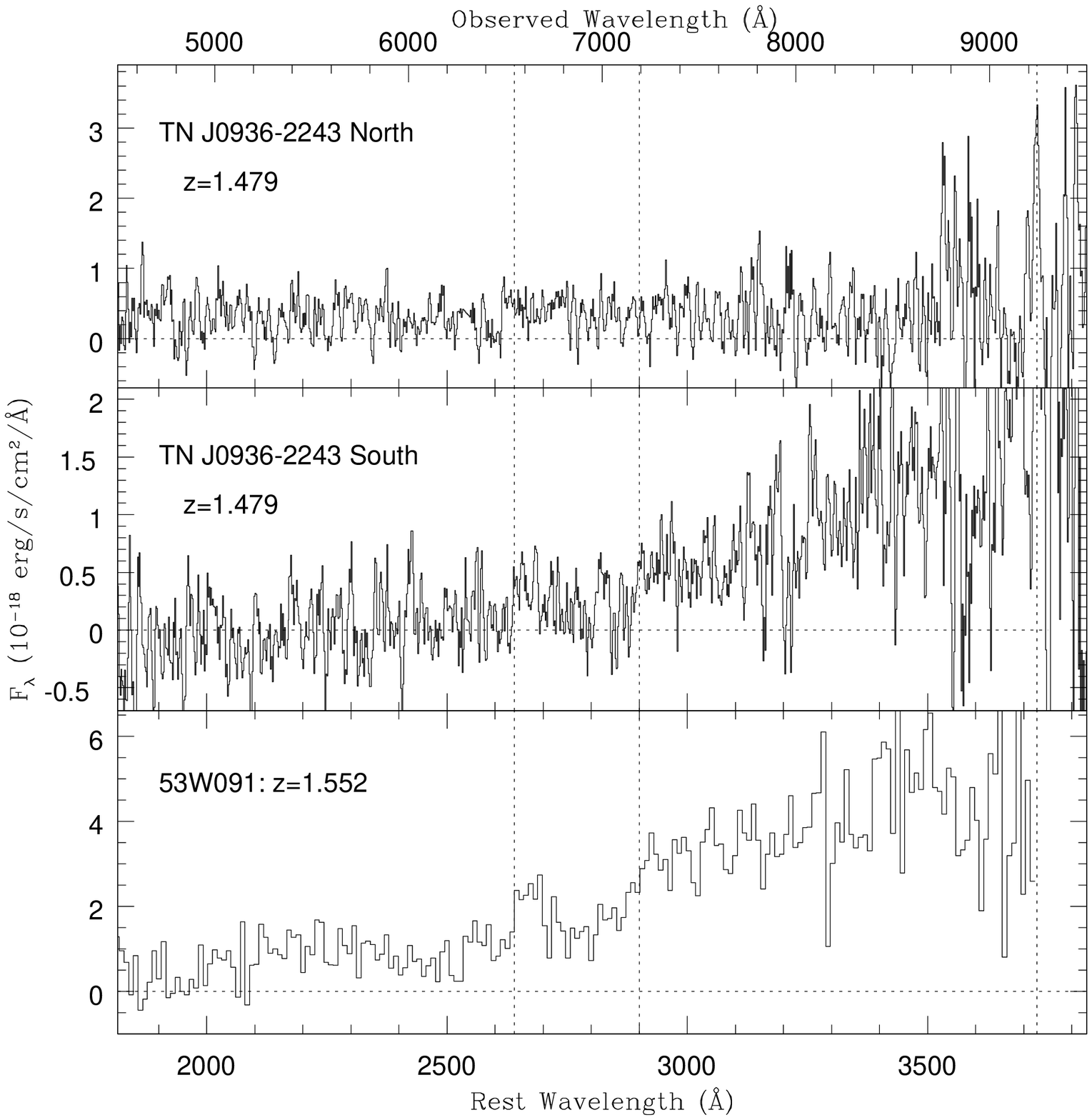,width=13cm}}
\figcaption[tn0936spec]{\small The top and middle panels show the spectra of the northern and southern components of TN~J0936$-$2243. Vertical dotted lines indicate the wavelengths of the 2640~\AA\ and 2900~\AA\ breaks and \OIIfull. For reference, we show the spectrum of LBDS~53W091 from \citet{spi97}. \label{tn0936spec}} 

The spectrum of the fuzzy northern object (Fig. \ref{tn0936spec}, top panel) shows a weak, relatively flat continuum with a single emission line at 9240~\AA, while the southern object (Fig. \ref{tn0936spec}, middle panel) shows a very red object, but no convincing emission lines. If we interpret the emission line in the northern object as [OII]~$\lambda$3727~\AA, and assume the same $z=1.479$ for the southern object, the latter displays the 2640~\AA\ and 2900~\AA\ continuum breaks. We therefore assume both objects are components of the same source, or companion objects.

The spectrum of the southern component closely resembles the old, red galaxies LBDS~53W091 at $z=1.55$ \citep[][Fig. \ref{tn0936spec}, bottom panel]{spi97} and LBDS~53W069 at $z=1.69$ \citep{dun99}. TN~J0936$-$2243 is $\sim 3 \times$ brighter than LBDS~53W091 (which has $S_{1412}=22.1$~mJy and $\Theta < 3\farcs 7$; Windhorst, van Heerde \& Katgert 1984\nocite{win84}), has a similar spectral index, and is twice as large. However, LBDS~53W069 ($S_{1412}=3.7$~mJy and $\Theta < 5\farcs 1$ is $\sim 20$ times fainter at 1.4~GHz than TN~J0936$-$2243.
While a deeper spectrum would be needed to measure the exact strengths of the 2640~\AA\ and 2900~\AA\ breaks in the southern component, the comparison with LBDS~53W091 suggests that the break is of similar magnitude or larger, implying an age of the stellar population of $\simgt$3~Gyr. Further evidence for an old age comes from the very red $R-K$ color.

The identification of TN~J0936$-$2201 with the southern $K=18.3$ component puts this object on the main trend in the Hubble $K-z$ relation at $z=1.479$. If the diffuse northern component is part of the same physical system, the total magnitude could even mount to $K=18.0$ in an 8\arcsec\ diameter aperture.
Because the radio luminosity of TN~J0936$-$2201 is not particularly weak ($L_{365}=9.4 \times 10^{36}$~erg~s$^{-1}$~Hz$^{-1}$), and no strong contribution from AGN emission is seen in the spectrum, this argues against a significant direct contribution from a non-stellar AGN dominated component to the K-band magnitudes \citep{eal97,lac00}.

\subsection{Quasars}
Since the initial literature search in paper~I\nocite{deb00a}, two more sources in our USS sample have been identified as quasars according to the literature. We list these sources together with two quasars presented in this paper in table \ref{quasars}. We fitted the continuum slope $\alpha_{opt}$ from line-free regions of the optical spectra ($f_{\nu} \propto \nu^{\alpha_{opt}}$) and measured the radio core-to-lobe flux density ratio $R$ for three objects. Our $R$ values are only rough approximations, based on 1.4~GHz or 4.8~GHz maps; deeper and higher resolution radio maps are needed to confirm the high core-dominance in these quasars.
We find that the three measured optical spectral indices are steeper than the median $\alpha_{opt} \simeq -1.0$ found from the MRC quasar sample \citep{bak99}.
\begin{table}
\caption{Quasars identified from our USS sample}
\tiny
\begin{tabular}{llrrrrrrrl}
\hline
Source & $z$ & $S_{1400}$ & $\alpha_{radio}$ & LAS & R$^a$ & $m_R$ & $m_K~^b$ & $\alpha_{optical}$ & Identification \\
 &  & mJy &  & \arcsec &  & mag & mag &  & \\
\hline
WN~J0717+4611 & 1.462 & 108 & $-$1.30 &     6.2 &     0.6 & 21.6 & 16.8$\pm$0.5 & $-$3.9 &    WN~J0717+4611 \\
WN~J0923+4602 & 0.727 &  12 & $-$1.49 &    21.6 &     3.6 & 18.3 & $>$16        &\nodata &   RX J09232+4602 \\
WN~J1124+3228 & 1.525 &  39 & $-$1.31 &     219 &     3.6 & 17.4 & 15.7$\pm$0.2 & $-$1.9 & FIRST J1124+3228 \\
WN~J2313+4053 & 2.99  &  11 & $-$1.50 & \nodata & \nodata & 18.8 &      \nodata & $-$2.2 &    WN~J2313+4053 \\
\hline
\multicolumn{10}{l}{$^a$ R is the radio core-to-lobe flux density ratio.} \\
\multicolumn{10}{l}{$^b$ Determined from Lick Gemini observations or from 2MASS \citep{jar00}.} \\
\multicolumn{10}{l}{REFERENCES: RX J09232+4602:\cite{wu99}; FIRST J1124+3228: \cite{whi00}} \\
\end{tabular}
\label{quasars}
\end{table}

We note that the quasars in our sample are at slightly higher redshifts, and probably contain much steeper spectrum radio cores than the $\alpha_{core} \sim 0$ of the MRC quasars \citep{kap98}. 
The radio morphology of our sources \citep[see][]{deb98} shows much more diffuse radio lobes than the more classic FR~II morphologies of the MRC quasars \citep[see][]{kap98}. Such diffuse lobes could easily be over-resolved in the high-frequency radio maps used to construct our USS sample (which are of $\sim5 \times$ higher resolution than the low frequency maps). This missing flux would tend to over-estimate the steepness of the derived spectral index, and add sources with such morphologies in an USS sample.

\section{Conclusions}

We have determined new redshifts of 49 USS sources, 33 of which are part of our new sample of USS sources (paper~I\nocite{deb00a}). Considering only the spectra from this main USS sample, one third are radio galaxies at $z>3$, showing the high efficiency of our sample in finding very high redshift radio galaxies, when combined with a near-IR identification campaign. The $K-$band images allow the preferential selection of the faintest objects, which on the basis of the Hubble $K-z$ diagram are expected to have the highest redshifts. This combined USS and faint $K-$band magnitude selection technique explains why we find similar numbers of sources from $z=0$ to $z=4$ from our sample, despite the expected decrease in the fraction of the most distant and luminous radio galaxies in a sample with a ten times lower flux limit than previous efforts.

We have also found four quasars from our USS sample, of which at least three have very red optical continuum slopes, suggesting that such red quasars have been missed by previous quasar samples.

We also find five sources without distinctive emission or absorption features. We exclude the possibility that these are stars or BL Lacs. Possible identifications include (i) radio galaxies with faint emission lines in the ``redshift desert'' at $1.5 \simlt z \simlt 2.3$, (ii) radio galaxies with an obscured AGN, which are dominated by a stellar continuum observed with insufficient S/N, and (iii) pulsars.

Three objects with $K\sim 20 - 21$ did not yield any optical continuum or line emission, even after 50$-$90~min integrations with the Keck telescope. These objects are possible $z>7$ targets.

This paper increases the number of known $z>2$ radio galaxies to 140 (including 26 at $z>3$, 6 at $z>4$, and 1 at $z>5$). In a companion paper \citep{deb00b}, we use these new emission line data to perform a statistical analysis of the emission line regions of radio galaxies from $z=0$ to $z=5.19$.

We are continuing our near-IR imaging campaign with the Keck, CTIO and VLT telescopes and follow-up spectroscopy with Keck and VLT to obtain a statistically significant sample of USS sources with a range of radio sizes and flux densities. This will allow us to further refine our high redshift selection techniques, and to increase the number of known $z>4$ radio galaxies for detailed follow-up studies of the formation and evolution of massive galaxies.

\acknowledgments

We are grateful for the excellent help provided by the staff of the
Lick, Keck, ESO, and WHT telescopes. We thank Arjun Dey for his help
during one of the Keck runs, Sally Laurent-Muehleisen and Nahum Arav
for useful discussions, Mike Brotherton for providing the electronic
version of the FIRST~J1124+3228 quasar, and Hy Spinrad for his
comments on the manuscript. We are grateful to the anonymous referee for
constructive comments that have improved the paper. This publication
makes use of data products from the Two Micron All Sky Survey, which
is a joint project of the University of Massachusetts and the Infrared
Processing and Analysis Center/California Institute of Technology,
funded by the National Aeronautics and Space Administration and the
National Science Foundation. The authors made use of the NASA/IPAC
Extragalactic Database (NED) which is operated by the Jet Propulsion
Laboratory, California Institute of Technology, under contract with
the National Aeronautics and Space Administration. The work by CDB,
WvB, WdV and SAS at IGPP/LLNL was performed under the auspices of the
U.S. Department of Energy by University of California Lawrence
Livermore National Laboratory under contract No. W-7405-Eng-48.  This
work was supported in part by the Formation and Evolution of Galaxies
network set up by the European Commission under contract ERB FMRX--
CT96--086 of its TMR programme.

\newpage

\psfig{file=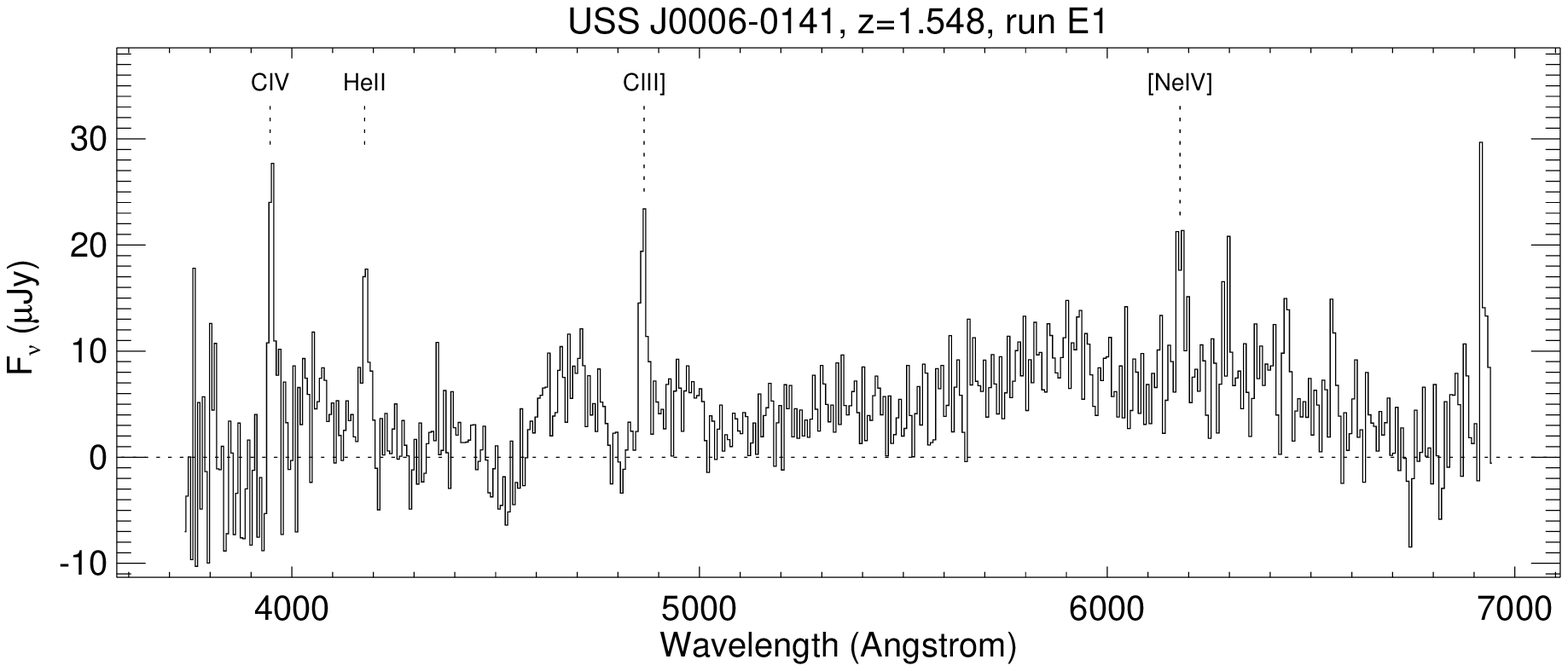,height=6cm}
\psfig{file=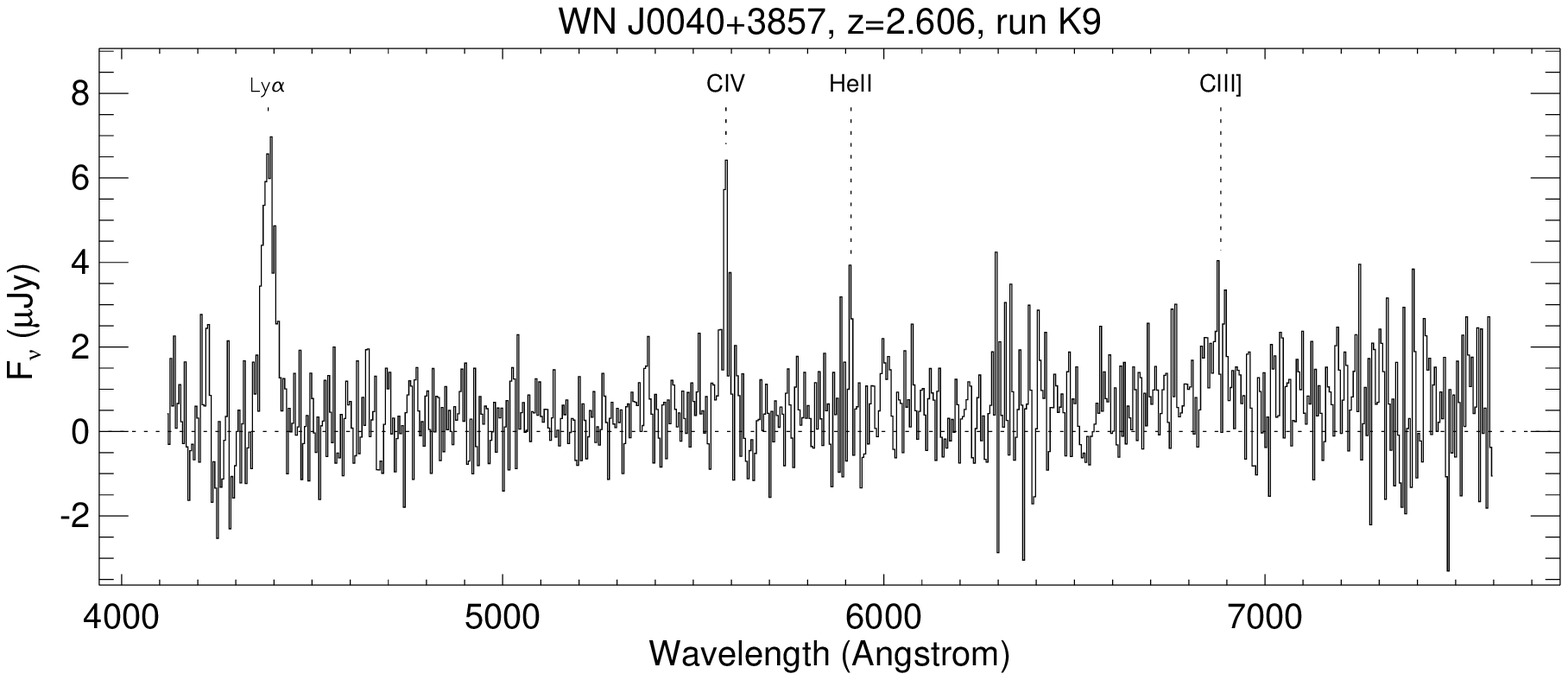,height=6cm}
\psfig{file=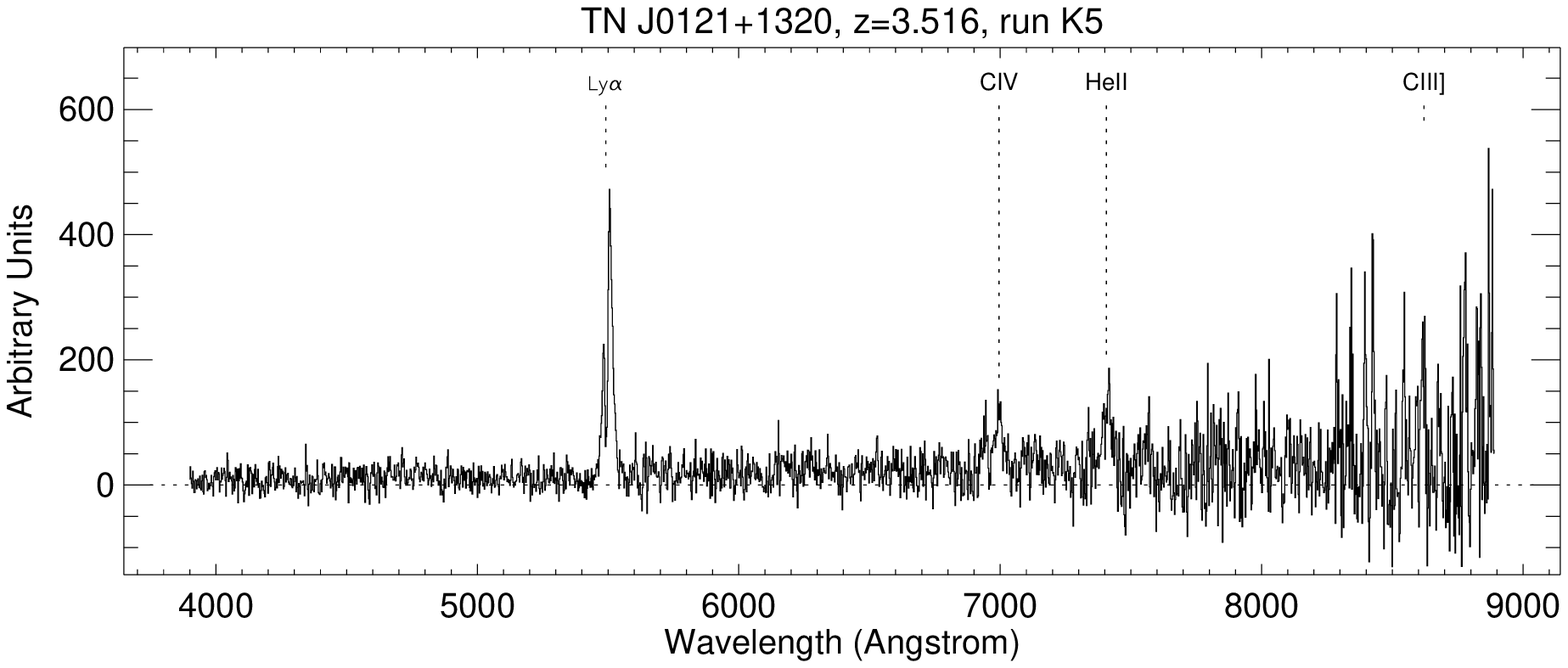,height=6cm}
\psfig{file=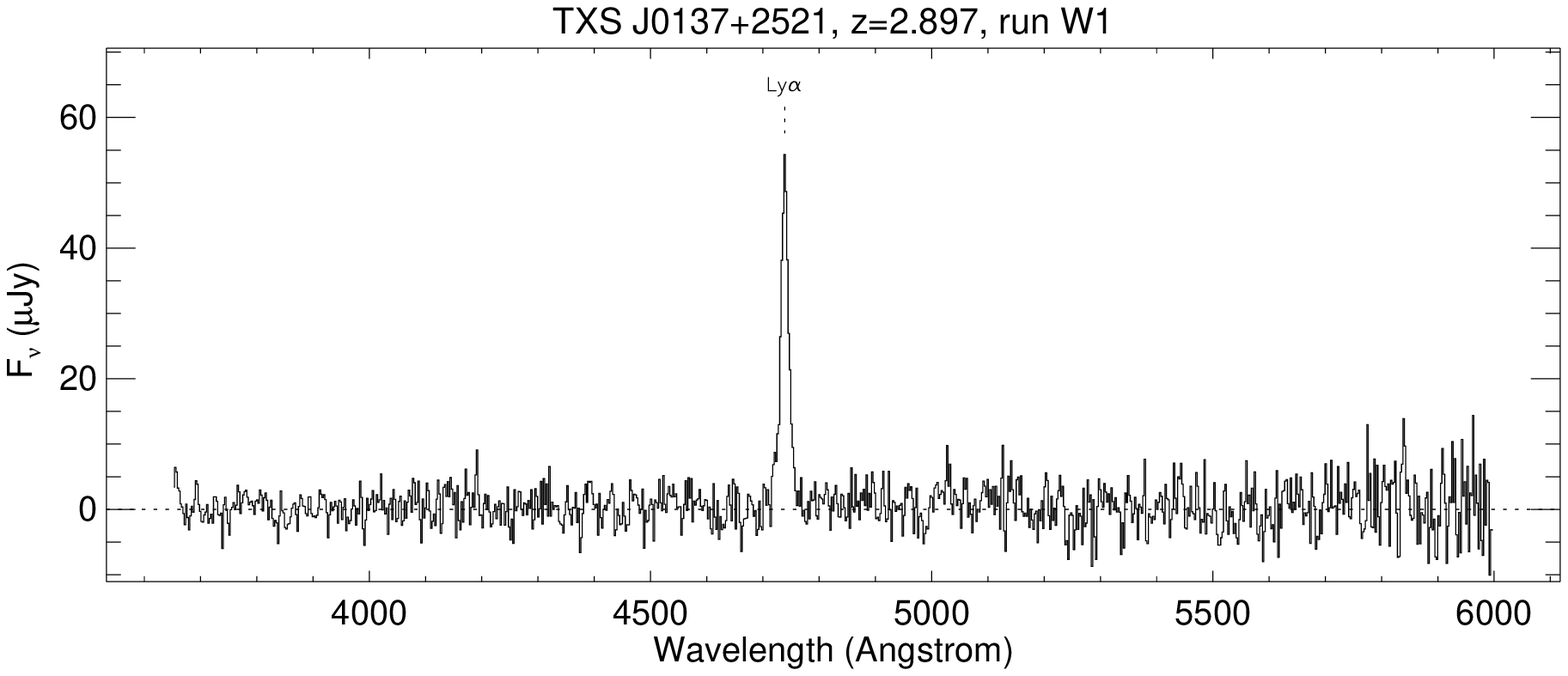,height=6cm}
\psfig{file=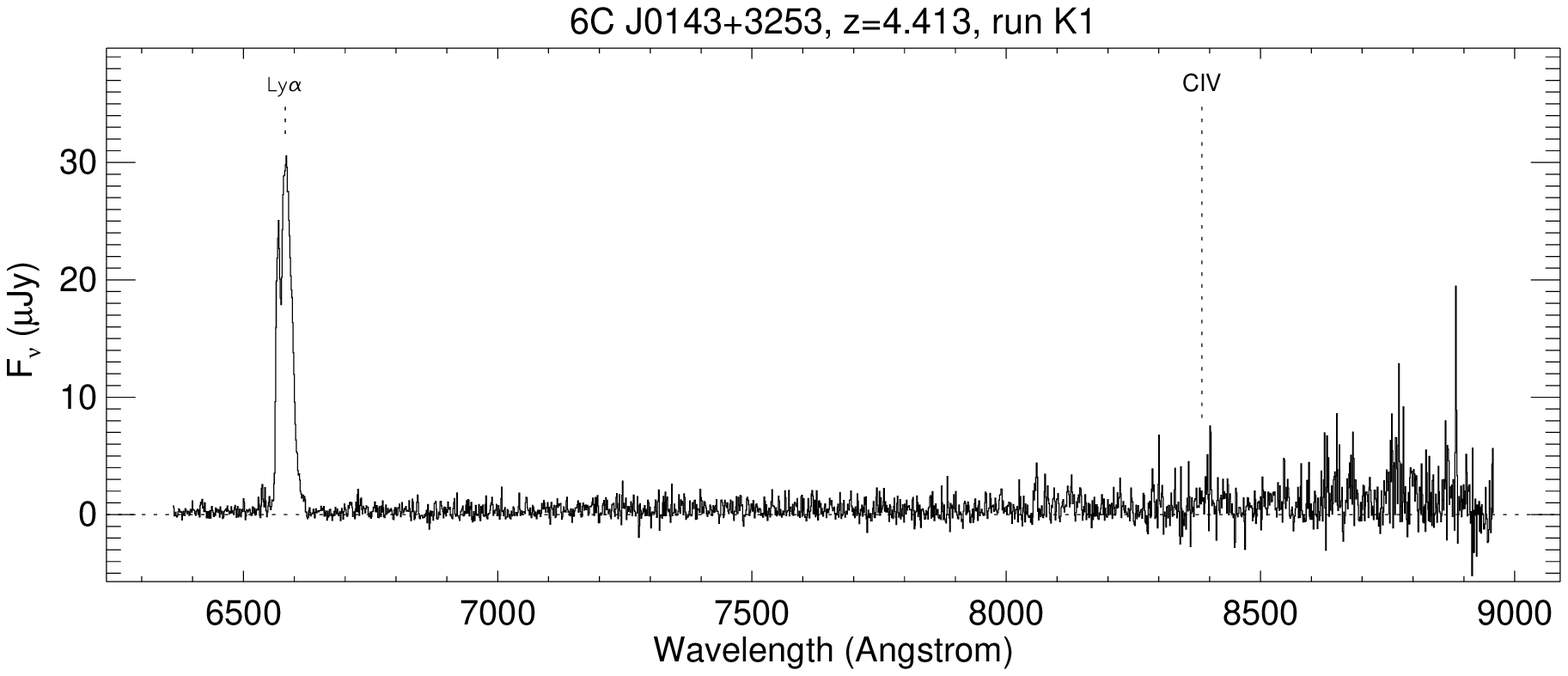,height=6cm}
\psfig{file=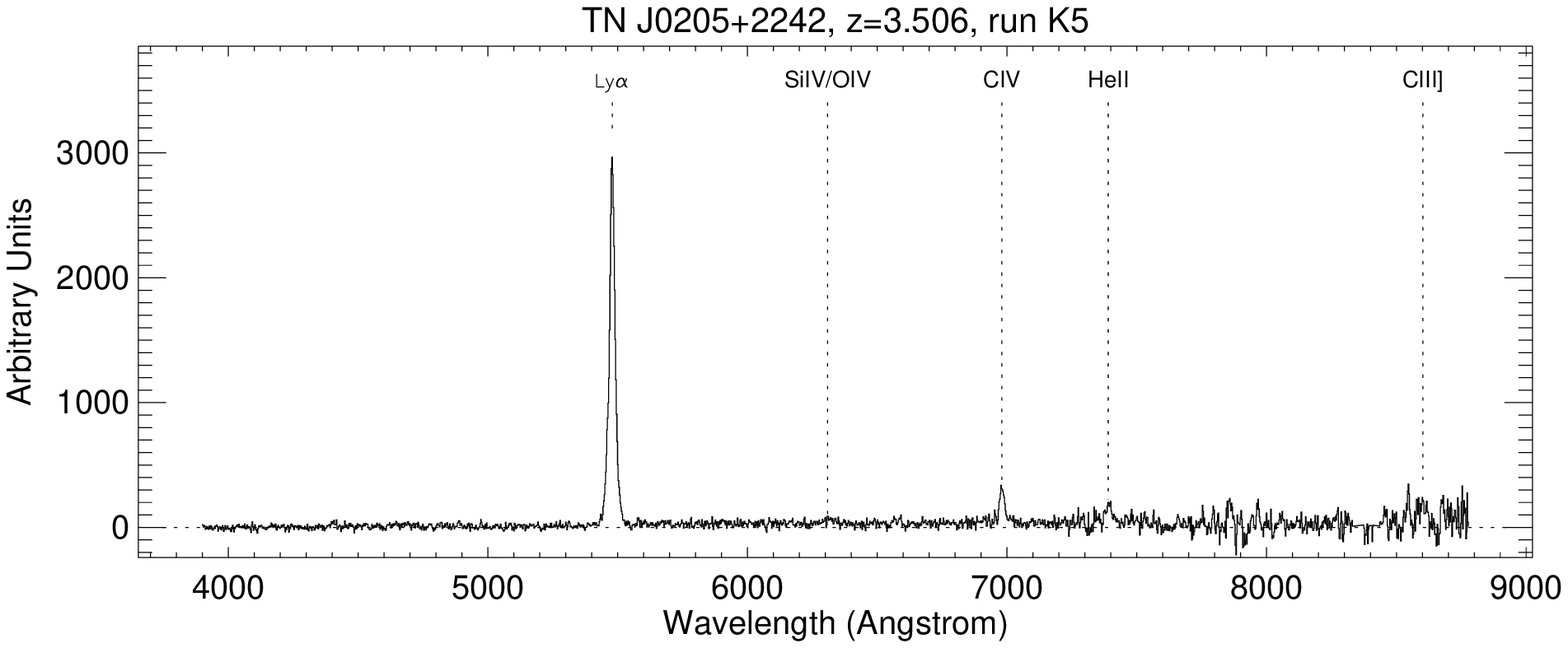,height=6cm}
\psfig{file=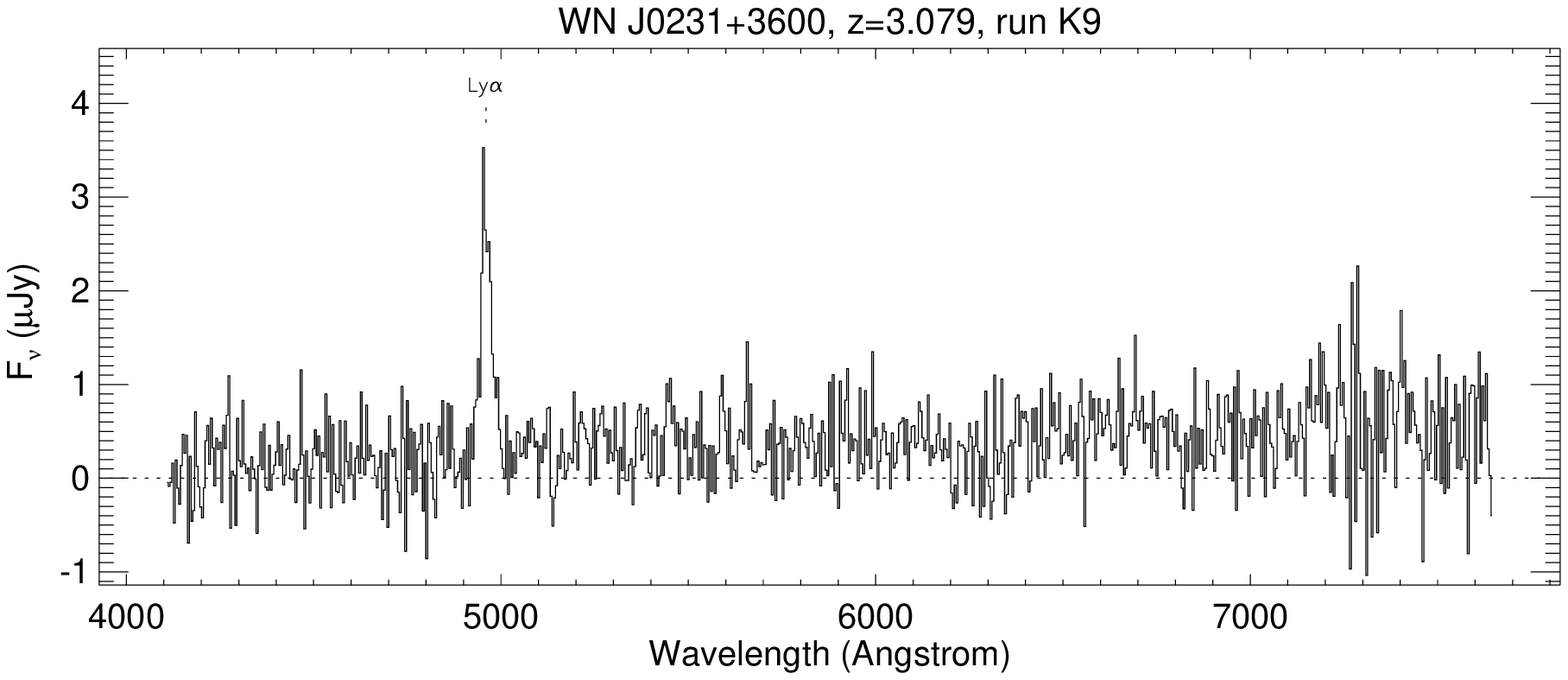,height=6cm}
\psfig{file=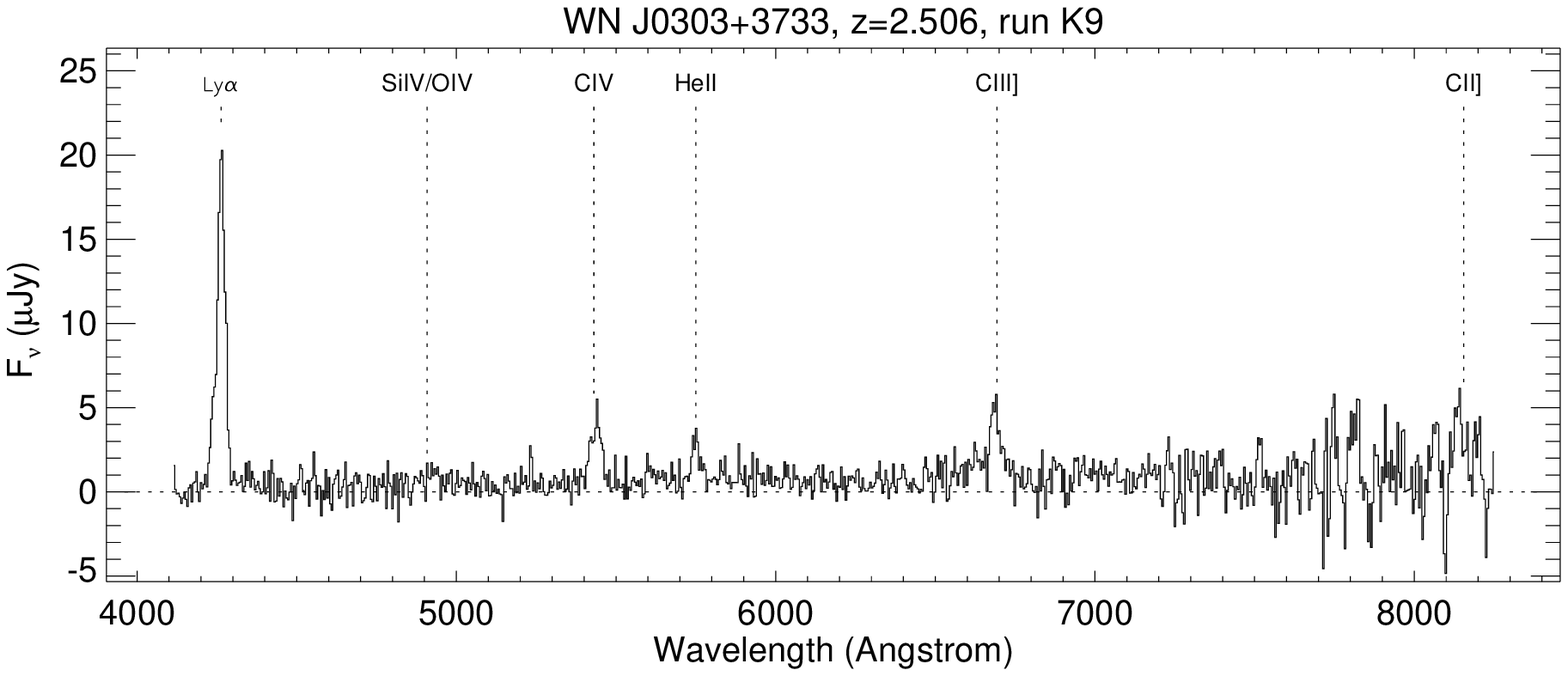,height=6cm}
\psfig{file=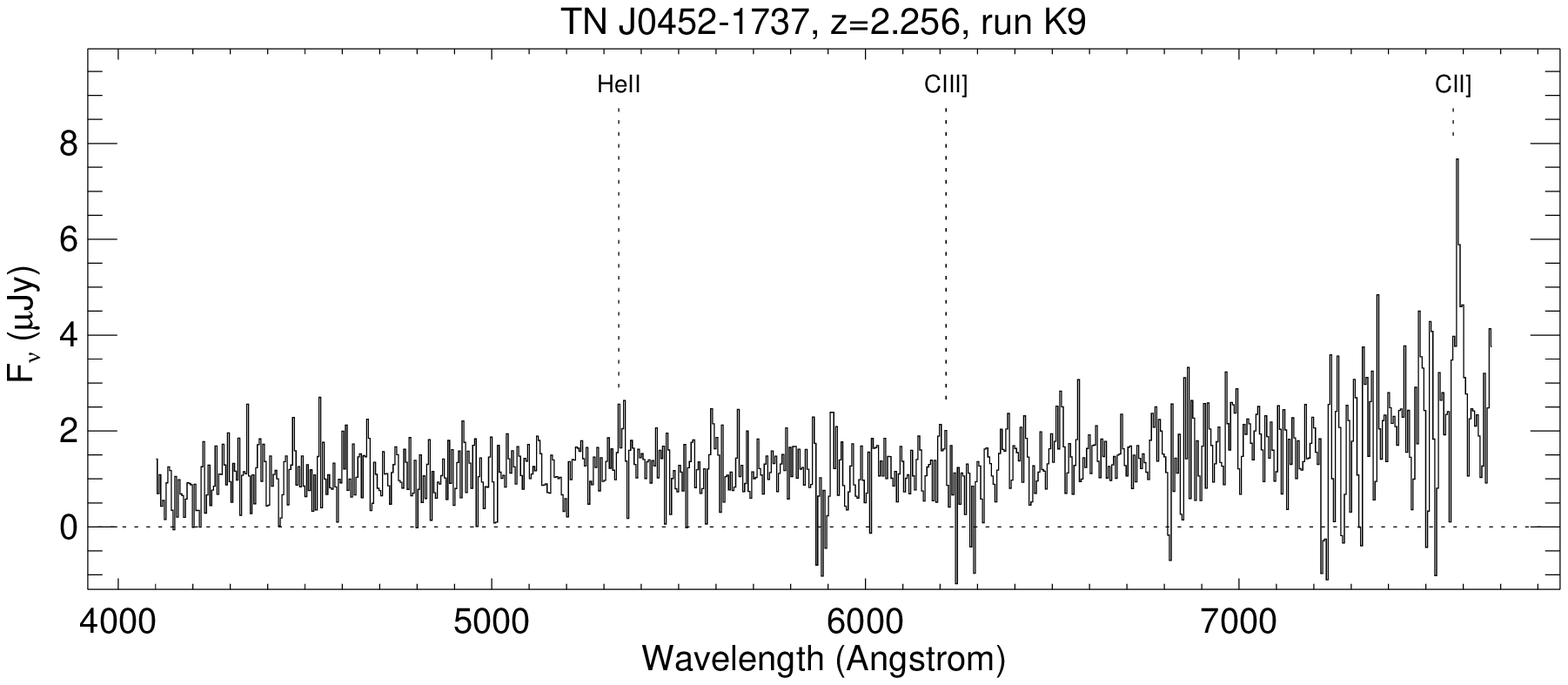,height=6cm}
\psfig{file=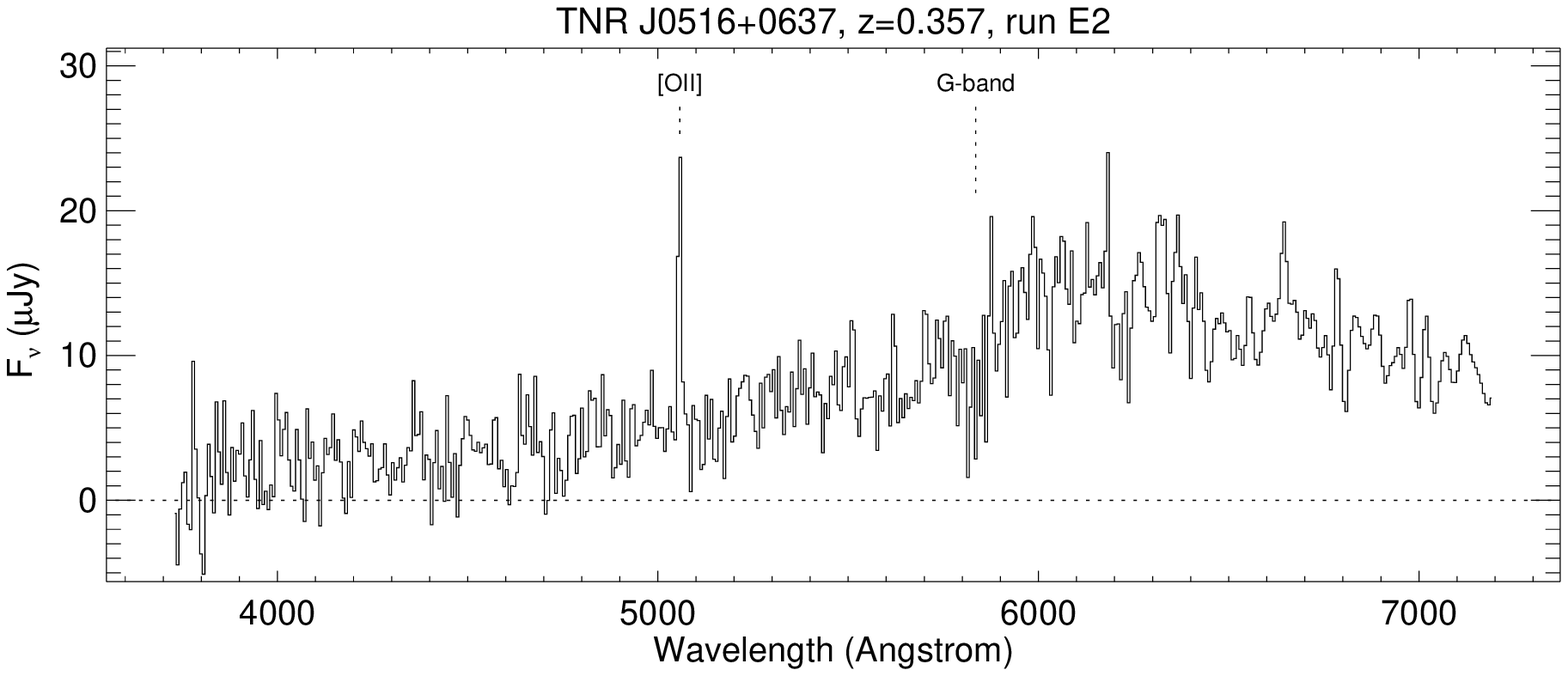,height=6cm}
\psfig{file=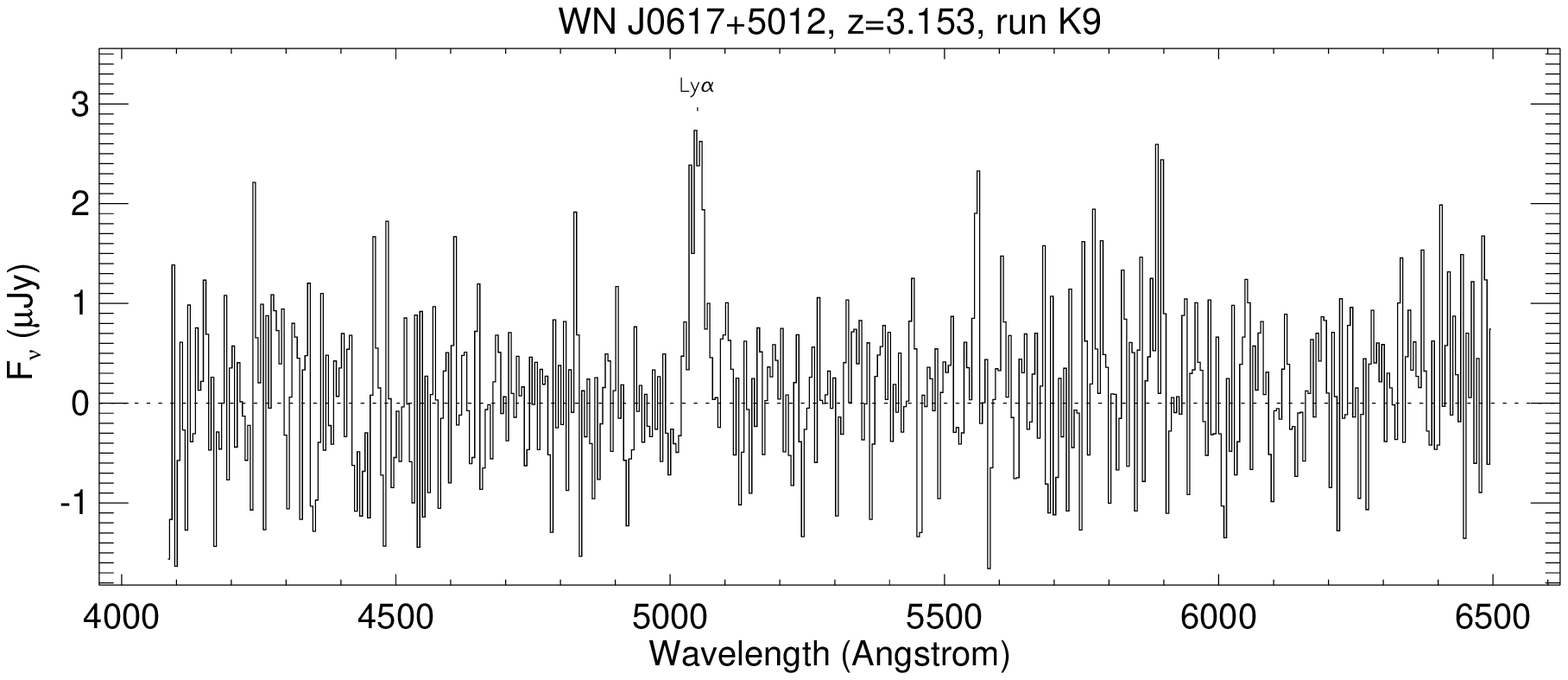,height=6cm}
\psfig{file=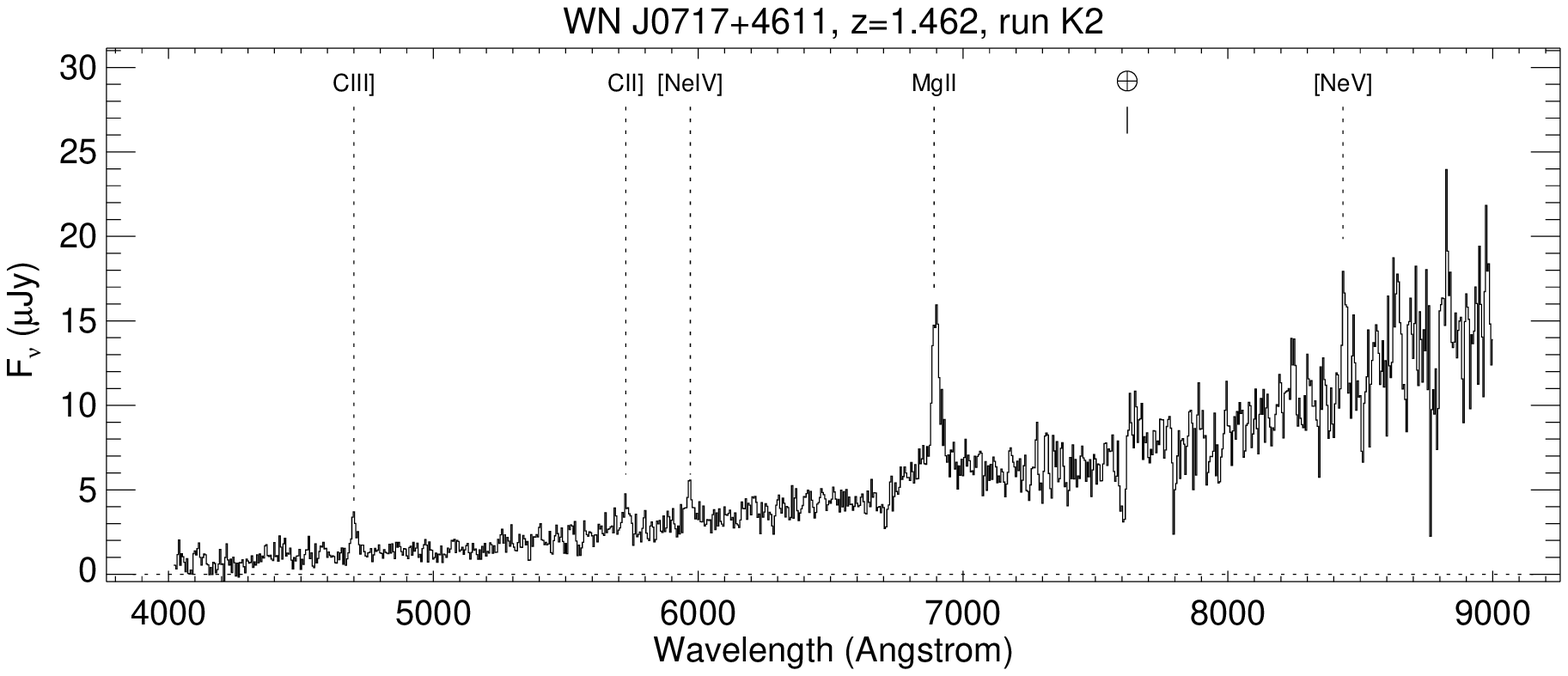,height=6cm}
\psfig{file=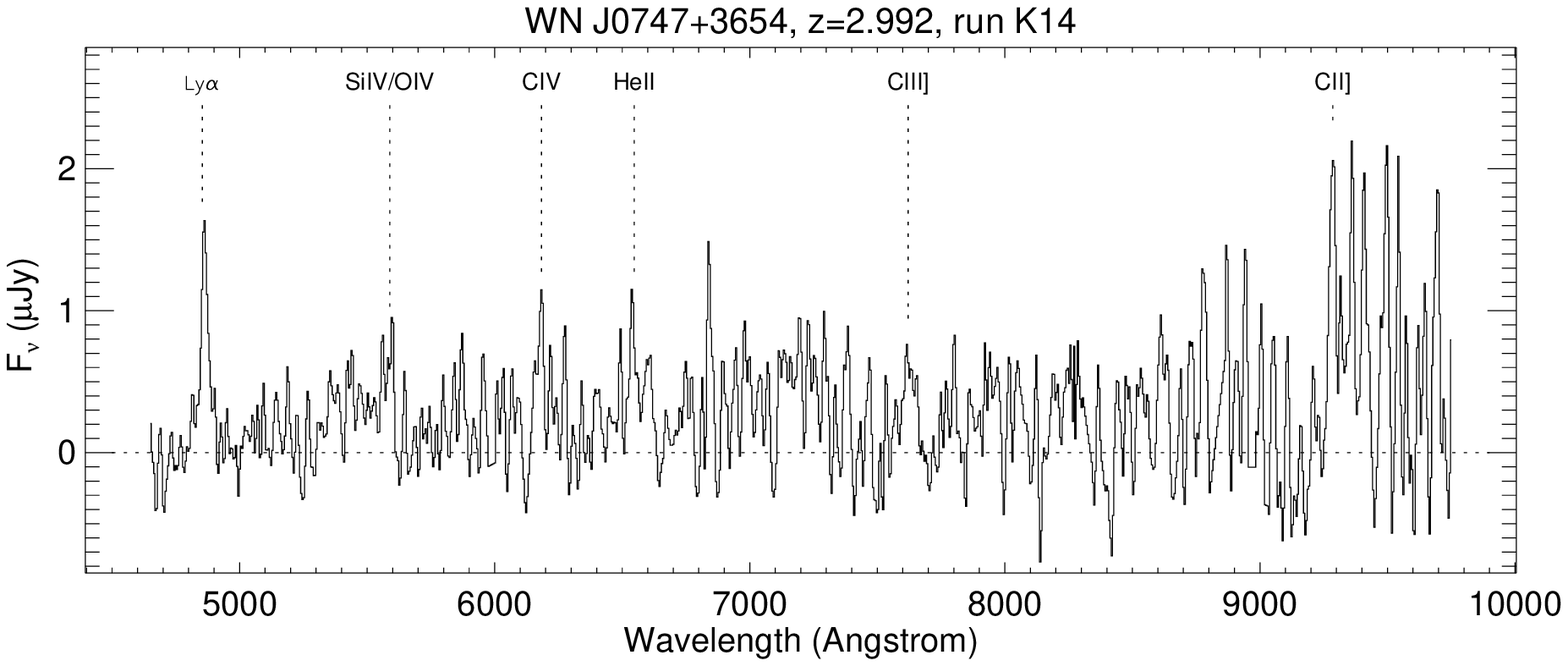,height=6cm}
\psfig{file=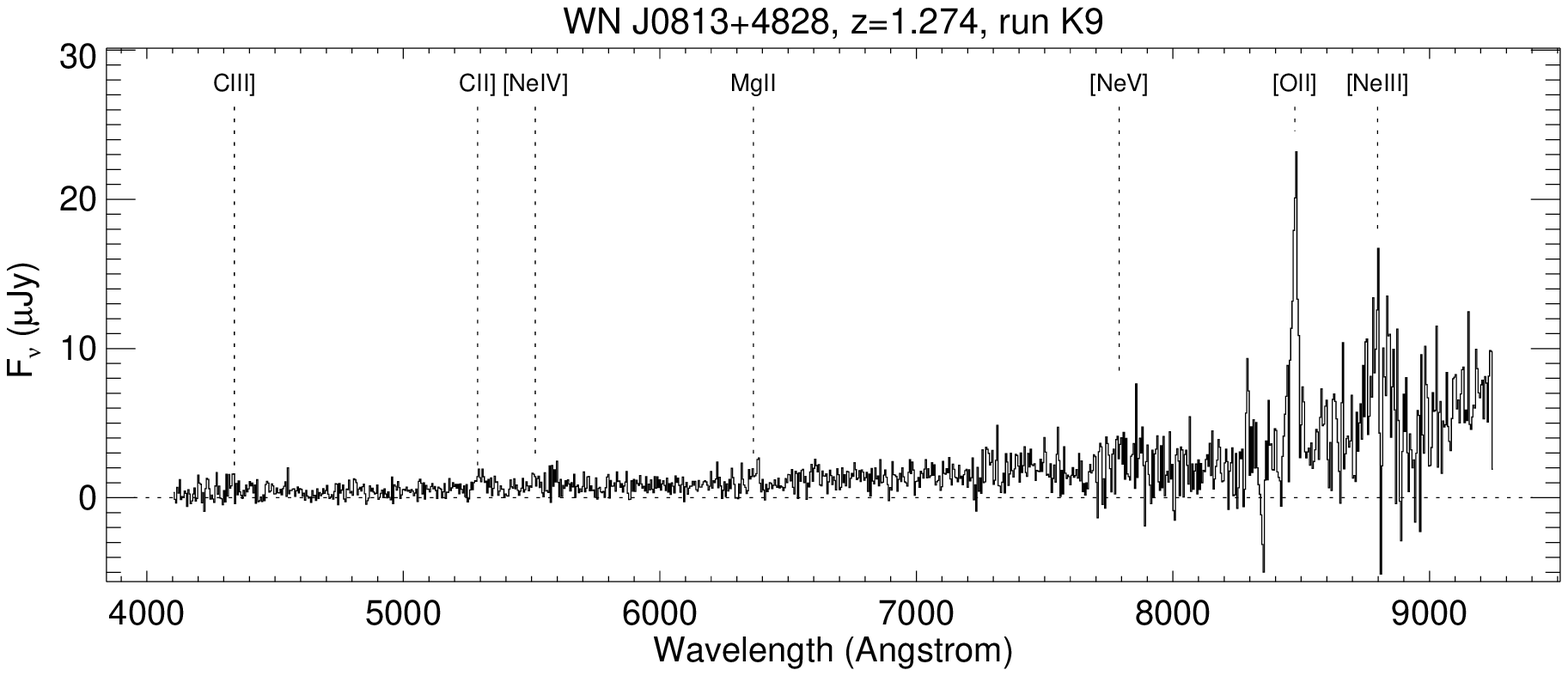,height=6cm}
\psfig{file=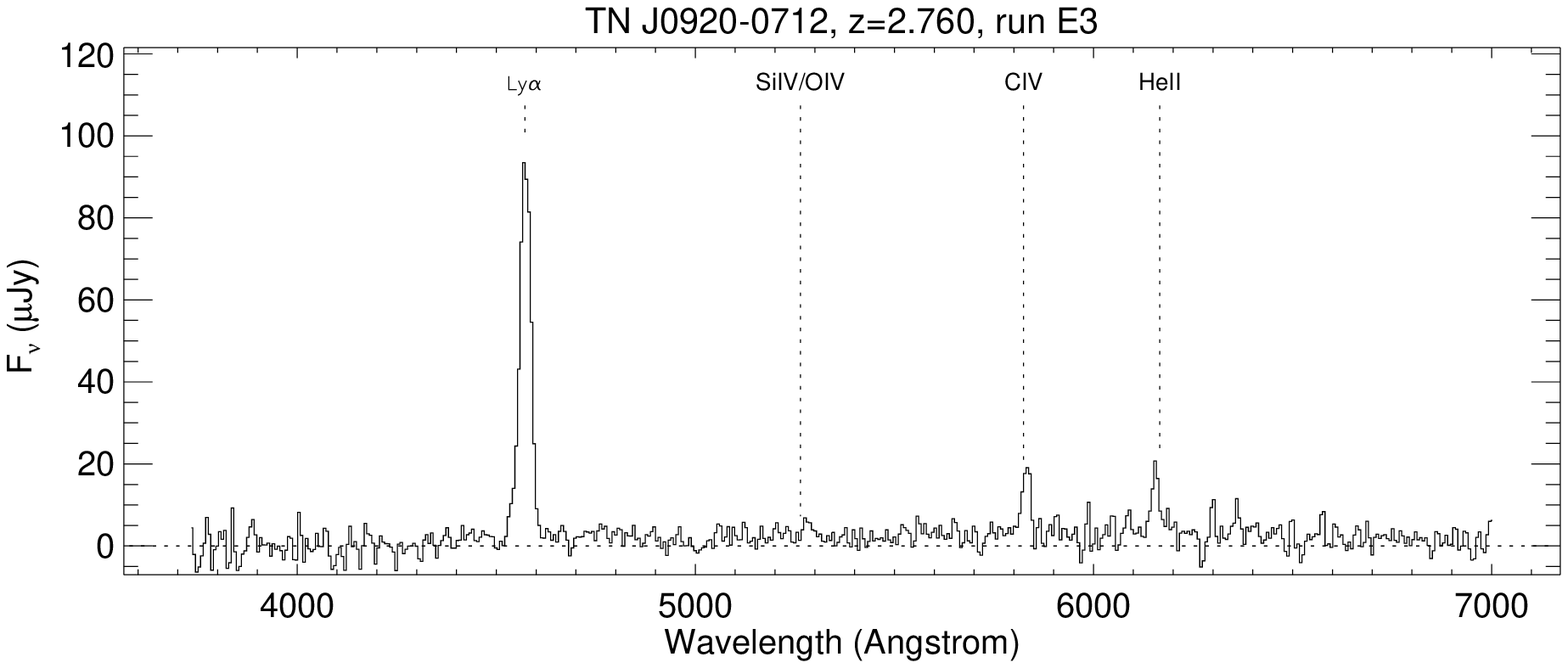,height=6cm}
\psfig{file=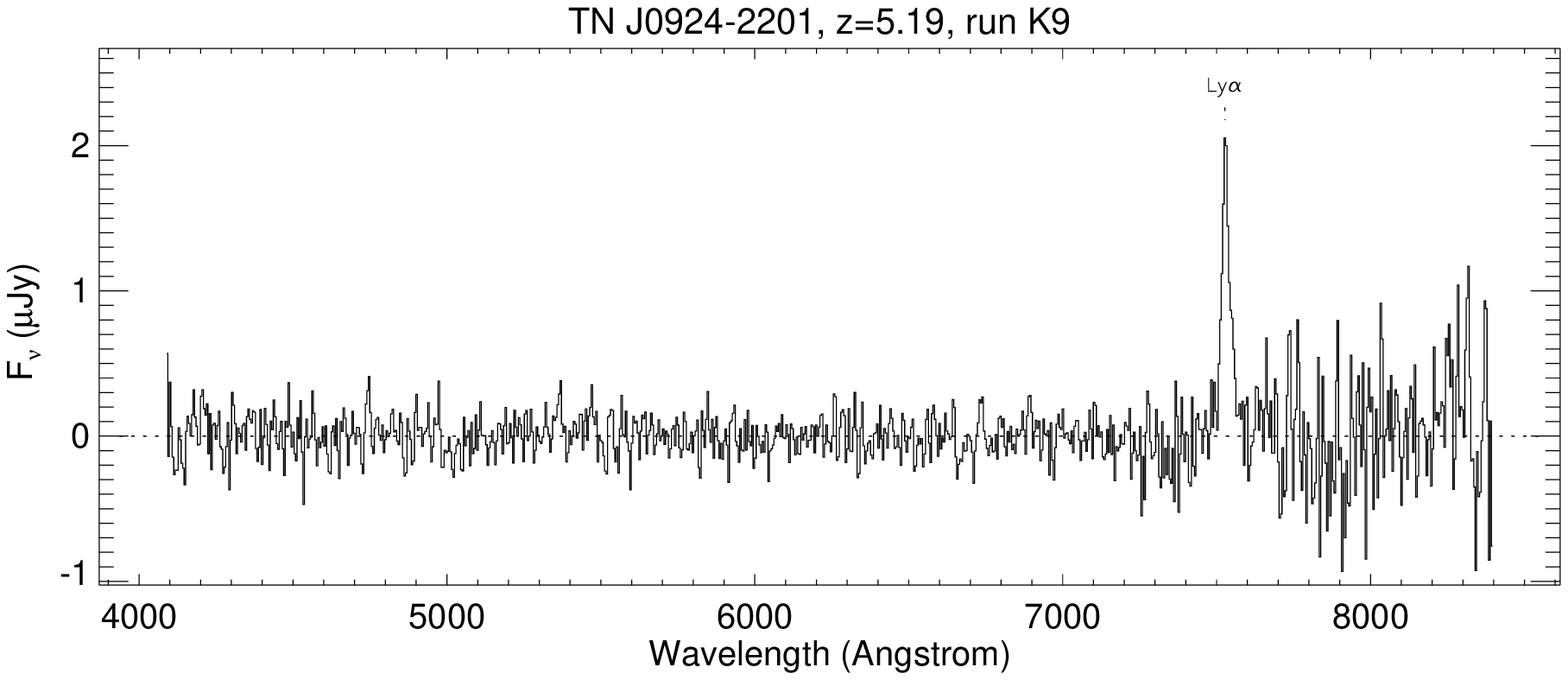,height=6cm}
\psfig{file=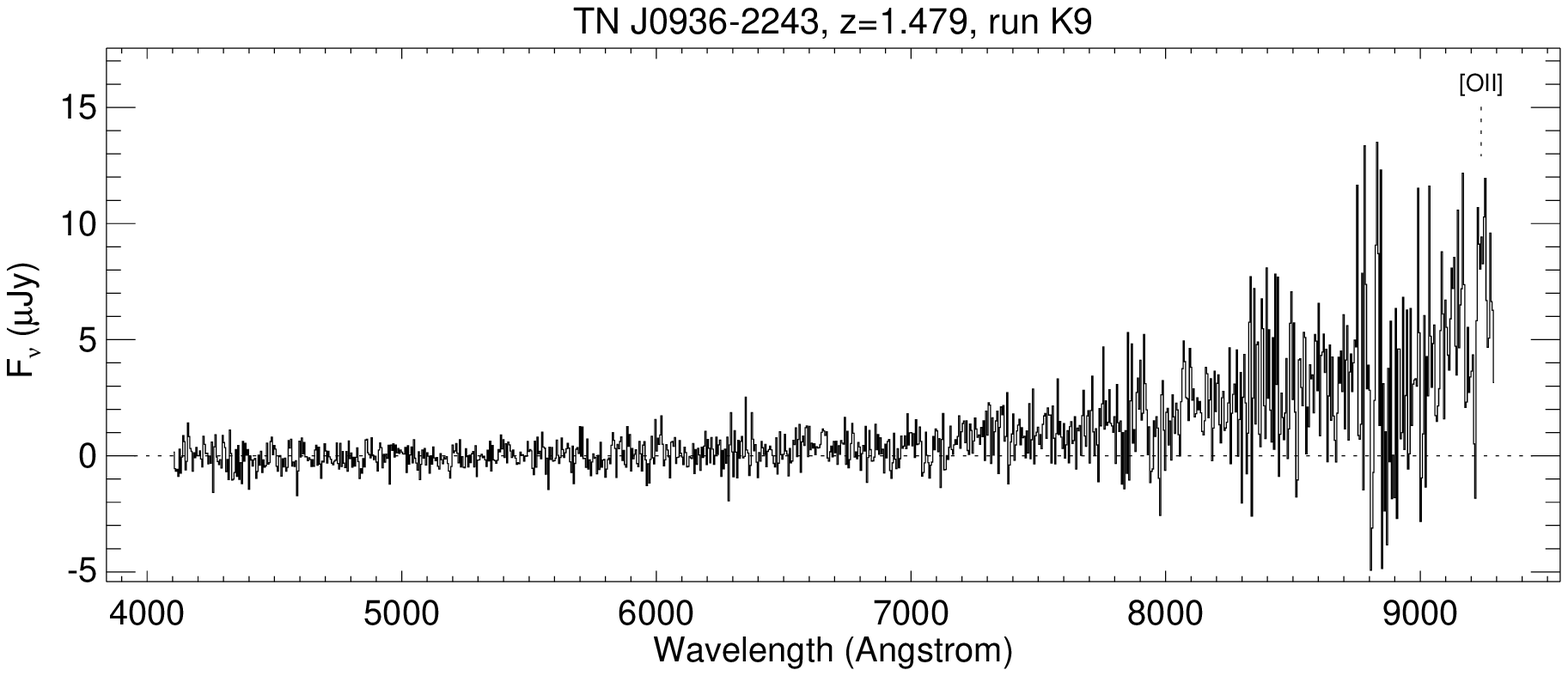,height=6cm}
\psfig{file=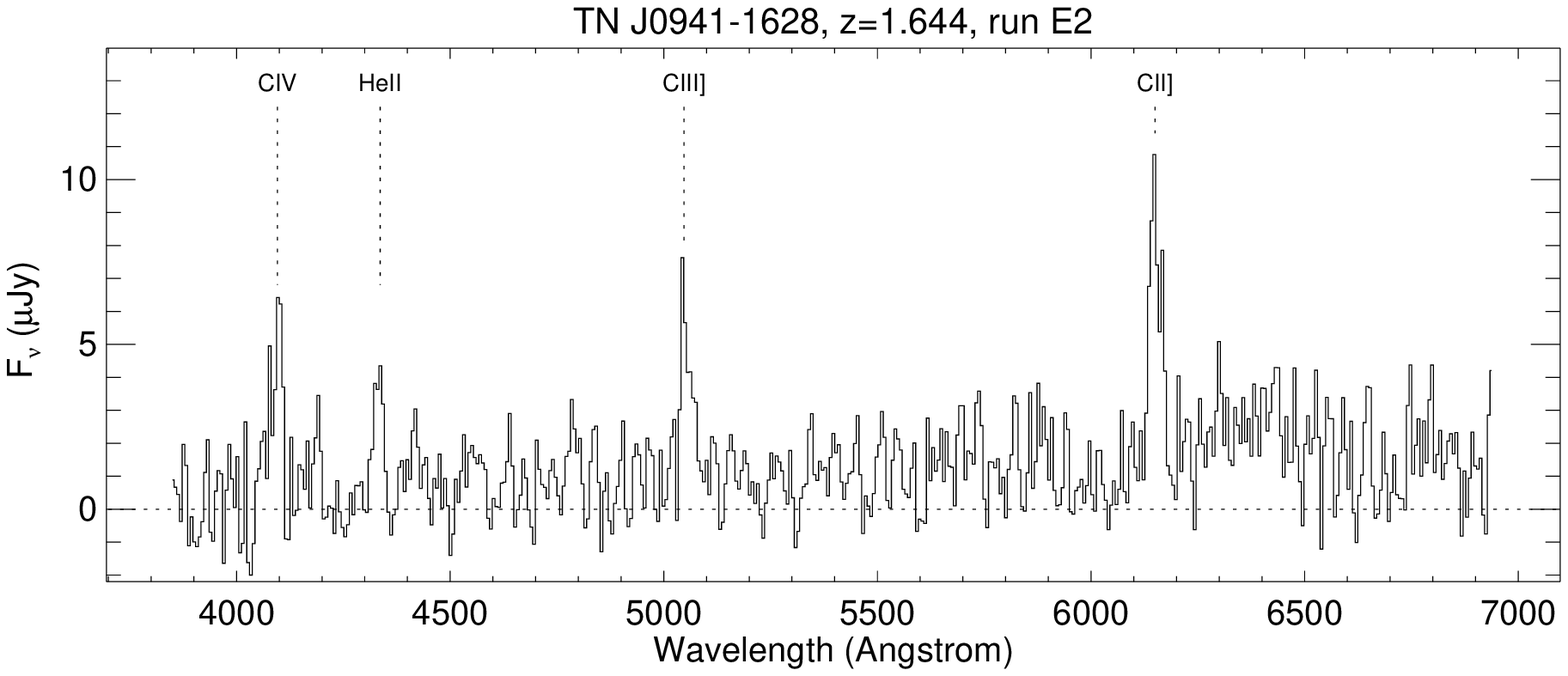,height=6cm}
\psfig{file=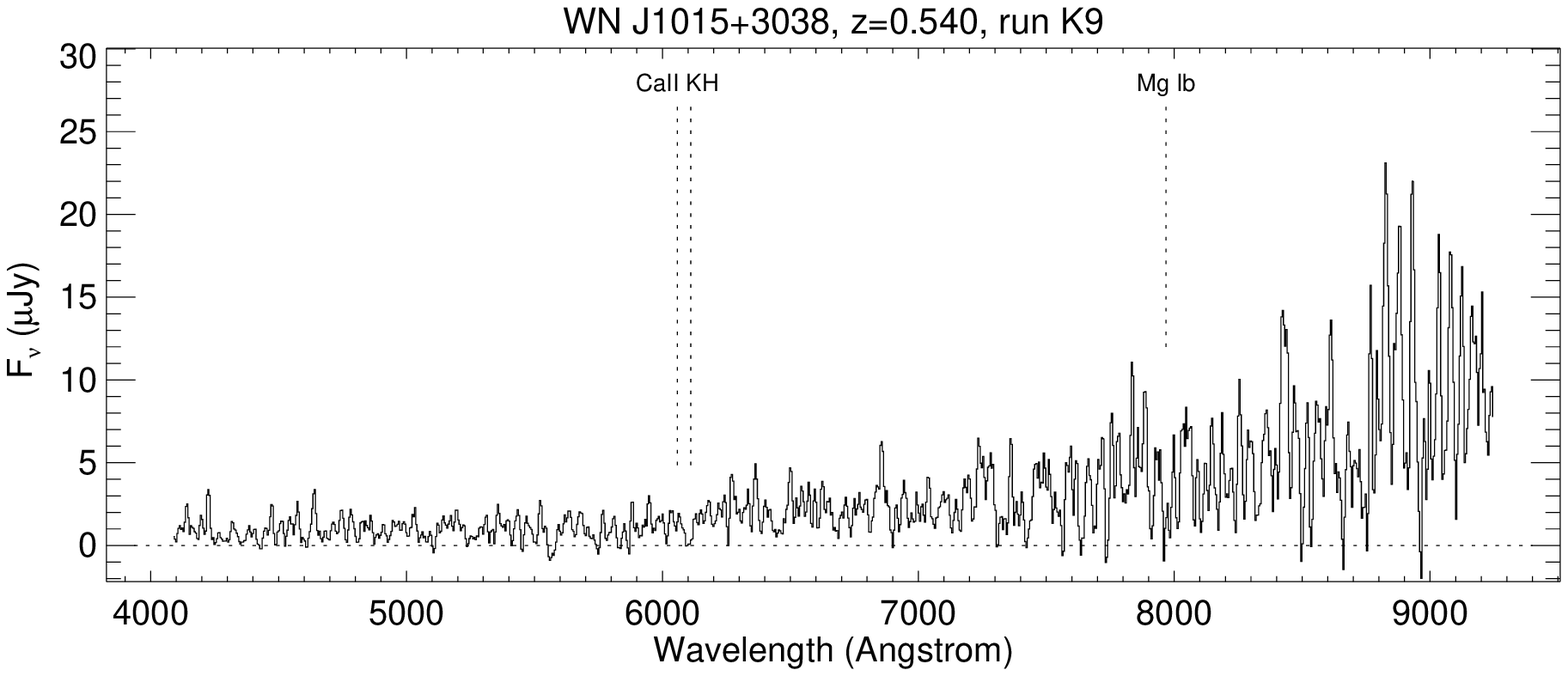,height=6cm}
\psfig{file=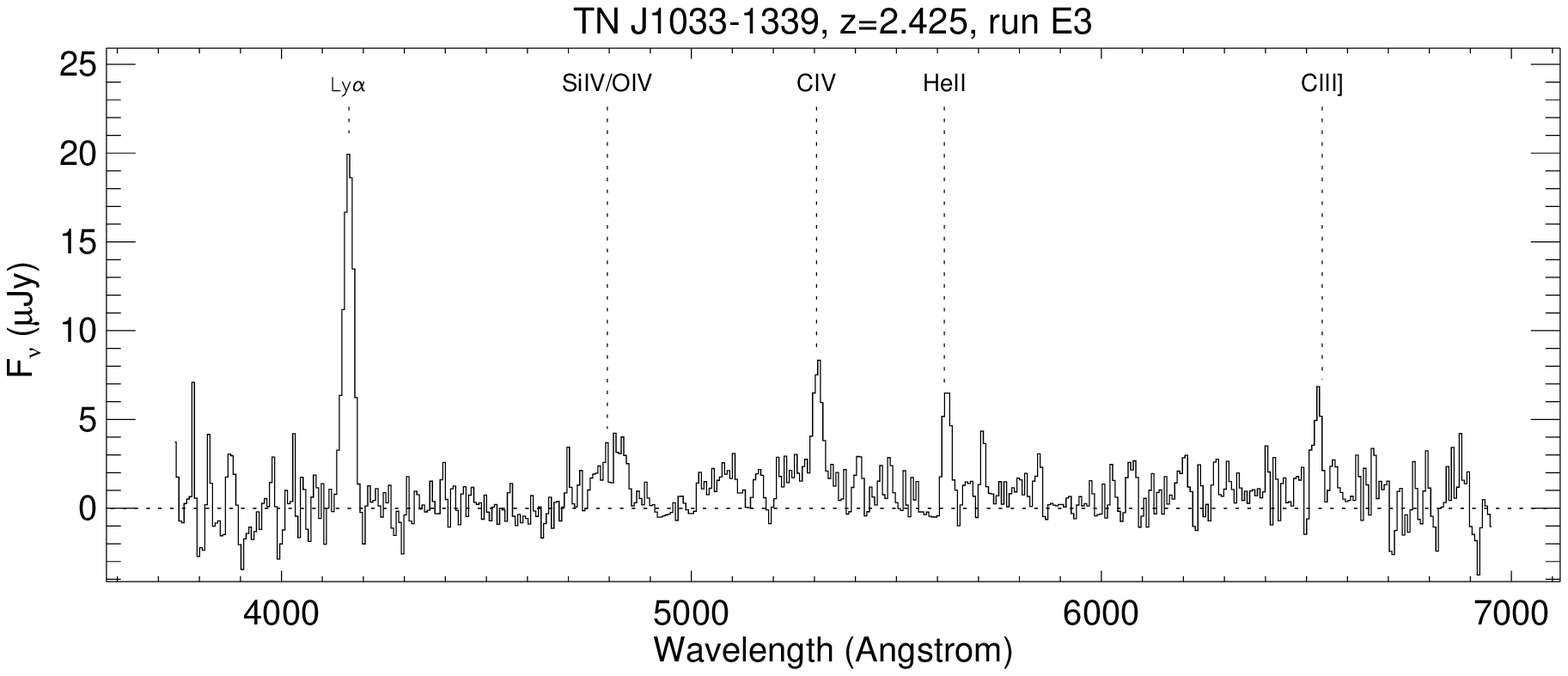,height=6cm}
\psfig{file=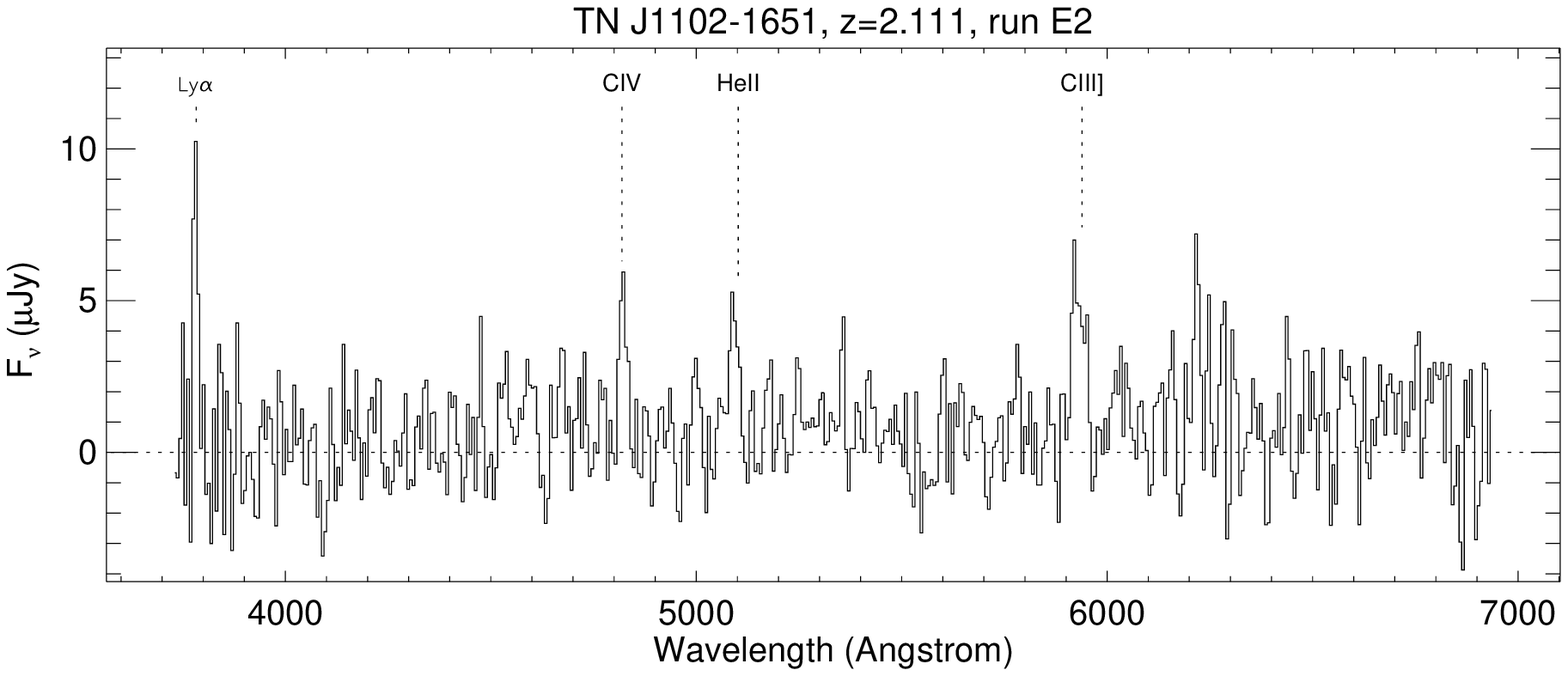,height=6cm}
\psfig{file=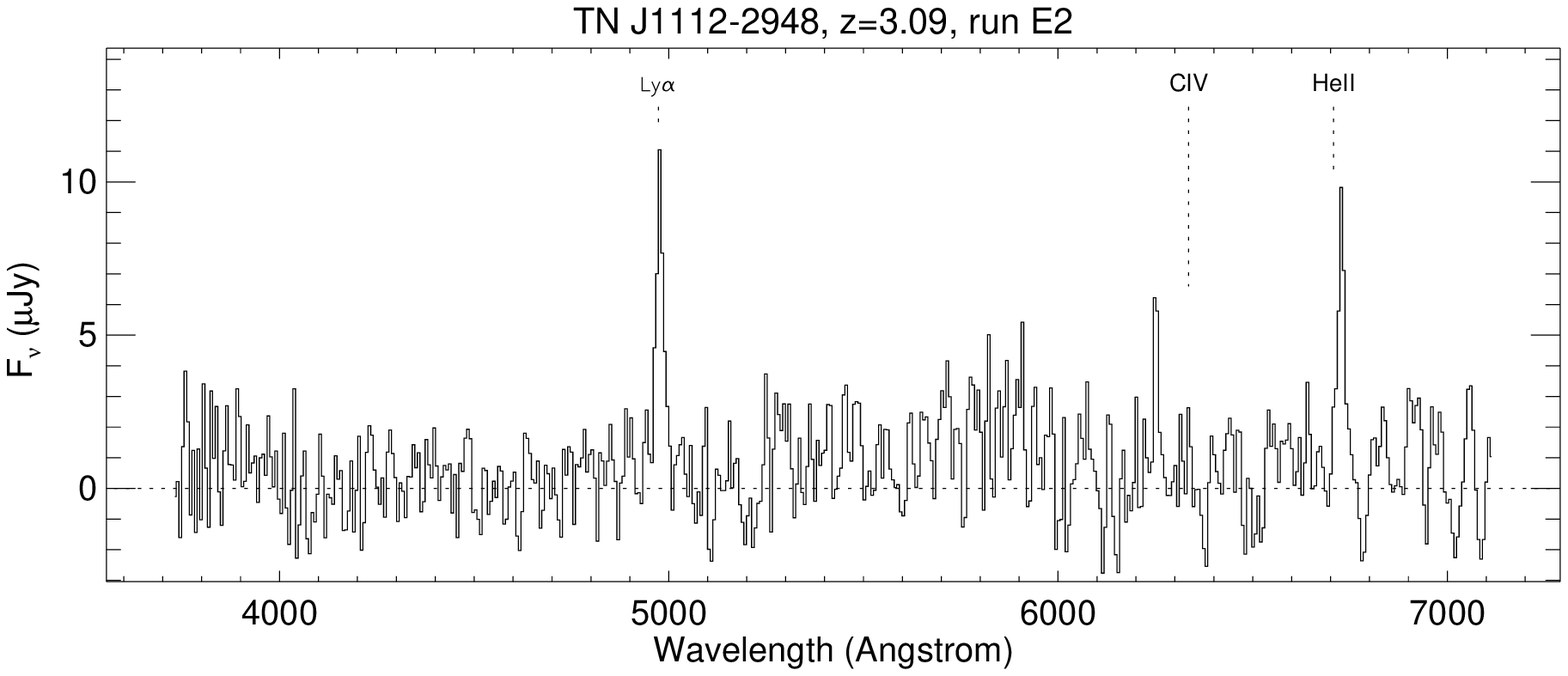,height=6cm}
\psfig{file=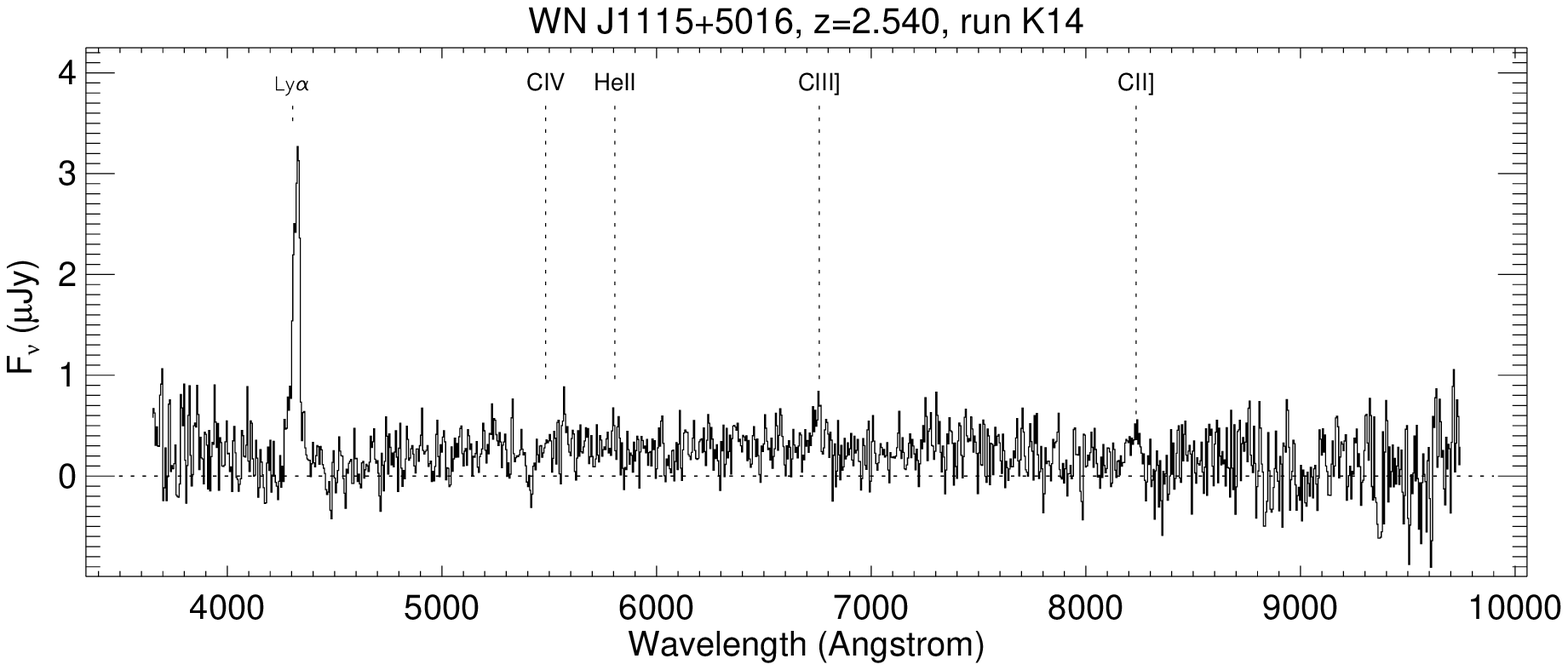,height=6cm}
\psfig{file=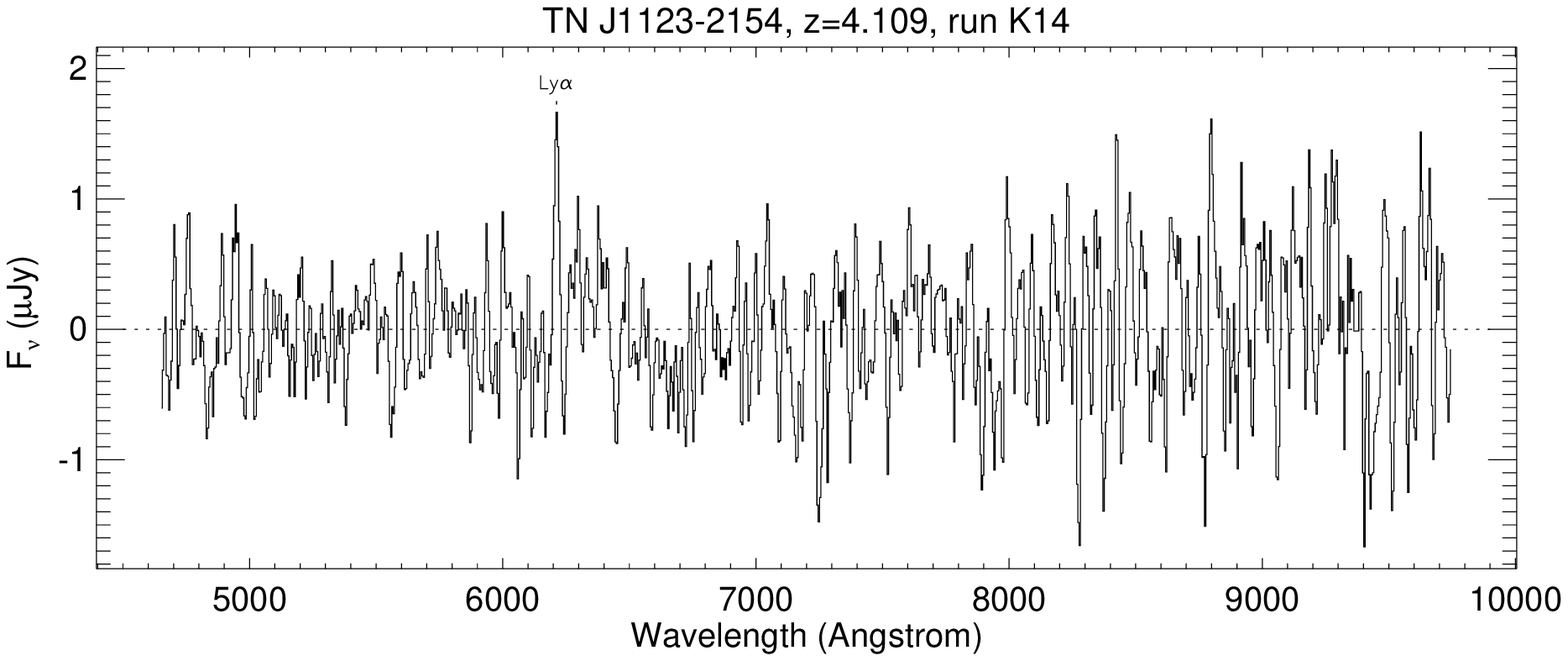,height=6cm}
\psfig{file=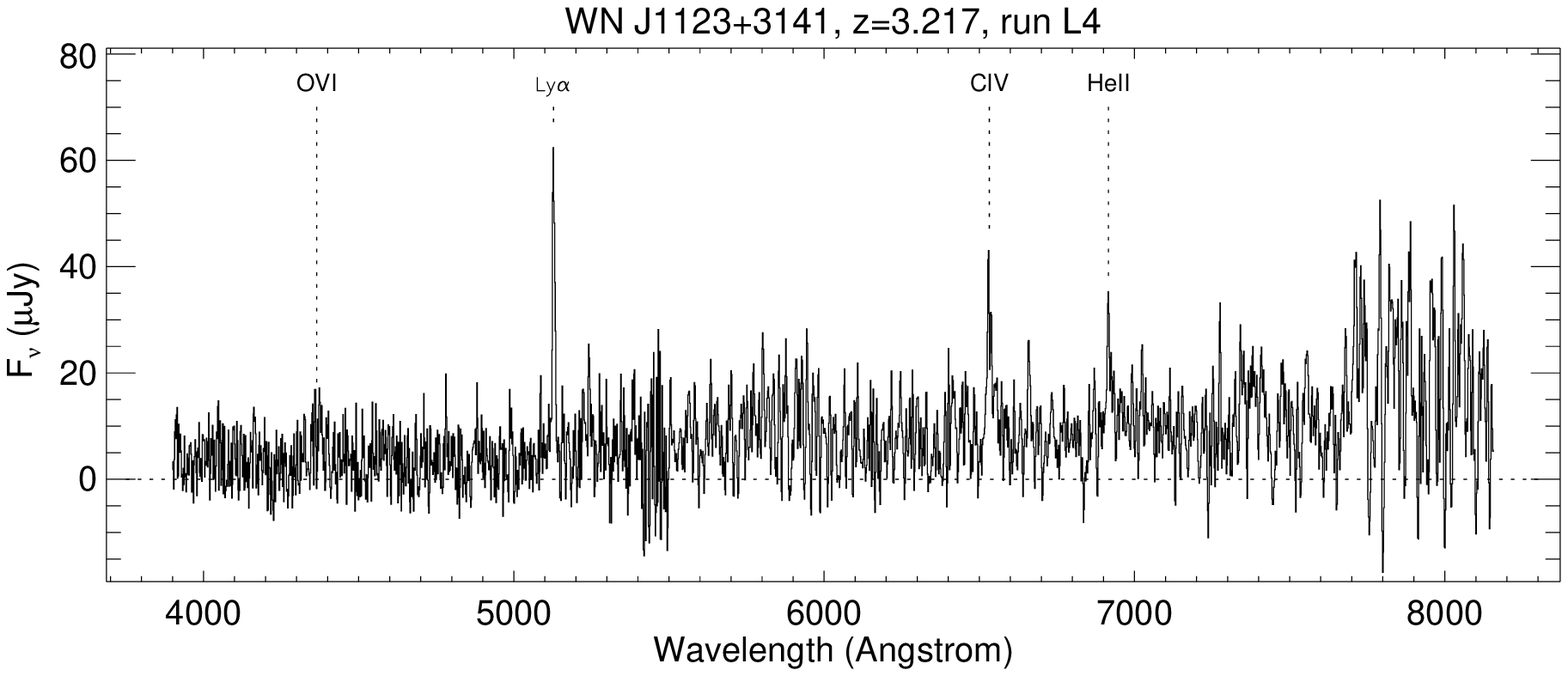,height=6cm}
\psfig{file=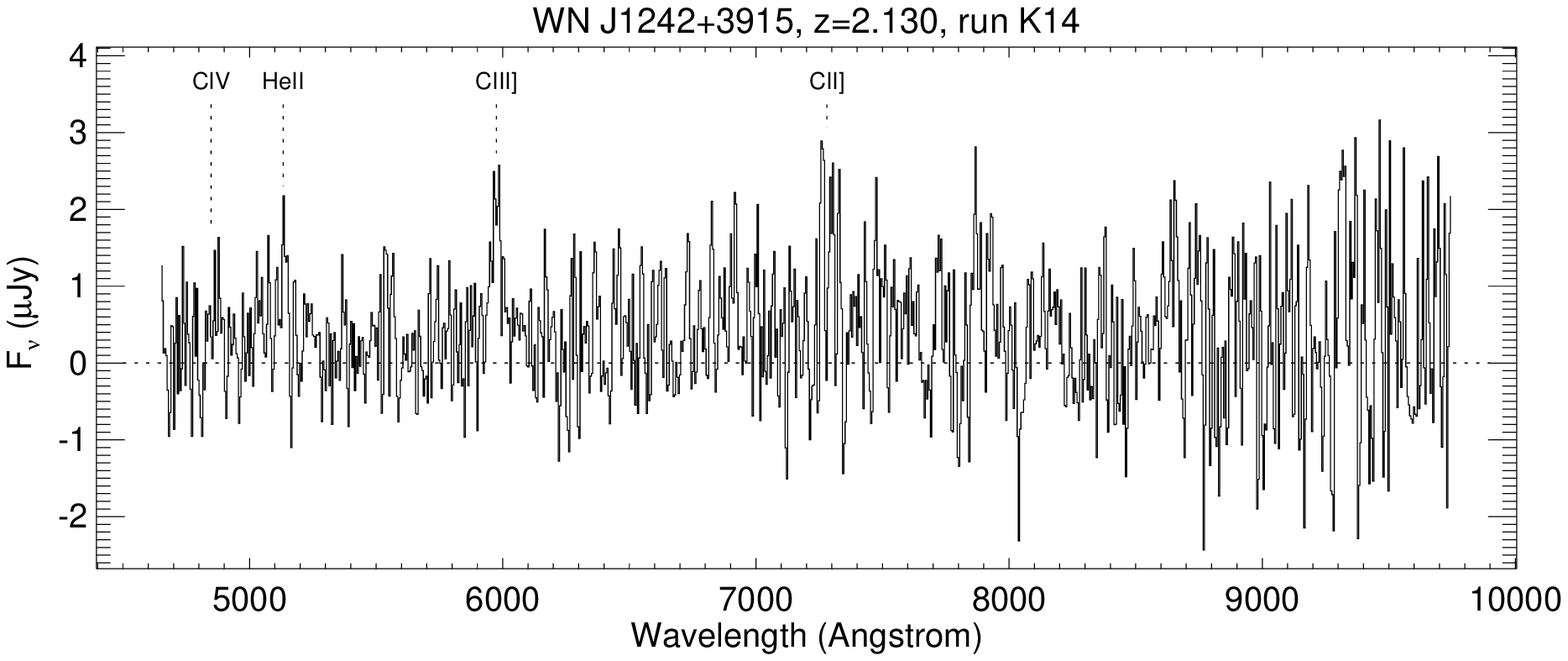,height=6cm}
\psfig{file=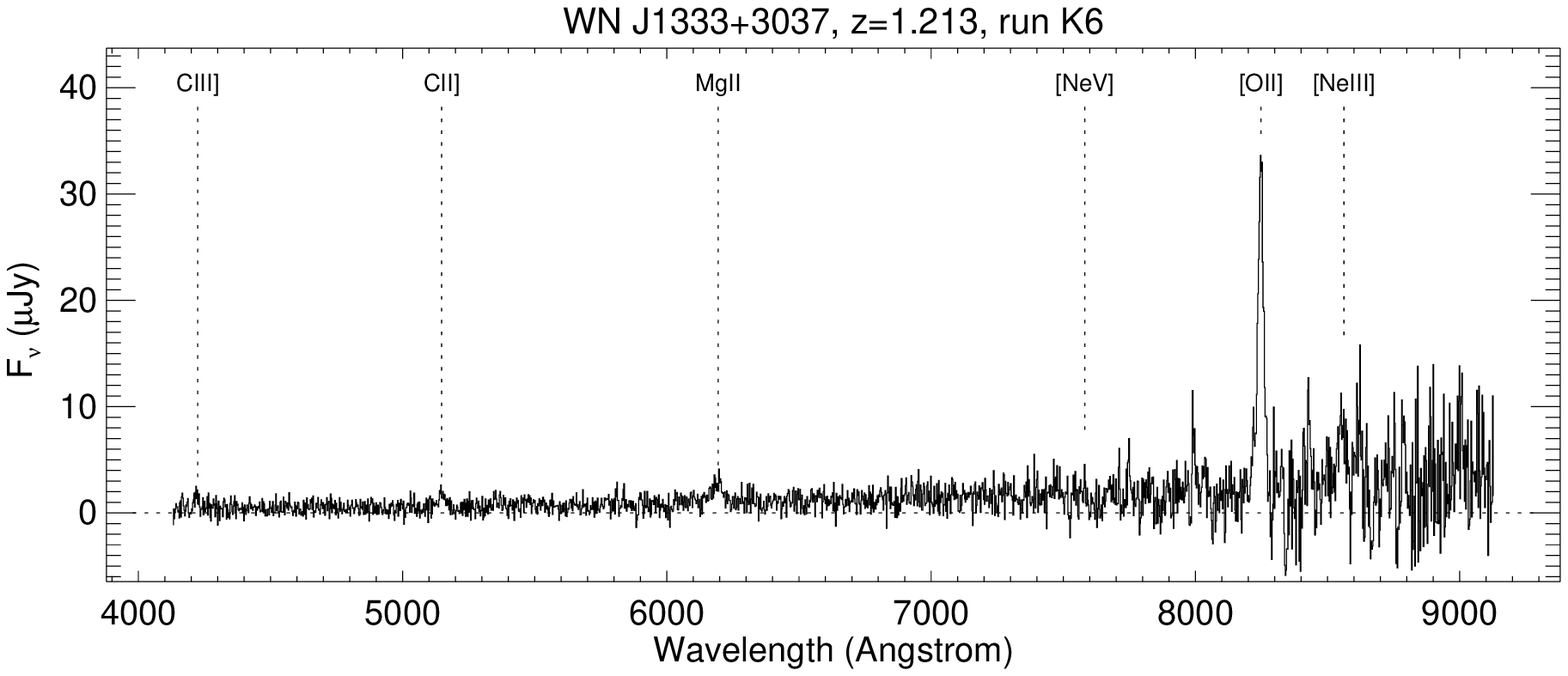,height=6cm}
\psfig{file=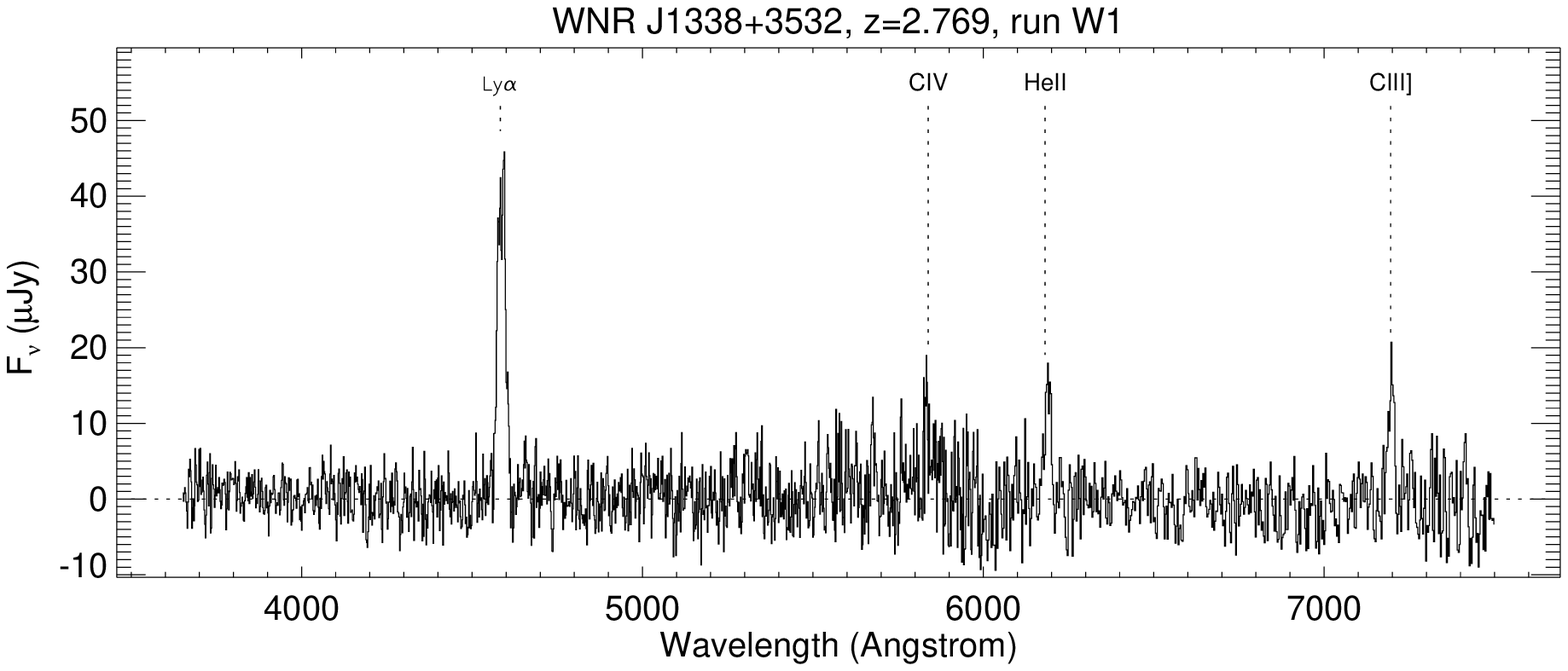,height=6cm}
\psfig{file=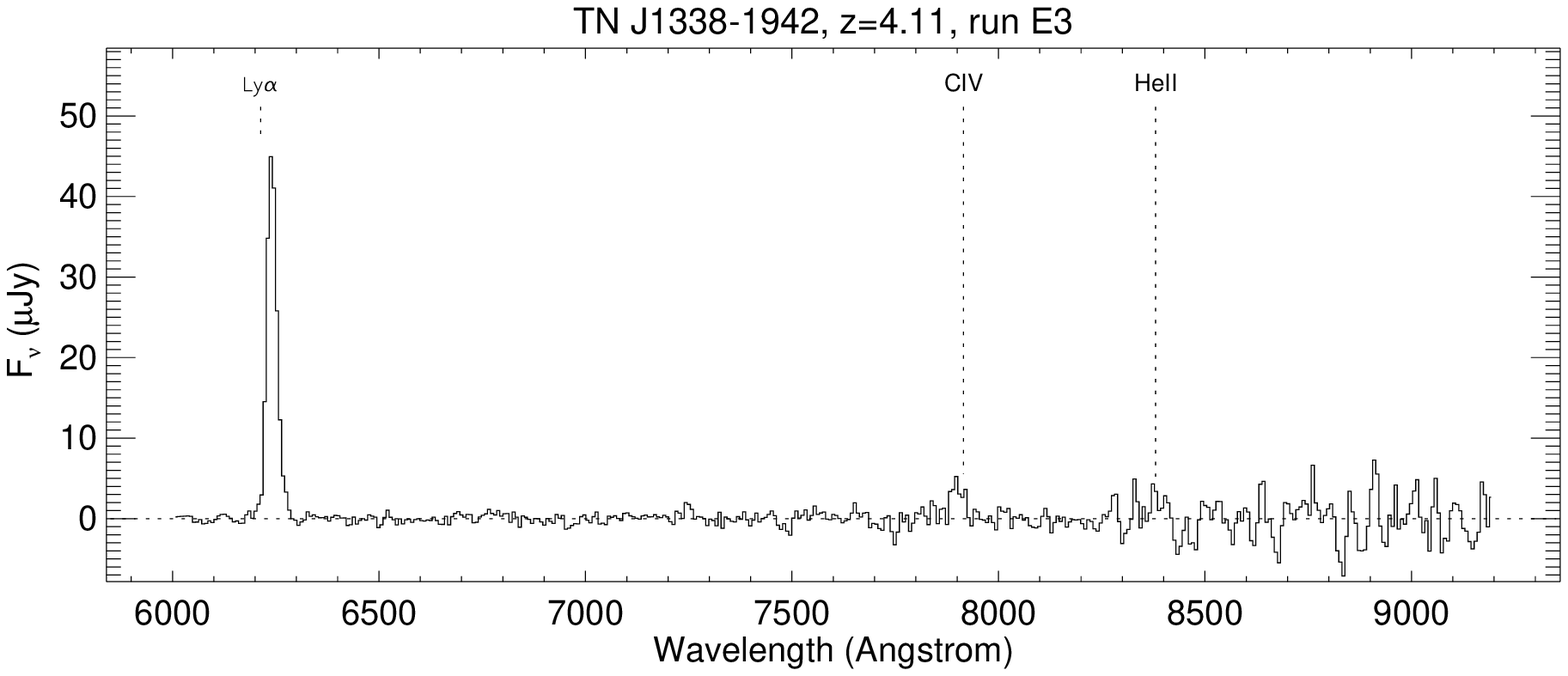,height=6cm}
\psfig{file=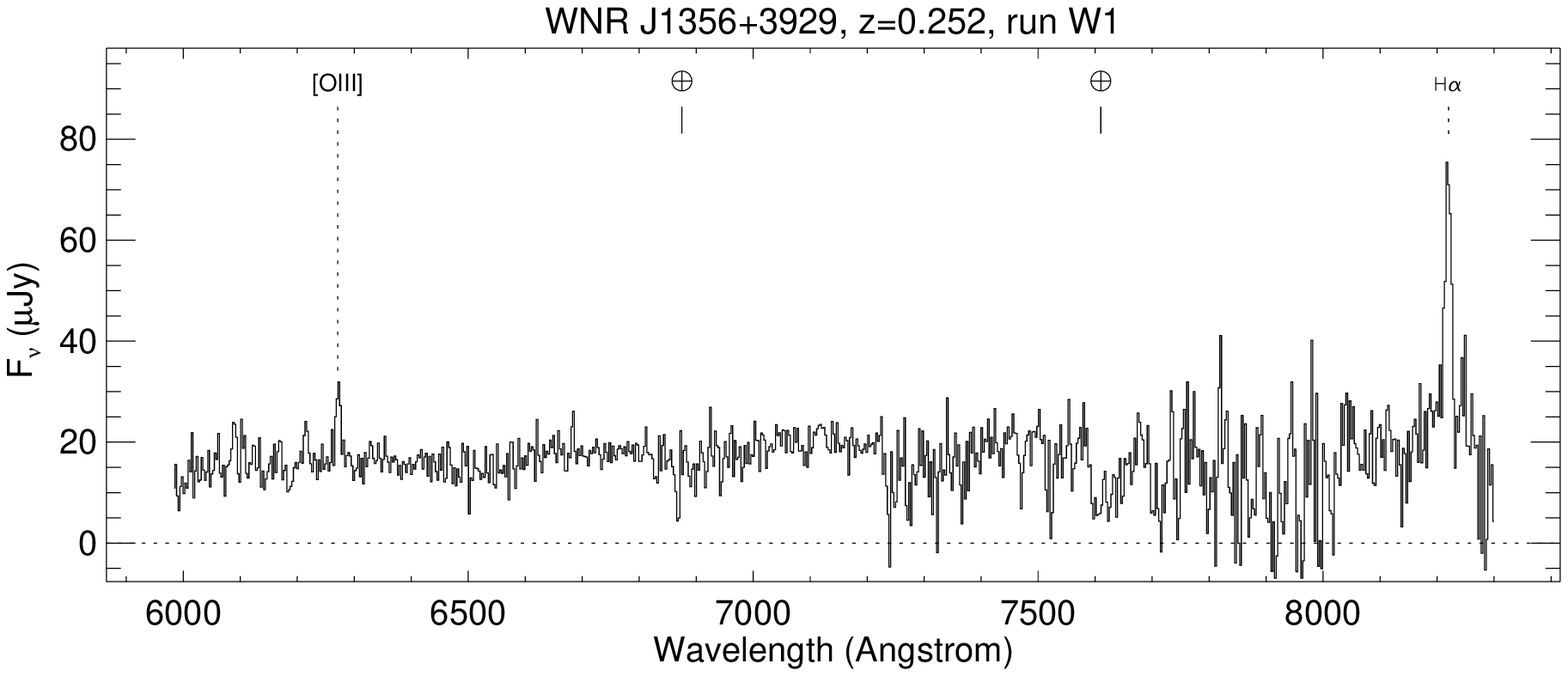,height=6cm}
\psfig{file=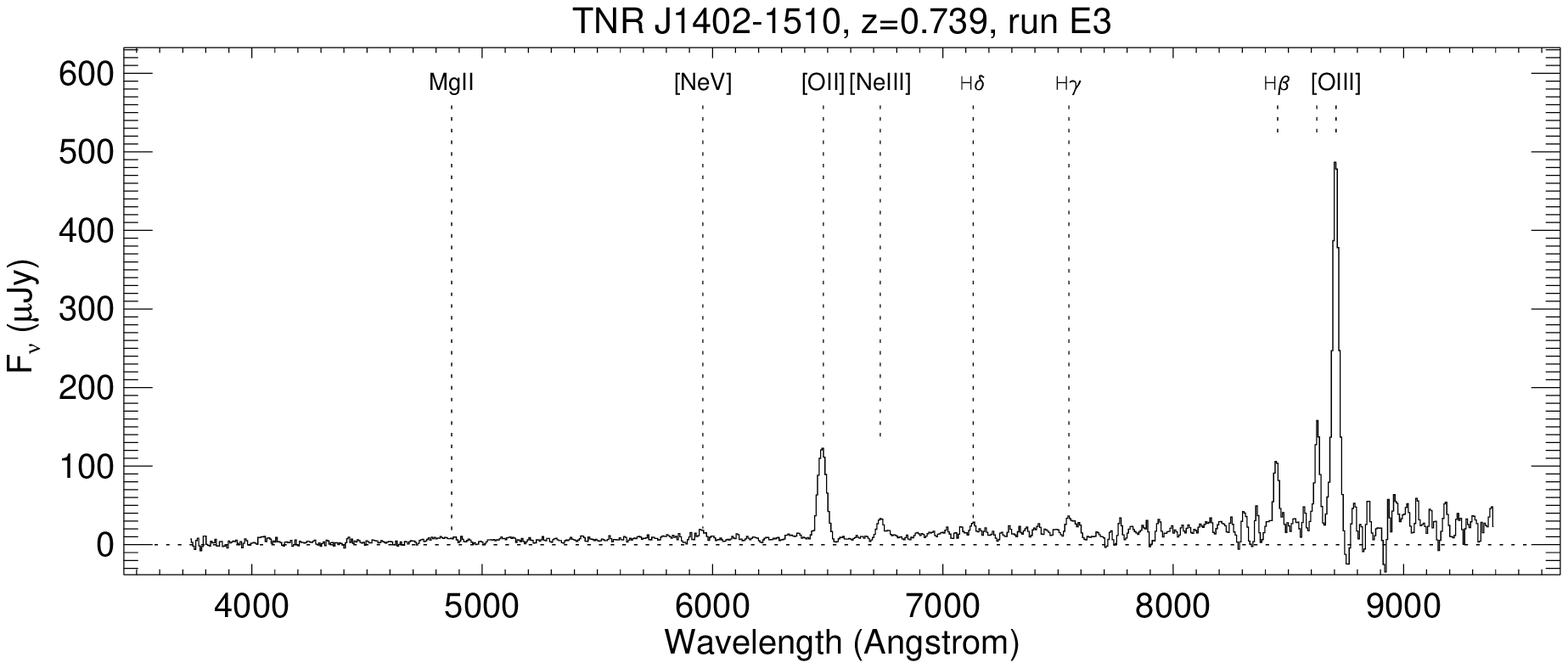,height=6cm}
\psfig{file=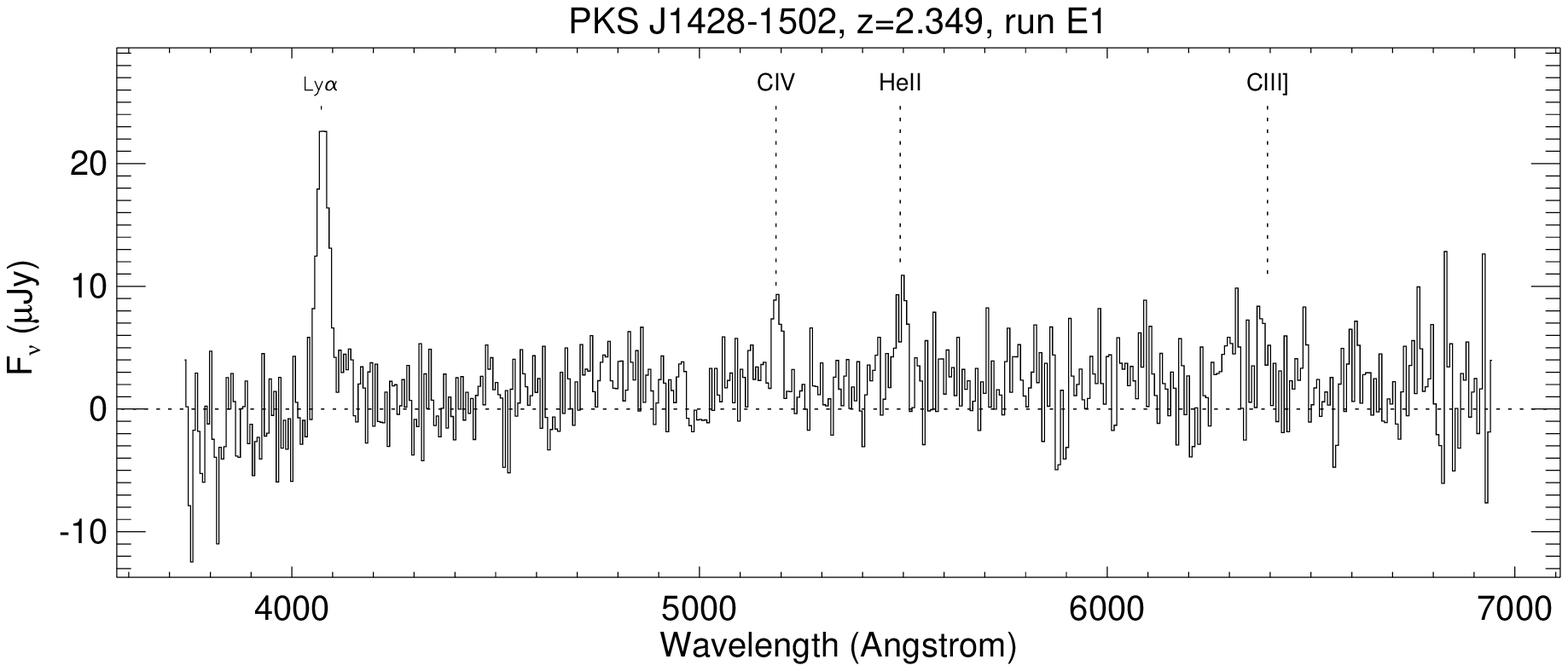,height=6cm}
\psfig{file=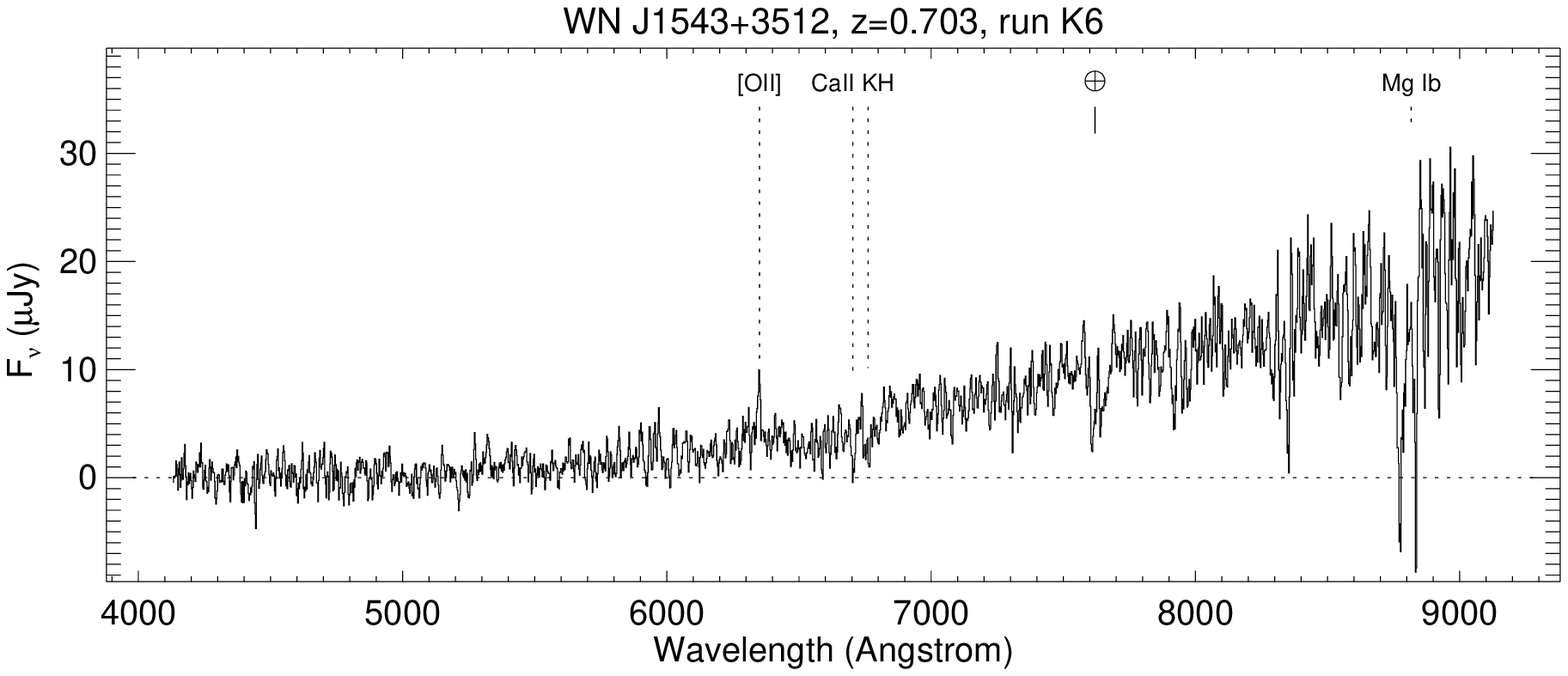,height=6cm}
\psfig{file=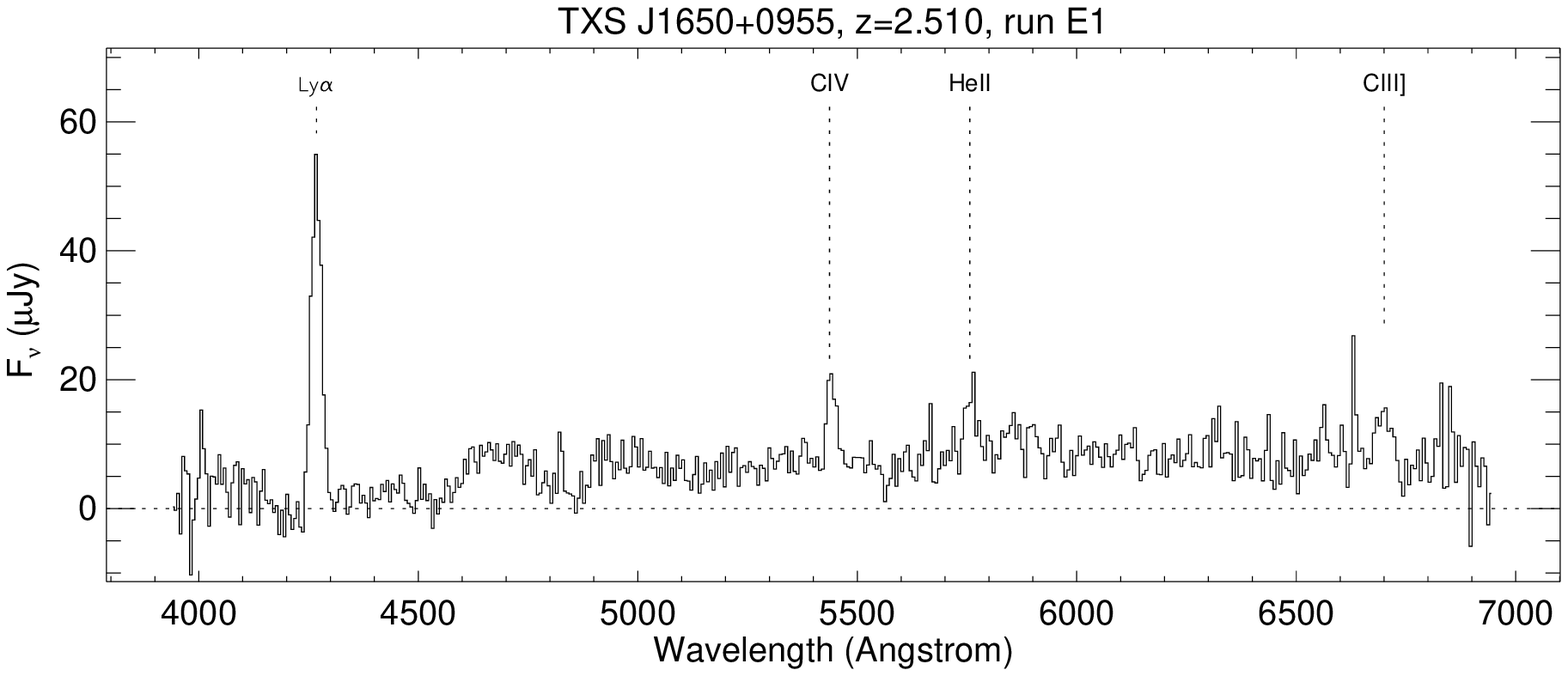,height=6cm}
\psfig{file=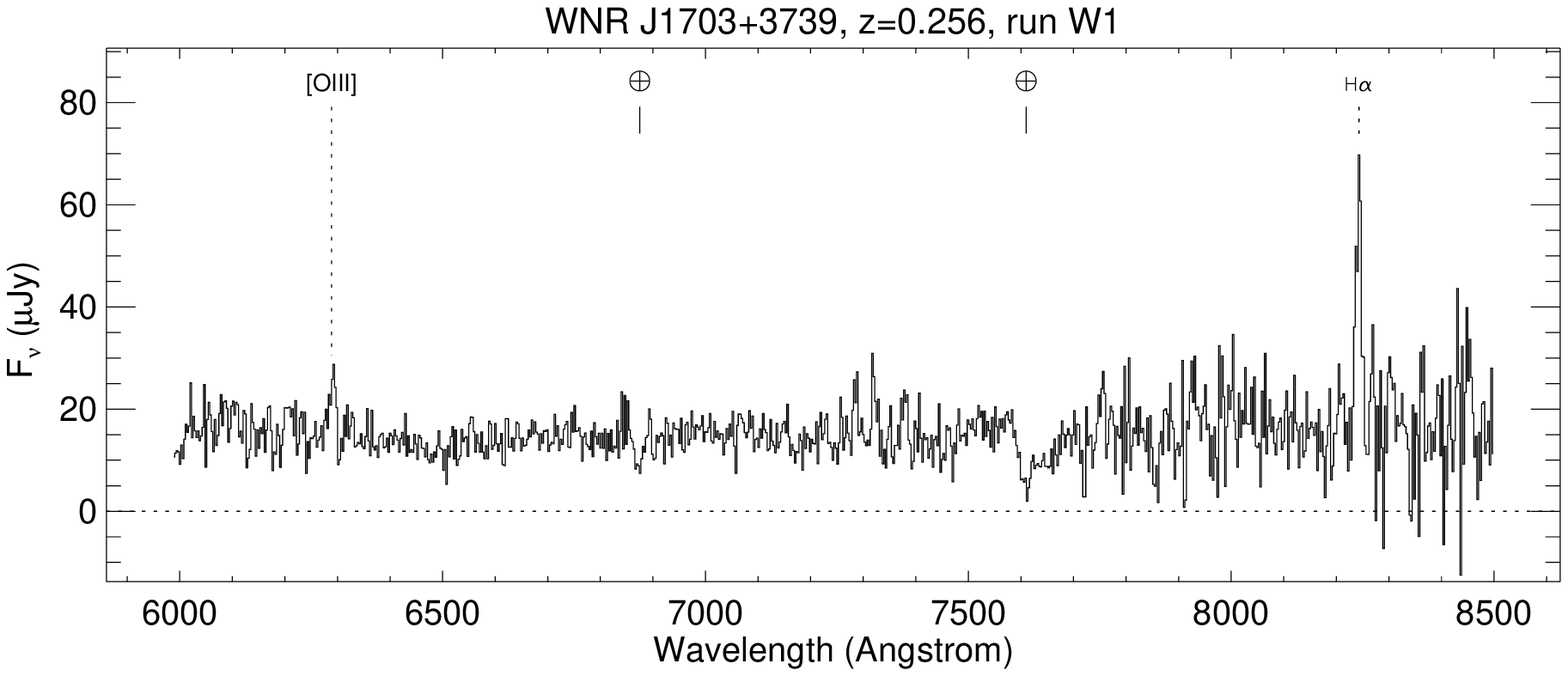,height=6cm}
\psfig{file=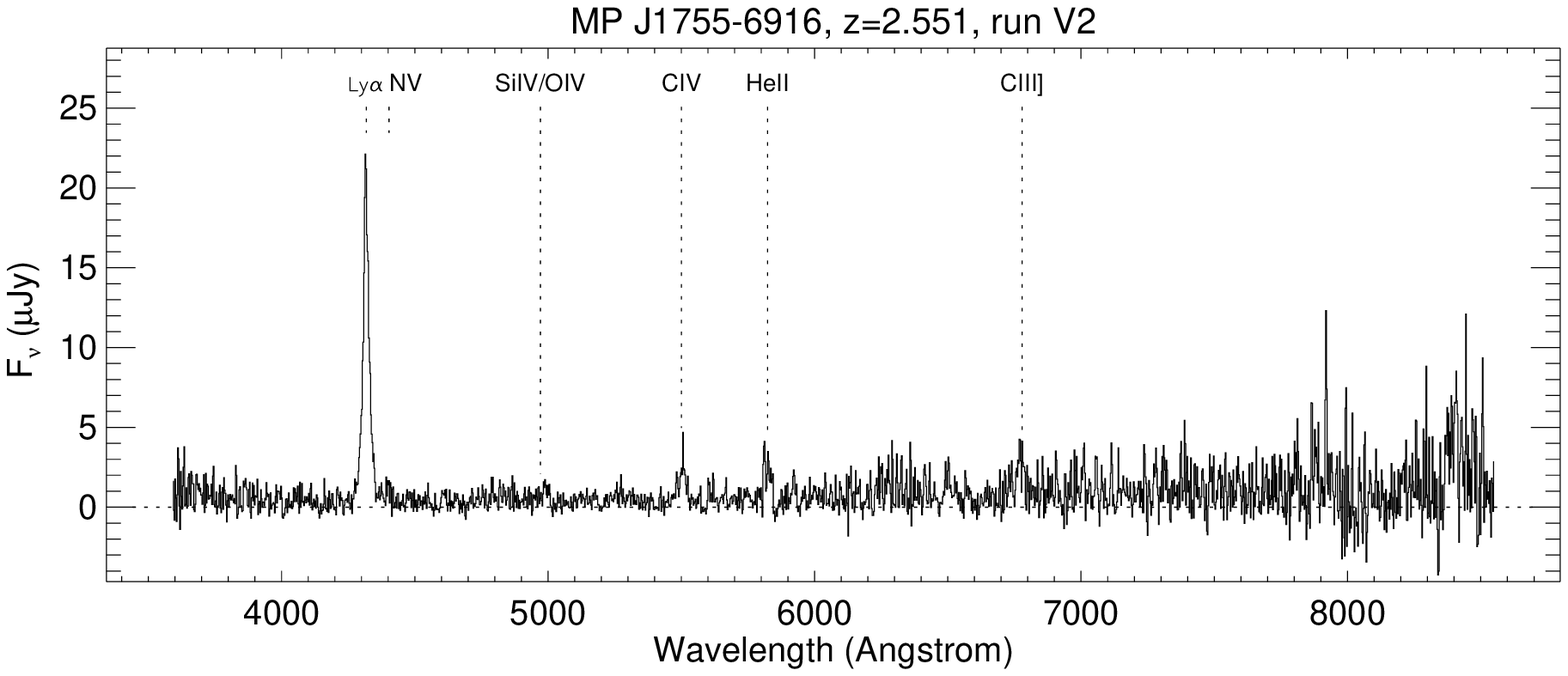,height=6cm}
\psfig{file=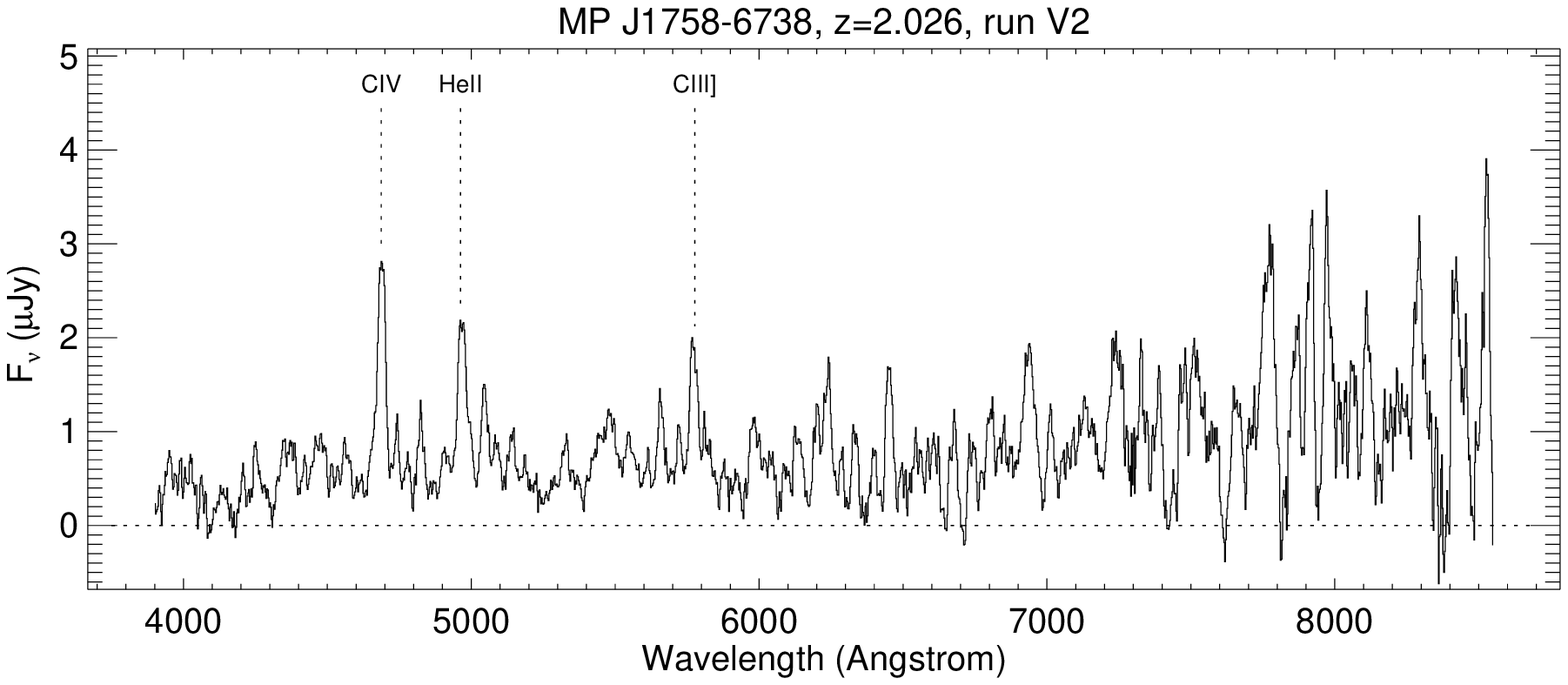,height=6cm}
\psfig{file=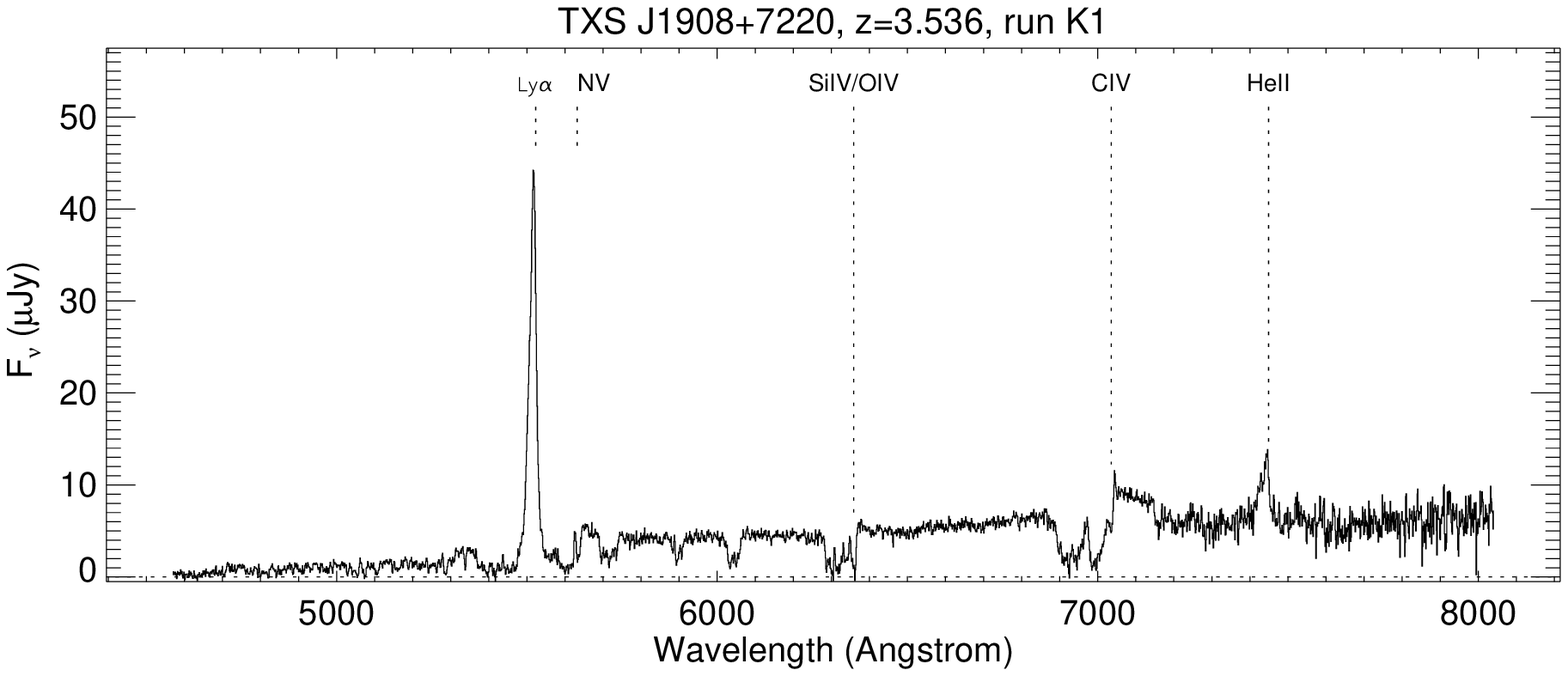,height=6cm}
\psfig{file=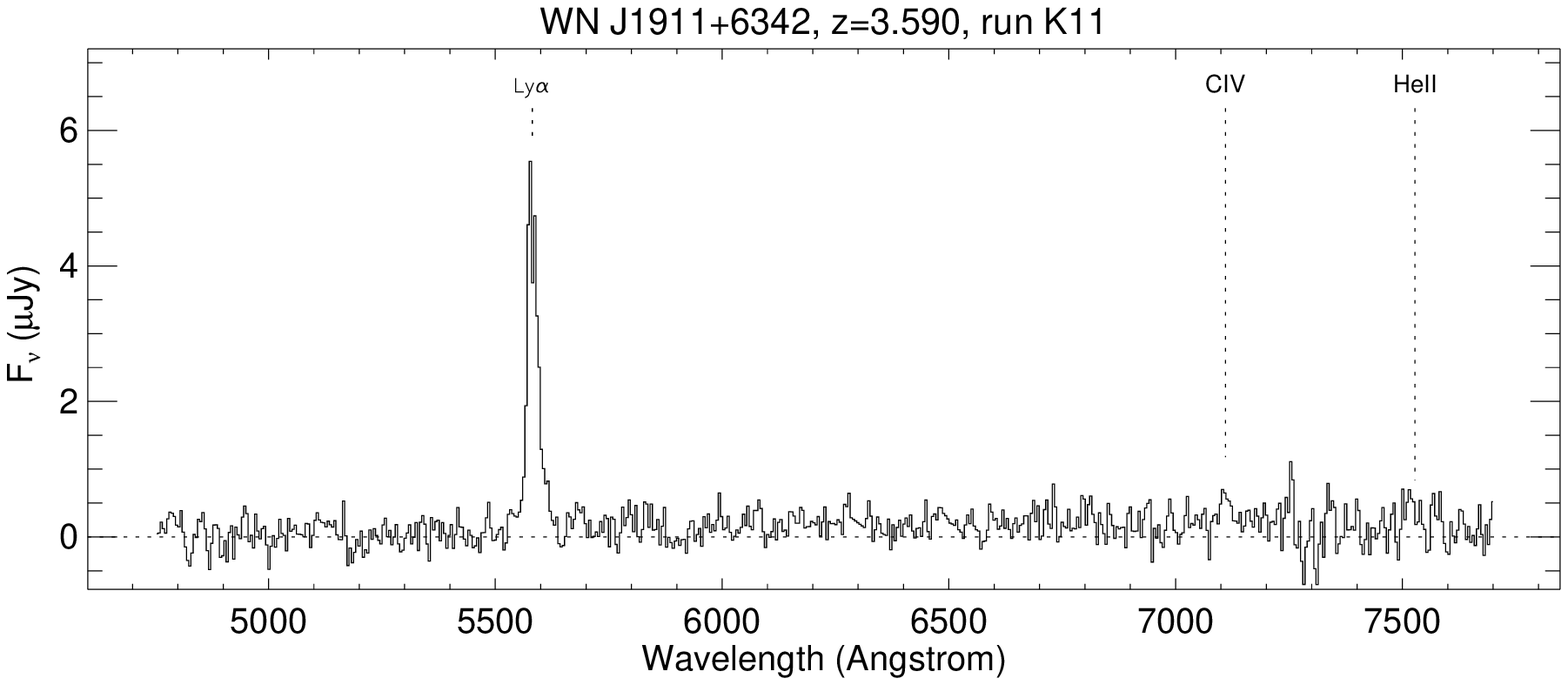,height=6cm}
\psfig{file=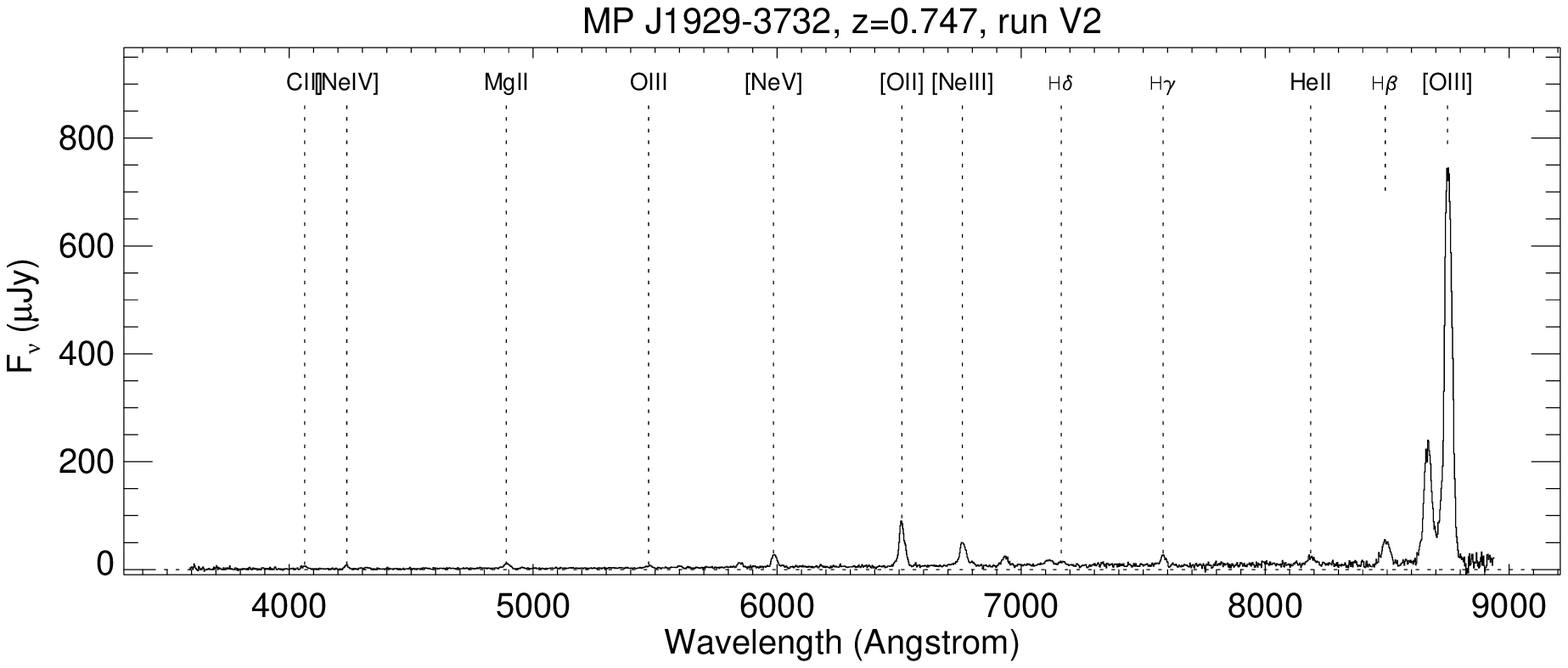,height=6cm}
\psfig{file=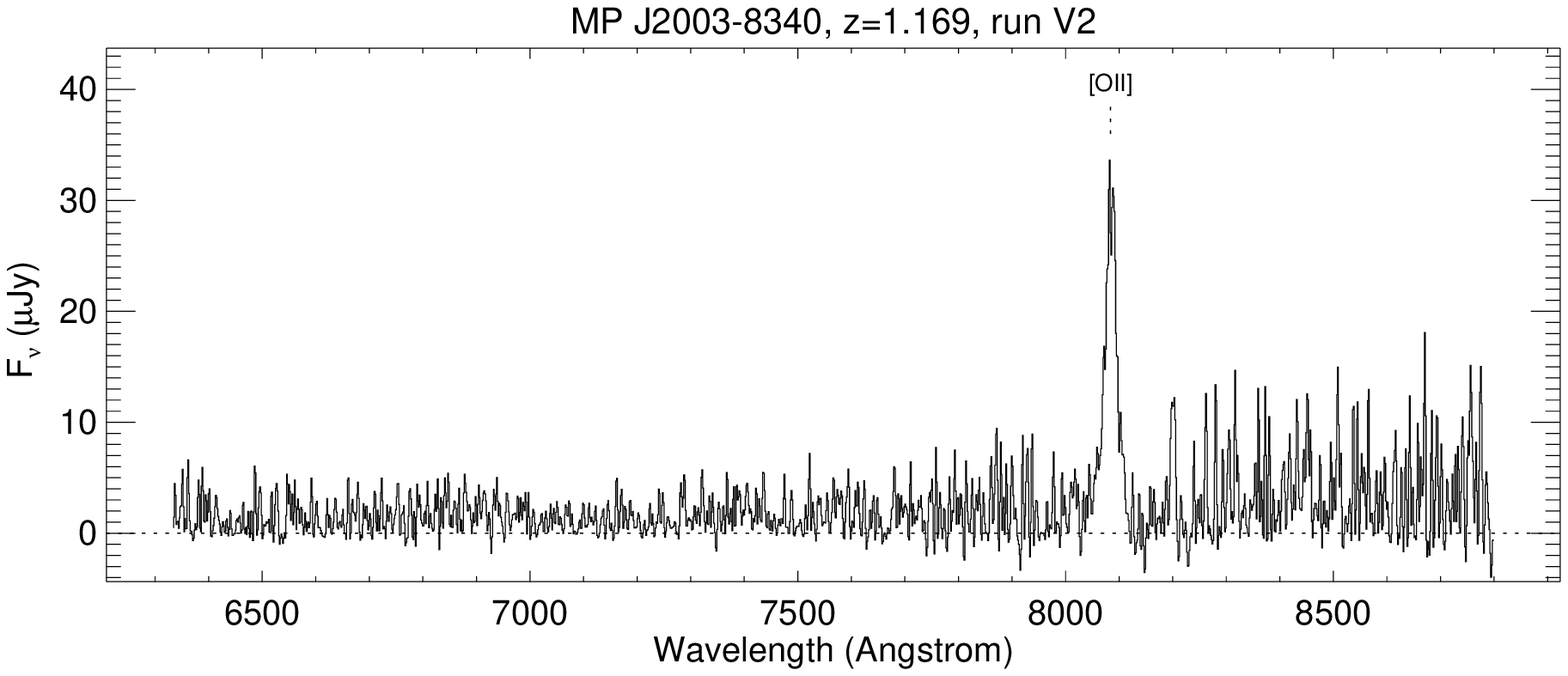,height=6cm}
\psfig{file=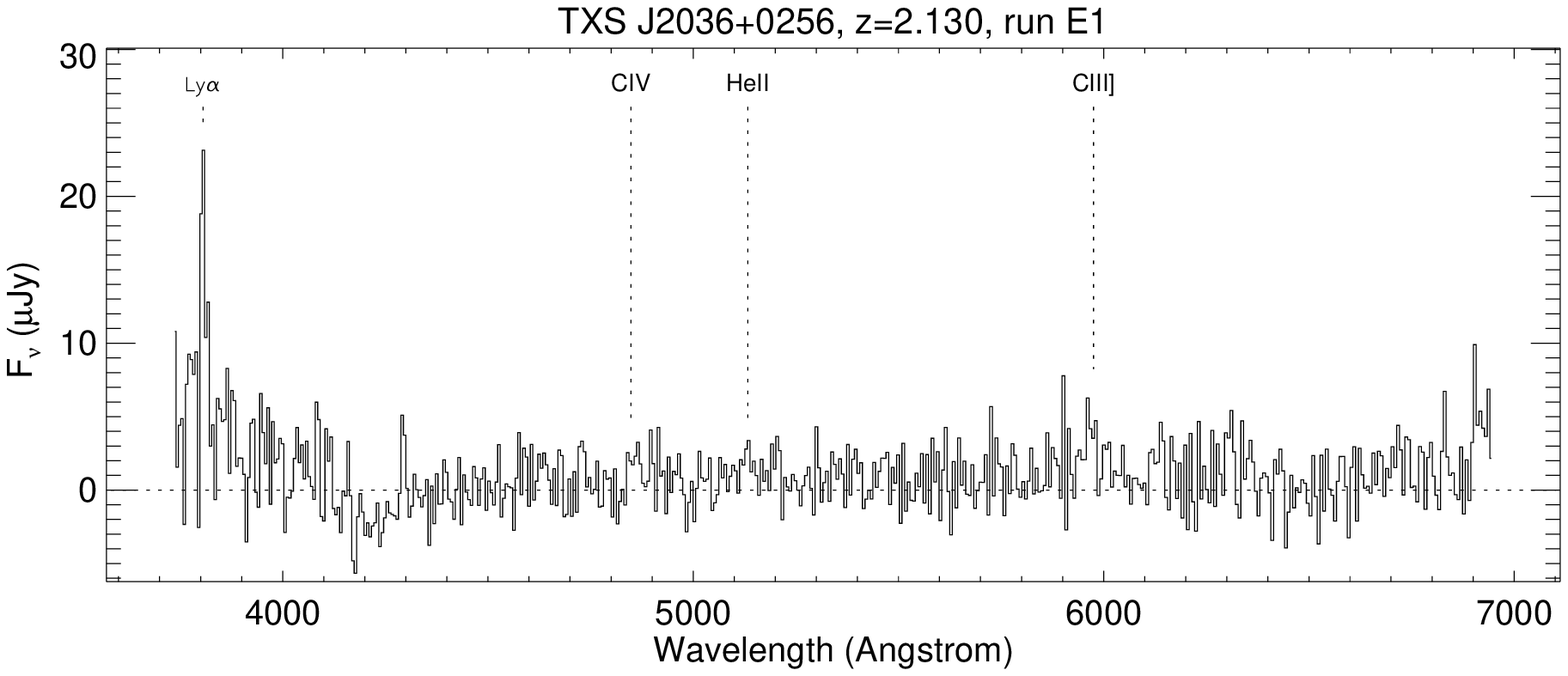,height=6cm}
\psfig{file=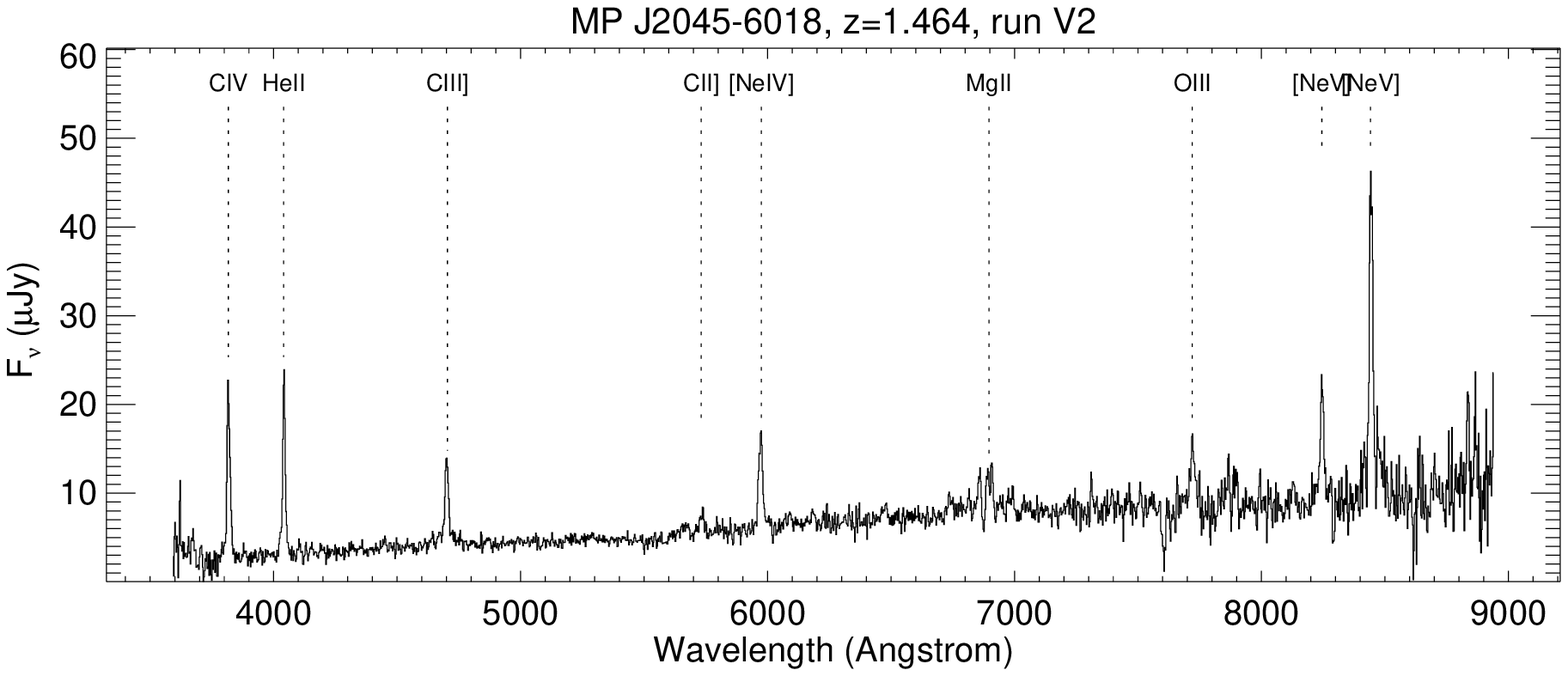,height=6cm}
\psfig{file=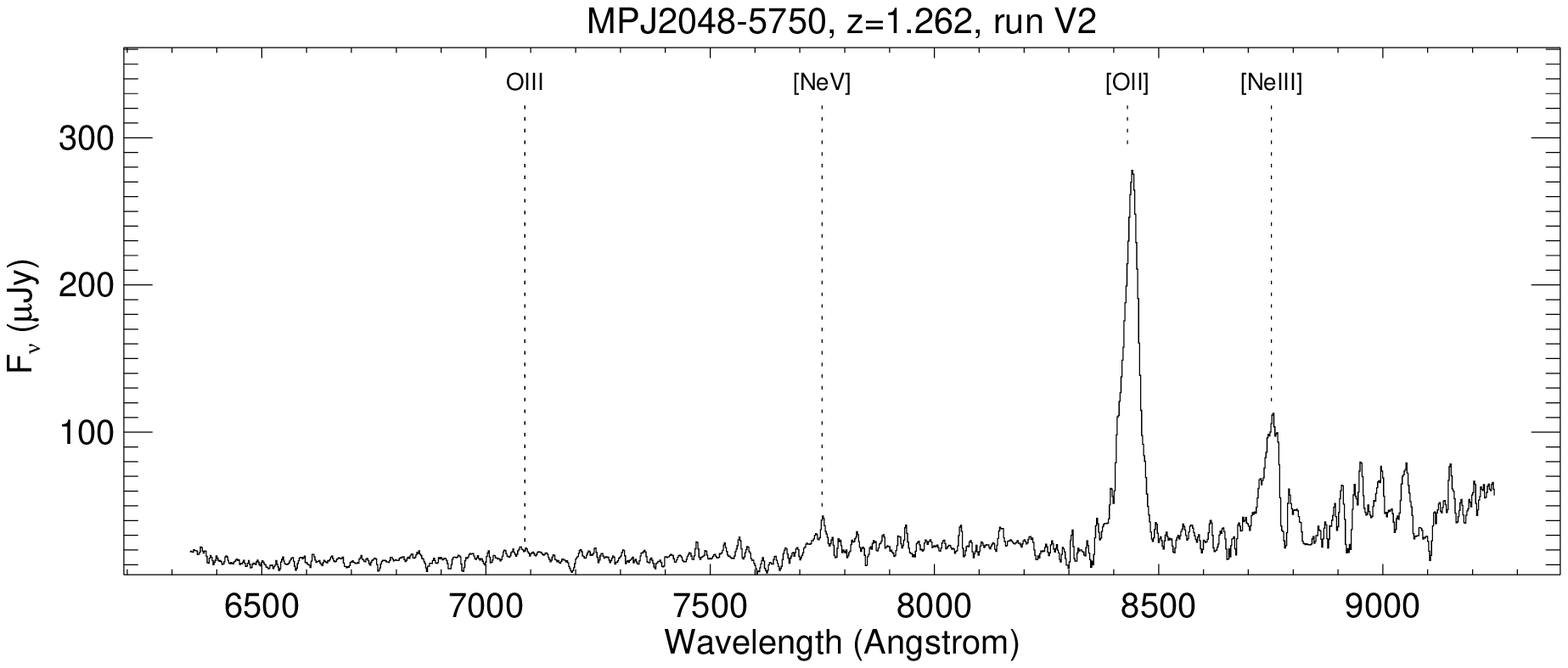,height=6cm}
\psfig{file=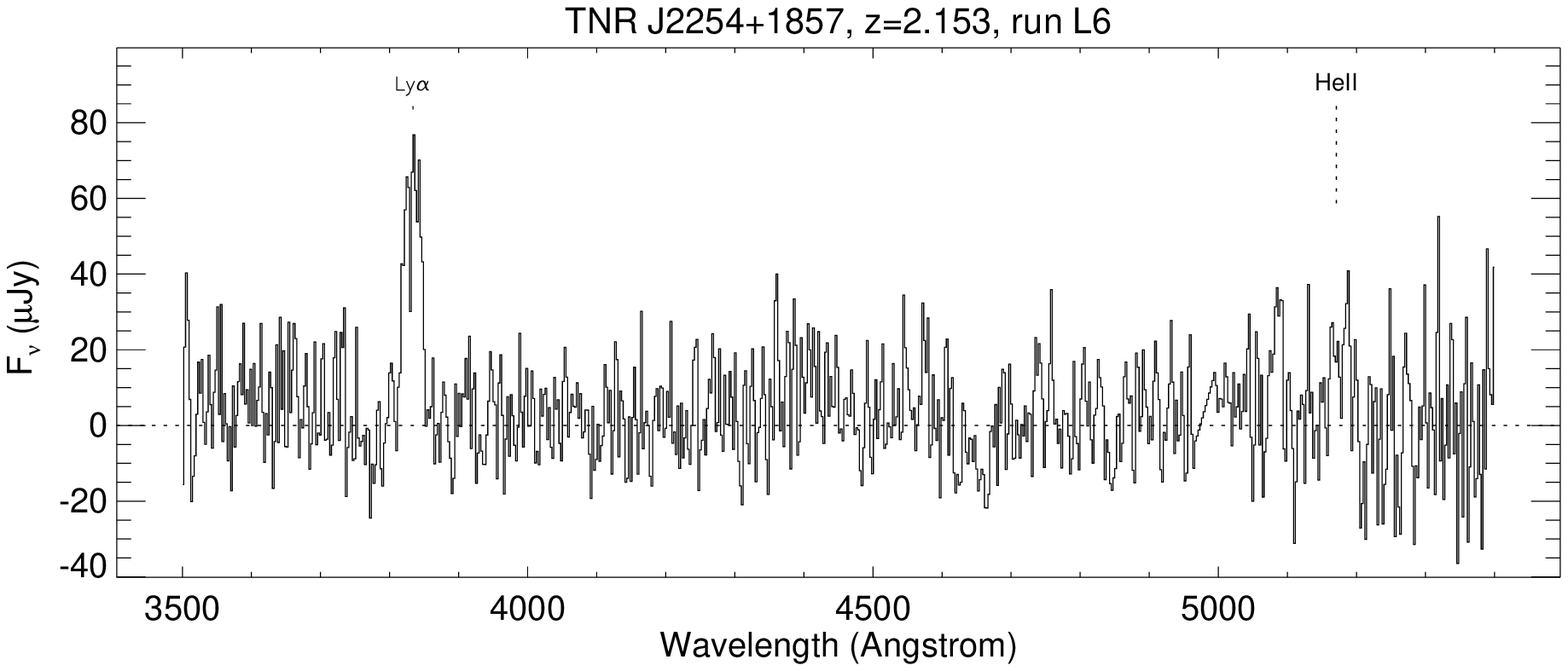,height=6cm}
\psfig{file=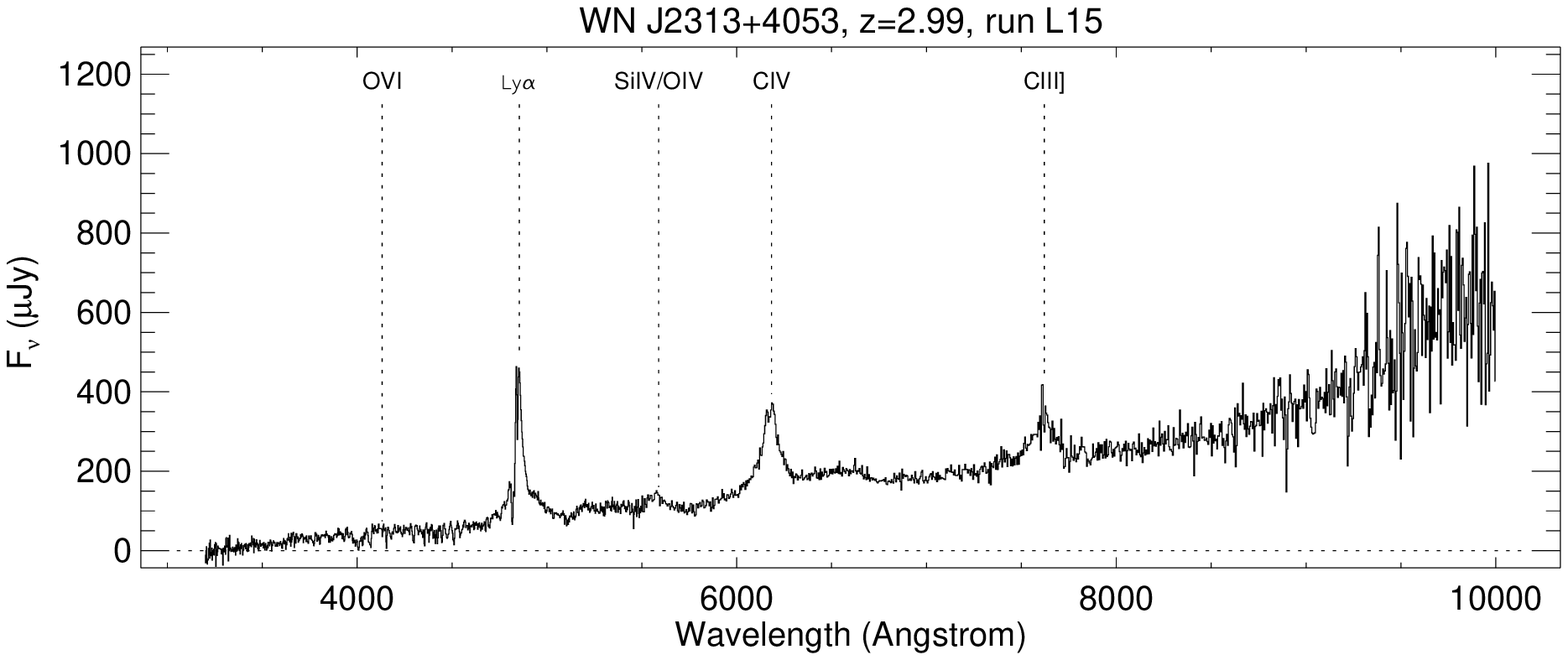,height=6cm}
\psfig{file=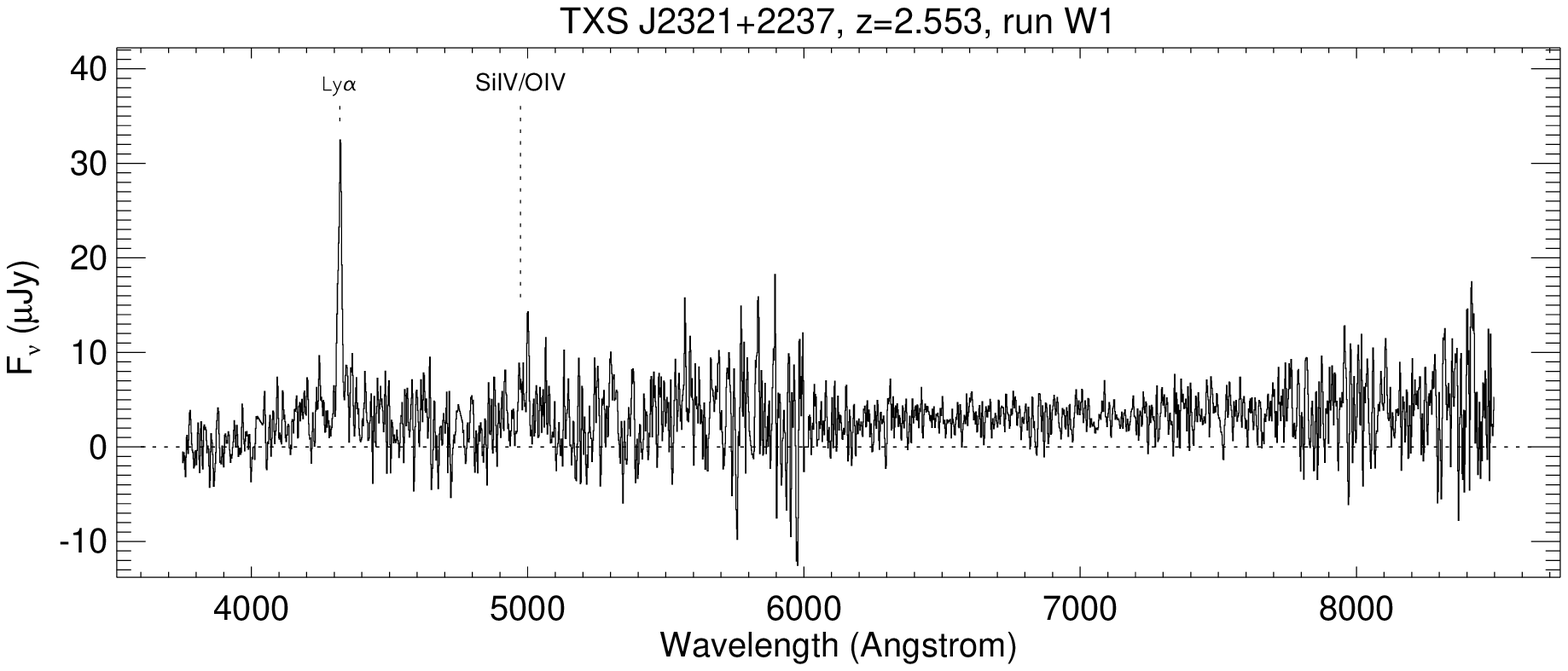,height=6cm}
\psfig{file=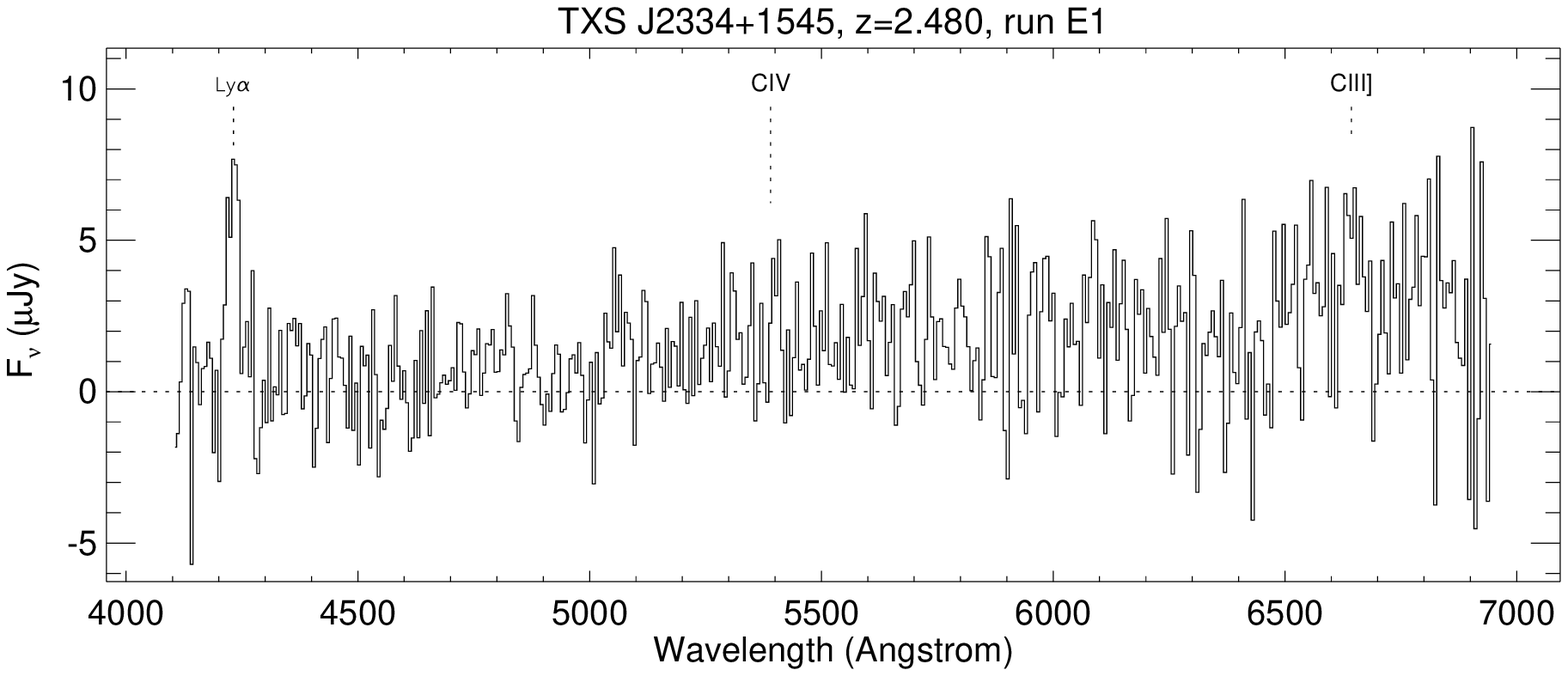,height=6cm}
\psfig{file=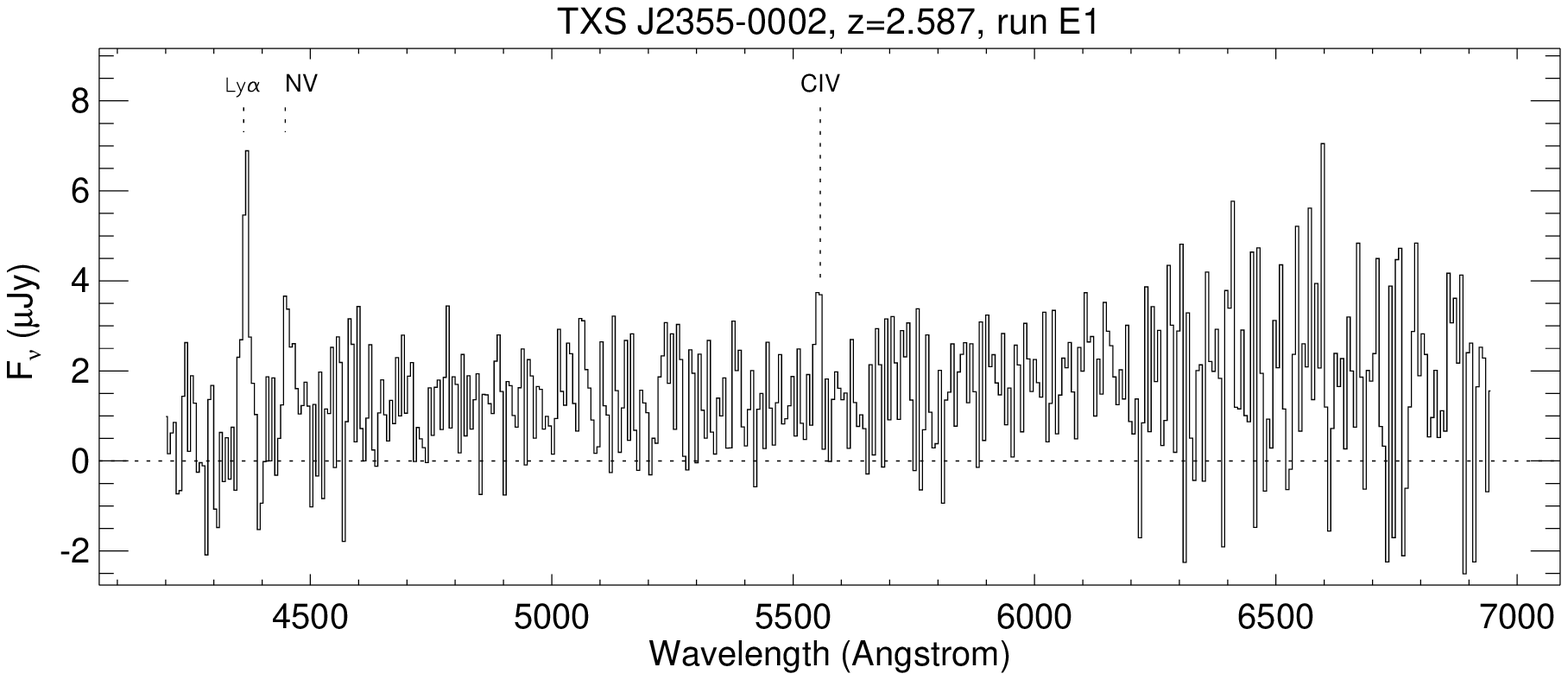,height=6cm}
\addtocounter{figure}{-8}
\figcaption[spectra]{\small Spectra of sources from the USS samples with prominent features indicated. The positions of the vertical dotted lines indicate the predicted observed wavelength of the lines at the redshift quoted on top of each individual spectrum, and not the wavelength of the fitted peak. Differences between the feature and the dotted line thus show velocity shifts of the lines. The source name, redshift, and observing run (see table \ref{specobs}) are shown on top of each spectrum. Telluric OH absorption is indicated by $\oplus$. \label{spectra}} 

\newpage

\section{Appendix}
\appendix
\addtocounter{figure}{7}
\psfig{file=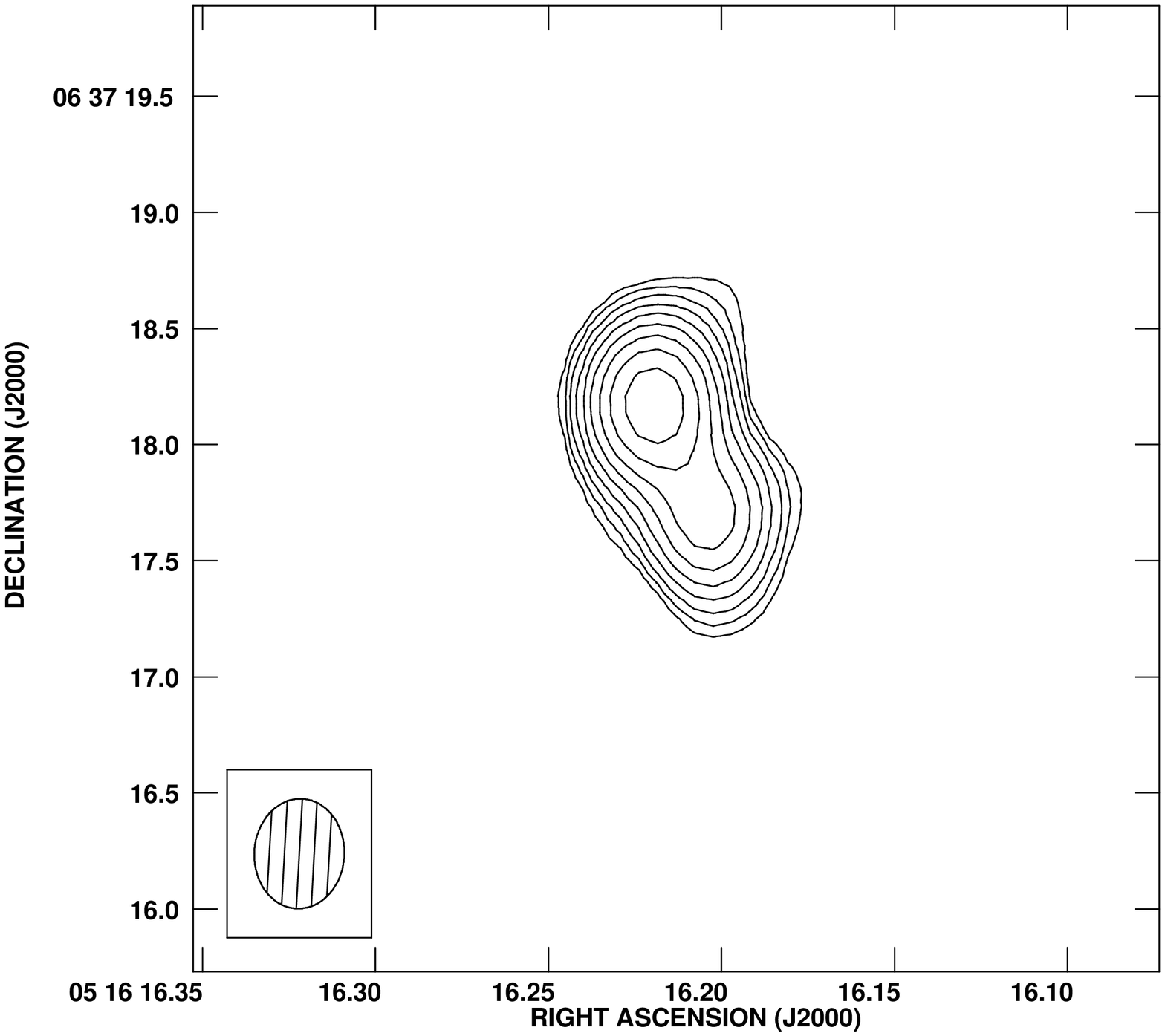,width=16cm,angle=-90}
\figcaption[TNR0516+0637.PS]{\small VLA map of TNR J0516+0637. The contour scheme is a geometric progression in $\sqrt 2$, which implies a factor 2 change in surface brightness every 2 contours. The first contour level is at $3\sigma=0.26$~mJy/beam. The VLA beam is indicated as a hatched ellipse in the lower left corner. }

\psfig{file=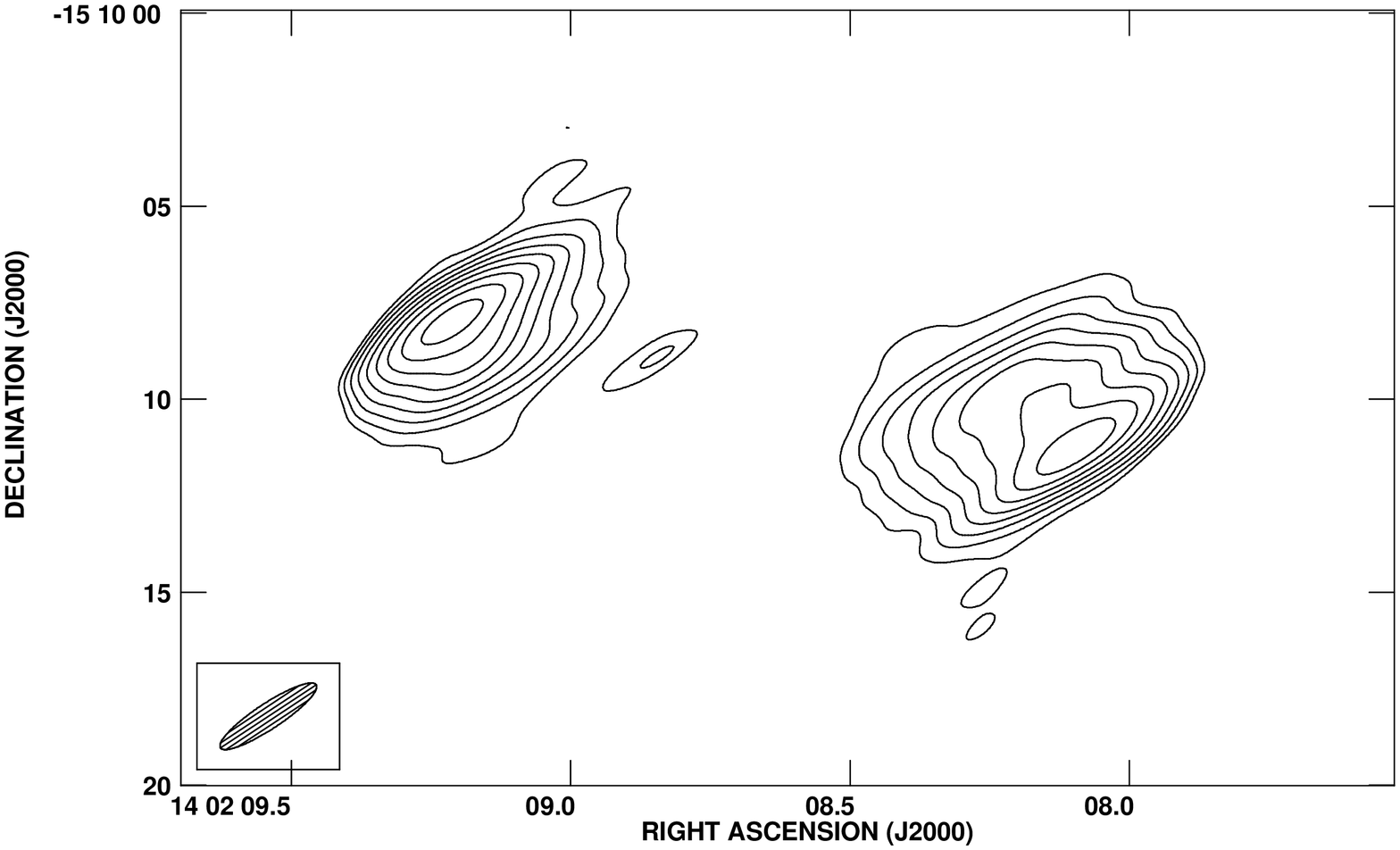,width=16cm,angle=-90}
\figcaption[TNR1402-1510]{\small VLA map of TNR J1402$-$1510. The contour scheme is as for TNR J0516+0637, but with $3\sigma=1.5$~mJy/beam.}

\psfig{file=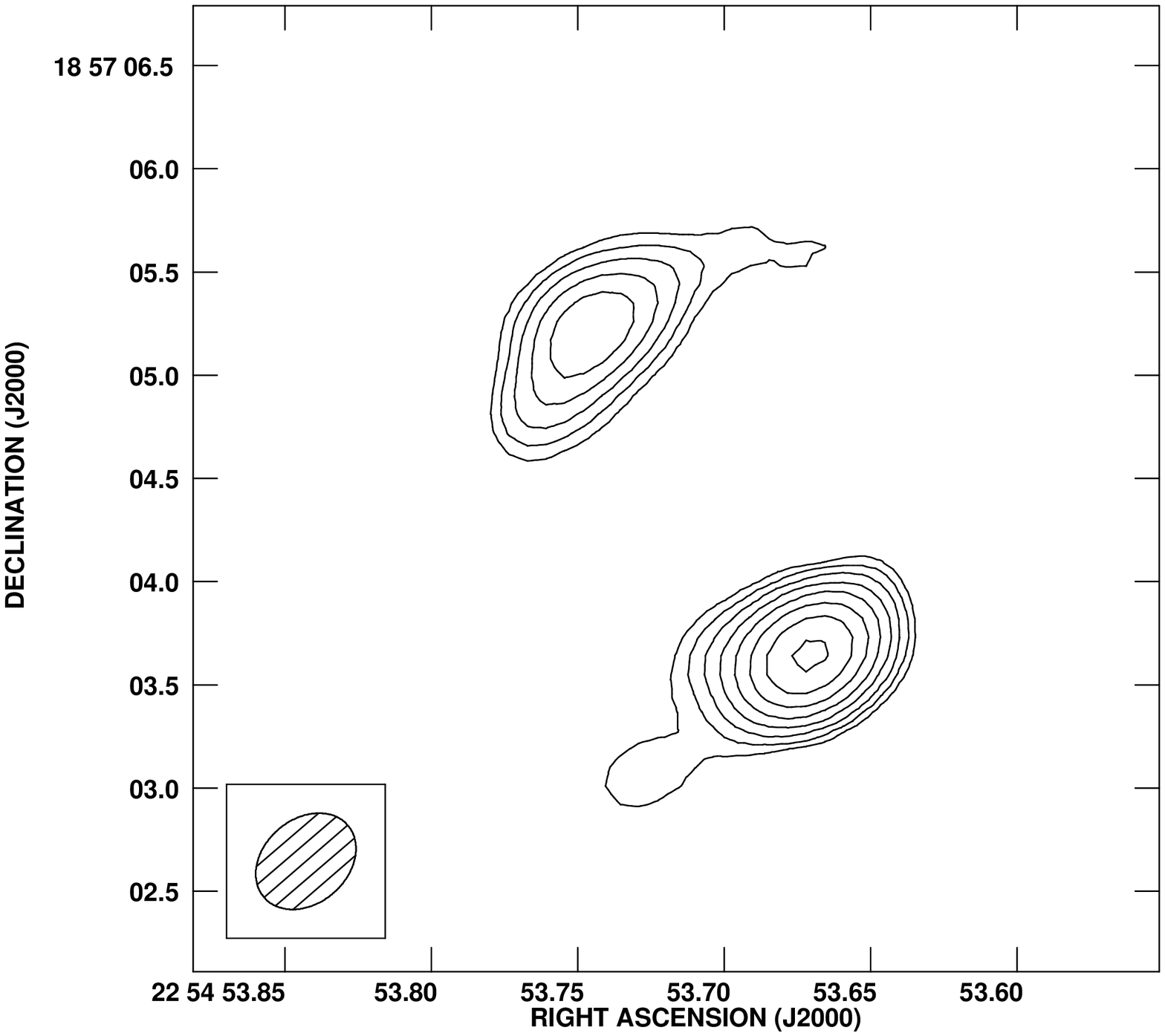,width=16cm,angle=-90}
\figcaption[TNR2254+1857.PS]{\small VLA map of TNR J2254+1857. The contour scheme is as for TNR J0516+0637.}

\end{document}